\documentclass{article}
\usepackage{graphicx} % Required for inserting images
\usepackage{authblk}
\usepackage{here}
\usepackage{amsmath}
\usepackage{multirow}
\usepackage{subcaption}
\title{A High-Speed CGH Calculation Method for Mirror Images on Bézier Surfaces using Optical Path Length Minimization}
\author[1]{KODAI ONO\thanks{ono.kodai.t8@elms.hokudai.ac.jp}}
\author[1]{SEOK KANG}
\author[1]{YUJI SAKAMOTO}

\affil[1]{Graduate School of Information Science and Technology, Hokkaido University, Kita 14, Nishi 9, Kita-ku, Sapporo, Hokkaido, 060-0814, Japan}
\date{December 2025}

\begin{document}

\maketitle

\begin{abstract} 
Rendering reflections in curved mirrors is crucial for enhancing the realism in computer-generated hologram (CGH), yet it poses a fundamental challenge due to the unique computational principles of CGH. Conventional methods using Bézier clipping are computationally prohibitive, and a previously proposed mirror surface subdivision method suffered from the computation time increasing with mirror curvature. To address these limitations, this paper proposes a novel calculation method based on Fermat's principle that directly and efficiently determines the reflection point by minimizing the optical path length from a point light source to a hologram pixel via the mirror surface, using Newton's method for optimization. Experimental results demonstrate that this method significantly reduces computation time compared to previous approaches. Furthermore, it enables the rendering of multiple reflections from several mirrors, a capability that was challenging for conventional techniques.
\end{abstract}

\section{Introduction}

 Computer-generated hologram (CGH) is a technology for calculating and recording optical interference patterns based on an object's three-dimensional information \cite{gabor1948new, waters1966holographic}. This technology enables the safe and easy reconstruction of virtual objects without requiring complex optical systems or chemicals. As the interference patterns are digital data, they can be easily replicated and transmitted. Furthermore, rapidly switching between multiple patterns enables the creation of holographic videos. To enhance the realism of 3D images generated by CGH, the incorporation of rendering techniques from computer graphics (CG) is crucial. Fundamental CG rendering techniques, such as hidden surface removal and shading, have long been developed to generate realistic images \cite{catmull1974subdivision, 10.1145/360825.360839}, and more advanced methods like physically based rendering (PBR), real-time ray tracing, and neural rendering have recently emerged, allowing for more physically accurate and sophisticated representations \cite{dodik2022path, koskela2019blockwise, 10.1145/3503250}. However, applying these techniques directly to CGH is challenging due to a fundamental difference in computational principles: CG typically generates a two-dimensional image from a single viewpoint, whereas CGH must provide smooth motion parallax that accommodates the viewer's eye movements. Indeed, while research is actively exploring the use of neural networks to accelerate CGH calculations, completely reconstructing physically accurate and continuous depth information remains a significant challenge \cite{HosseinEybposh:20}.

Various rendering methods have been proposed to generate realistic 3D images that accommodate the motion parallax inherent in CGH. Ichikawa et al. proposed a ray-tracing method originating from the center of each elemental hologram, but this approach resulted in discontinuous parallax \cite{Ichikawa:13}. To overcome this issue, Watanabe et al. enabled ray tracing on a pixel-wise basis on the hologram plane, achieving smooth and continuous motion parallax \cite{Watanabe:24}. Other research has focused on improving the accurate reconstruction of the optical properties of object surfaces. Yamaguchi et al. proposed a method to represent diverse reflection properties by applying CG reflection models such as those of Blinn and Torrance-Sparrow \cite{Yamaguchi:09}. Focusing on the representation of transparent objects, Nishi et al. proposed a method based on wave optics to simulate light refraction by switching wavelengths at media boundaries \cite{Nishi:25}. However, although Watanabe et al.’s method achieved continuous parallax and can represent reflections from a planar mirror, it does not provide a calculation method for non-planar mirror shapes. Overcoming this limitation is the next significant research challenge.

In the real world, curved mirrors are ubiquitous, ranging from those used in optical experiments (such as concave, convex, and cylindrical mirrors) to road reflectors, brass instruments, and even highly designed art objects. Therefore, the ability to render reflections in curved mirrors is essential for recreating realistic scenes. However, calculating curved mirror reflections in CGH involves inherent difficulties distinct from those in conventional CG. In conventional CG, the standard ray-tracing method is a viewpoint-based approach that traces rays from the observer; consequently, the intersection of a ray with the mirror surface is trivially determined as the reflection point \cite{10.1145/1198555.1198743}. In contrast, calculation for CGH requires finding the path of a light ray that travels from a specific point light source, reflects off the mirror surface, and arrives at a specific pixel on the hologram plane. While the start and end points of this path are known, the intermediate reflection point on the mirror surface is not. This search for an unknown reflection point constitutes the fundamental difficulty of calculating mirror images for CGH.

To address this challenge, Arai et al. proposed a method to approximately calculate reflection points on mirrors defined by Bézier surfaces, using an approach based on Bézier clipping \cite{10.1145/97880.97916, Arai:24}. However, their algorithm was complex and computationally expensive. Point-based CGH calculation is inherently suitable for GPU parallelization, as the computation for each pixel is independent. Although Arai et al. attempted to accelerate their implementation by using CUDA, it still did not reach practical processing speeds \cite{Lee:14}. In our own previous study, we proposed a ``mirror surface subdivision method'' that divides a Bézier surface into numerous small triangular polygons and applies Watanabe et al.'s fast algorithm for planar mirrors to each one \cite{10.1117/1.OE.64.7.075102}. While this method achieved significant acceleration compared to that of Arai et al., it introduced a new limitation: for surfaces with multi-directional curvature, such as convex and concave mirrors, maintaining accuracy requires increasing the number of subdivisions, which in turn leads to a proportional increase in computation time.

Therefore, in this study, we propose a novel physically based approach to resolve the trade-off between accuracy and computation time inherent in the mirror surface subdivision method. Our method employs a gradient-based search to efficiently find the point on the mirror surface that minimizes the optical path length from the point light source to the hologram pixel, in accordance with Fermat's principle \cite{goodman2005introduction, Born_Wolf_Bhatia_Clemmow_Gabor_Stokes_Taylor_Wayman_Wilcock_1999}. This approach aims to consistently reduce computation time, irrespective of the mirror's curvature. To validate our proposed method, we conduct optical experiments to evaluate the relationship between the accuracy of the resulting mirror image and the termination condition of the gradient method. The results demonstrate that our method can accurately compute images for various mirror types, including convex, concave, and cylindrical surfaces. Moreover, we show that it can render multiple reflections from several mirrors—a capability not achieved by conventional methods.

\section{Conventional Method}
\subsection{Point-based method for CGH calculation}

\begin{figure}[t]
\centering
\includegraphics[clip,width=10.0cm]{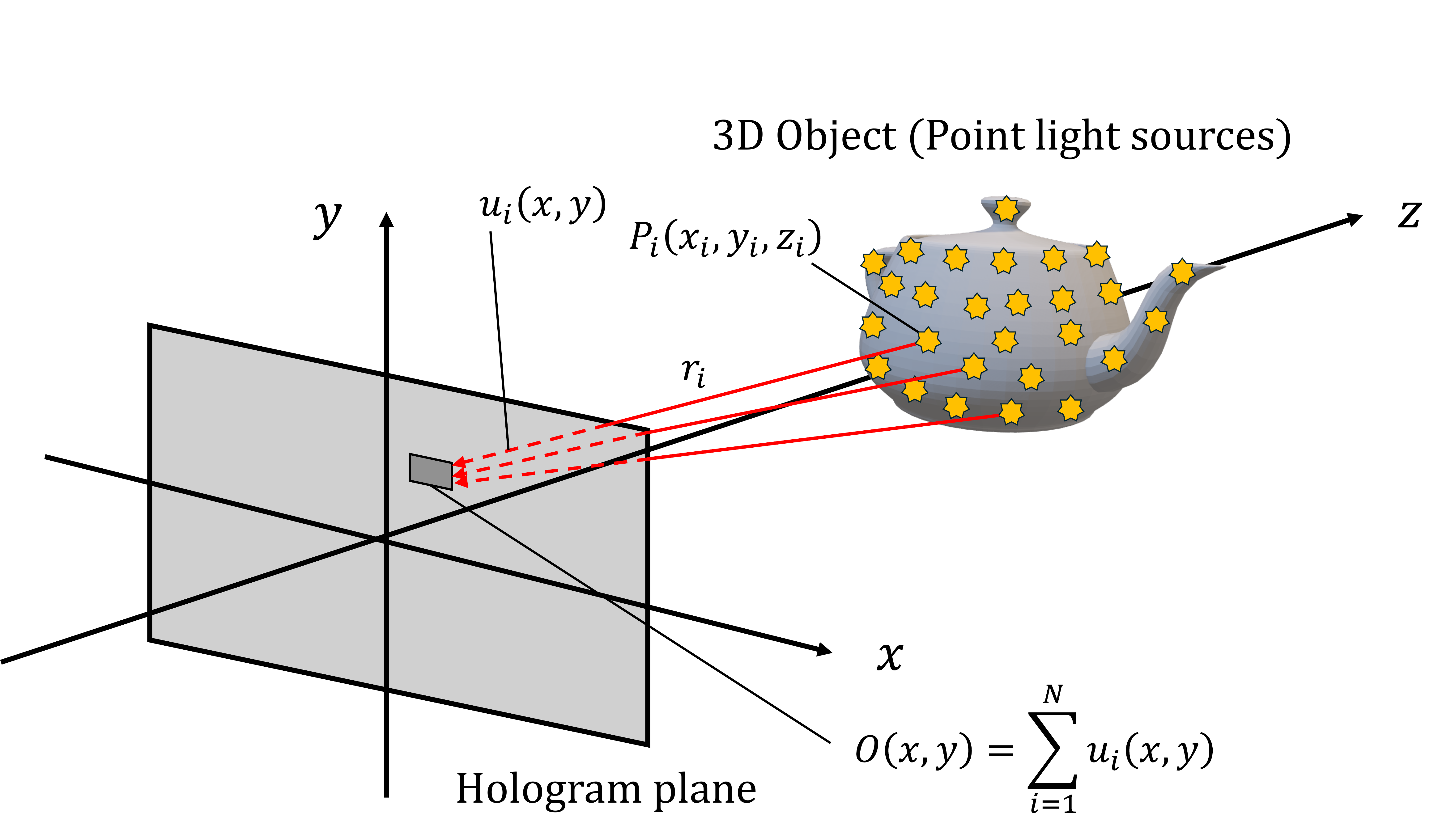}
\caption{Conceptual diagram of CGH calculation using the point-based method. The light waves here propagate from a point light source $(x_i, y_i, z_i)$ on the object surface to each pixel $(x, y)$ on the hologram plane.}
\label{fig:point-based_method}
\end{figure}

The point-based method approximates a three-dimensional object as a collection of numerous point light sources placed on the object surface, as shown in Fig. \ref{fig:point-based_method}. The light waves propagating from each point light source to the pixel on the hologram plane at coordinates $(x,y)$ are calculated and then superimposed to obtain the object light. While the point-based method offers the flexibility to represent arbitrary object shapes, it has a significant drawback in that the computational cost increases proportionally with the number of point light sources. If the coordinates of the $i$-th point light source are $(x_i, y_i, z_i)$, the complex amplitude $u_i(x,y)$ formed on the hologram plane by the spherical wave emitted from this point light source is expressed as

\begin{align}
u_i(x, y) &= \frac{a_i}{r_i}\exp\{-j(kr_i+\phi_i)\}\notag \\
 r_i &= \sqrt{(x-x_i)^2 + (y-y_i)^2 + z_i^2}\notag \\
 k &= \frac{2\pi}{\lambda}\notag,
\end{align}
where $r_i$ is the distance from the point light source to each pixel on the hologram plane, $a_i$ is the amplitude of the point light source, $k$ is the wave number, $\phi_i$ is the initial phase of the point light source, $\lambda$ is the wavelength of light, and $j$ is the imaginary unit. For a total of $N$ point light sources, the complex amplitude distribution of the object light, $O(x,y)$, is the superposition of all individual waves, given by

\begin{equation*}
 O(x, y) = \sum^N_{i=1}u_i(x, y).
\end{equation*}

Finally, the hologram data is generated by simulating the interference between this object light, $O(x,y)$, and a reference light.

\subsection{Calculation of specular reflection from planar mirrors by pixel-wise ray tracing}

\begin{figure}[t]
\centering
\includegraphics[clip,width=10.0cm]{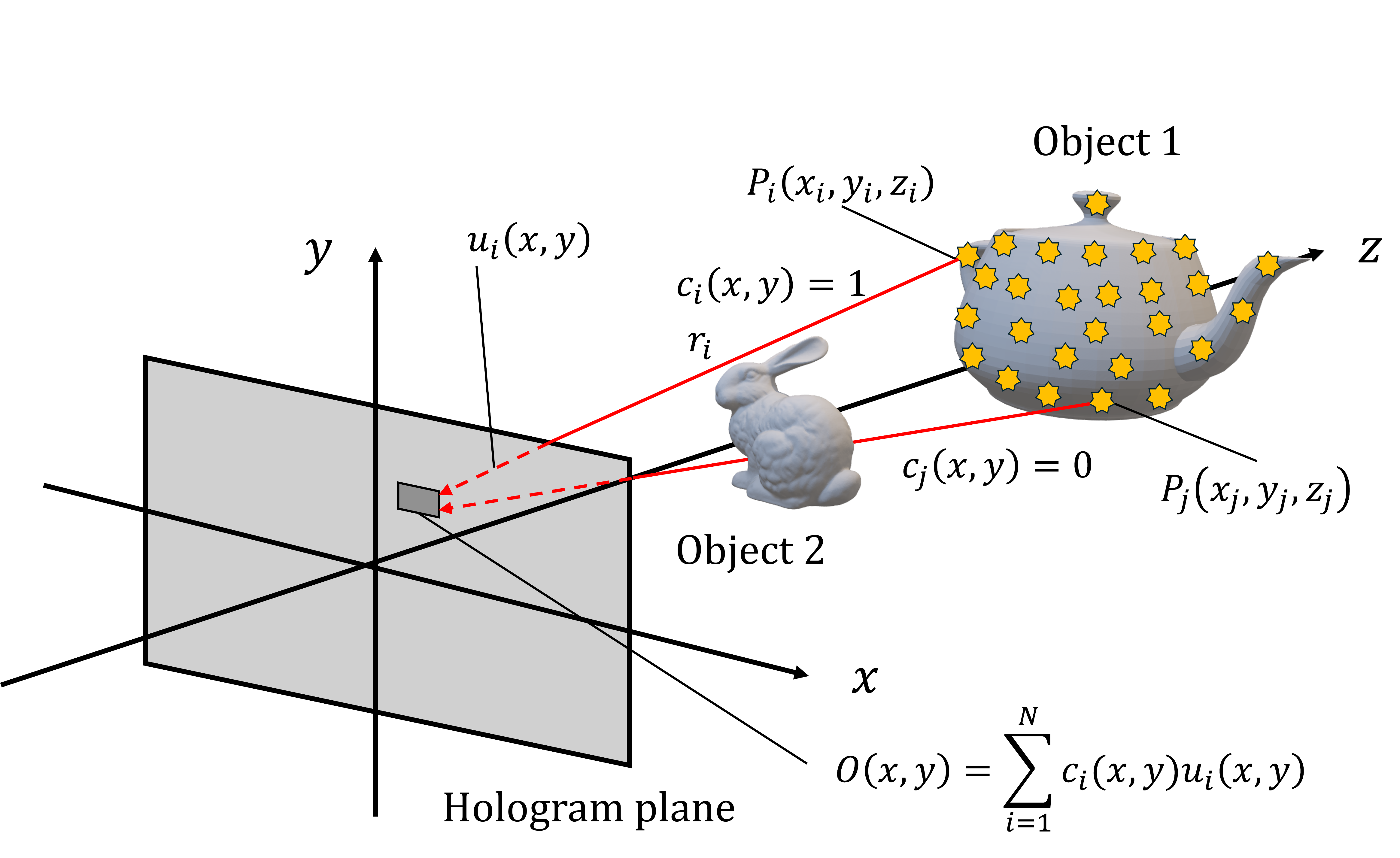}
\caption{Conceptual diagram of hidden surface removal in the point-based method that applies ray tracing. A ray is traced from a point light source placed on the object in the background (Object 1) towards a pixel on the hologram plane. A determination is made as to whether this ray intersects with the occluding object in front (Object 2), and the visibility coefficient $c_i$ is determined. If the ray intersects with the occluding object, $c_i=0$ and light wave propagation calculations from that point light source are not carried out. In contrast, if the ray is not occluded and reaches the pixel, $c_i=1$ and light wave propagation calculations are carried out.}
\label{fig:pixel-based_ray_tracing}
\end{figure}

\begin{figure}[t]
\centering
\includegraphics[clip,width=10.0cm]{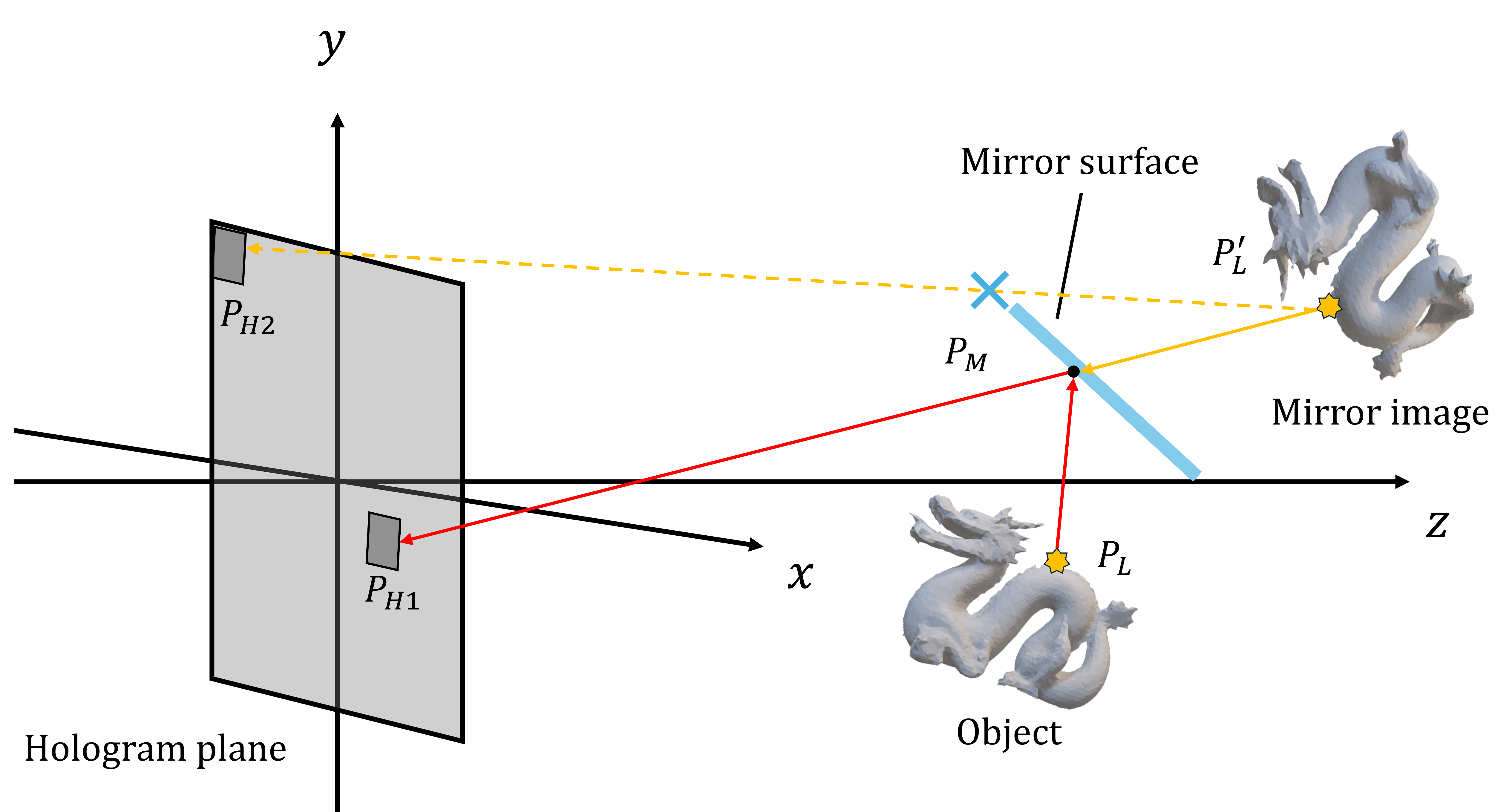}
\caption{Conceptual diagram of specular reflection calculations on a planar mirror. First, a point light source on the real object is replicated as a specular image point, at a position symmetrical to the planar mirror. Next, a determination is made as to whether the reflection point calculated from the line connecting this specular image point and the pixel is within the range of the mirror. If this point is outside the range of the mirror, no calculations to propagate light waves to that pixel are carried out. Furthermore, if the light path from the point light source via the reflection point on the mirror surface to the pixel is blocked by another object, no calculations are carried out, thereby accurately rendering only physically valid mirror images.}
\label{fig:planar_mirror}
\end{figure}

Conventional point-based methods struggle with physically correct hidden surface removal and the representation of optical phenomena such as specular reflection and refraction. To solve this problem, Watanabe et al. proposed a method that introduces ray tracing, a CG rendering technique, into CGH calculations \cite{Watanabe:24}. This method treats 3D objects as polygon models, and its basic principle is visibility determination using ray tracing. First, as shown in Fig. \ref{fig:pixel-based_ray_tracing}, a ray is traced from a point light source $P_i$ to each pixel on the hologram plane. It is determined whether this ray intersects with other polygons before reaching the pixel, and the result is reflected in the visibility coefficient $c_i(x,y)$ (1 if no intersection, 0 if intersection). Using this visibility coefficient, the object light $O(x,y)$ is calculated as

\begin{align*}
 O(x, y) &= \sum^{N}_{i=1}c_i(x,y)u_i(x, y).
\end{align*}

This process enables physically accurate hidden surface removal and the accompanying representation of continuous motion parallax.

This method can also be applied to the representation of specular reflection. To calculate the mirror image, as shown in Fig. \ref{fig:planar_mirror}, a virtual point light source, the specular image point, is first replicated at a position symmetrical to the mirror surface, and hidden surface removal is then carried out for the mirror image. This process ensures that only physically valid mirror images are rendered by carrying out a series of validation checks. First, it verifies that the reflection point lies within the physical boundaries of the mirror, which prevents unrealistic reflections from beyond its edges. Second, ray tracing is used to determine if the entire optical path—from the original light source to the hologram pixel via the reflection point—is occluded by other objects.

However, this method has notable limitations. The established technique is confined to planar mirrors, with no algorithm provided for curved surfaces. Furthermore, for curved mirrors, the equations for finding the reflection point become high-order polynomials, making it difficult to analytically solve for the positions of both the specular image point and the corresponding reflection point.

\subsection{Calculation of specular reflection from Bézier surfaces by Bézier clipping}

While Watanabe et al. enabled the calculation of reflections from planar mirrors, their method presented a challenge for curved mirrors, as the equations for finding the reflection point become high-order and difficult to solve analytically. Furthermore, with a curved mirror, the specular image point moves depending on the observer's viewpoint and is not fixed at a single location. To address these issues, Arai et al. proposed a numerical method that models a curved mirror as a Bézier surface and applies the Bézier clipping technique from the field of CG to compute the mirror image \cite{Arai:24}.

\begin{figure}[h]
\centering
\includegraphics[clip,width=7.0cm]{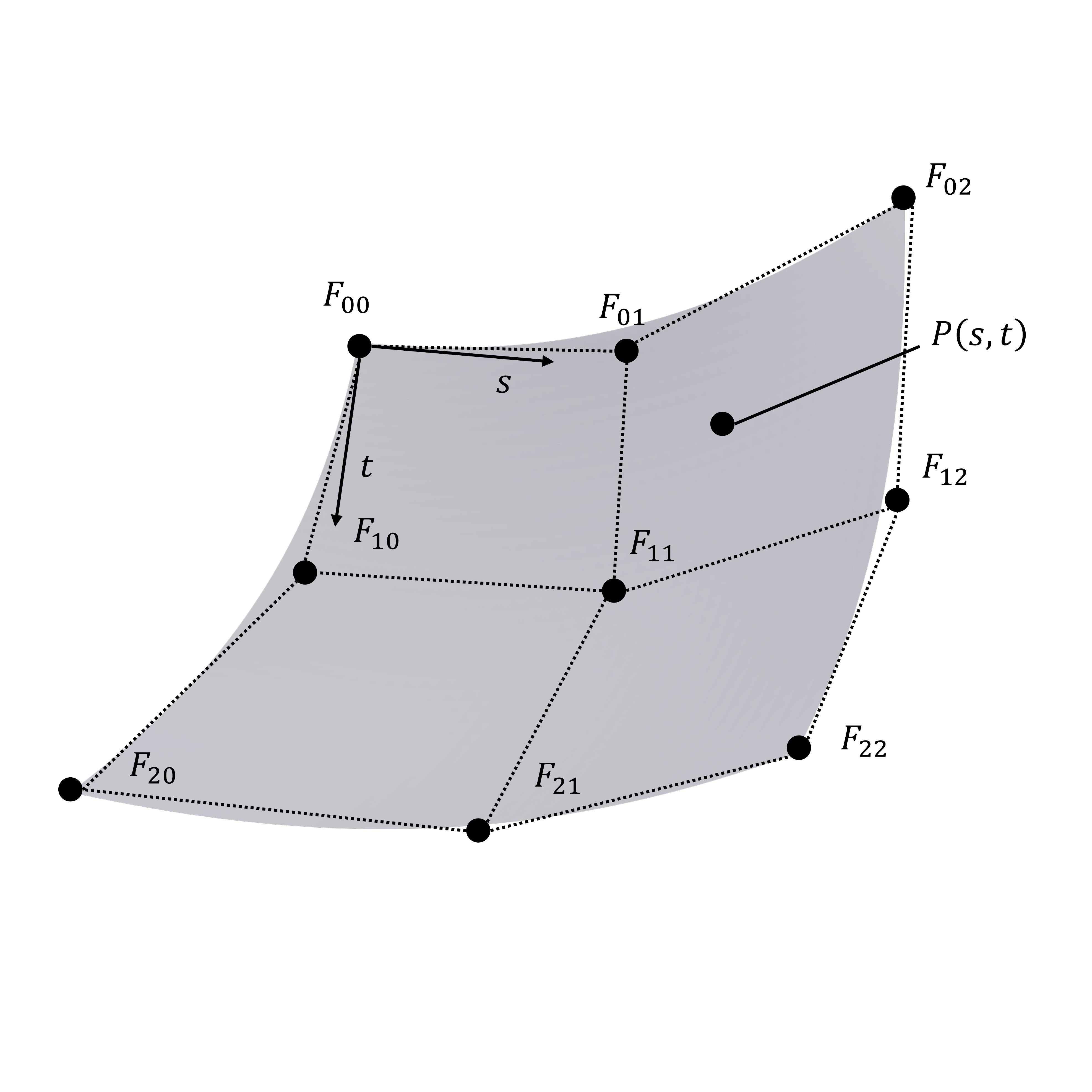}
\caption{Example of a quadratic $\times$ quadratic Bézier surface constructed using $3 \times 3$ control points $F_{ij}$. A Bézier surface is constructed as an $(n-1) \times (m-1)$ degree surface depending on the number of control points $n$ and $m$. Any point $P(s,t)$ on the surface is uniquely determined by two parameters $s$ and $t$ (usually $0\le s,t \le 1$) along the directions indicated by the arrows.}
\label{fig:Bezier_surface}
\end{figure}

A Bézier surface is a parametric surface whose shape is determined by a grid of control points, as shown in Fig. \ref{fig:Bezier_surface}. A point $P(s,t)$ on the surface is defined by 

\begin{equation}
P(s,t)={\sum_{i=0}^{n}{\sum_{j=0}^{m}{B_i^n(s)}B_j^m(t)F_{ij}}},
\end{equation}
where $F_{ij}$ are $n\times m$ control points arranged in three-dimensional space, and the arrangement of these points forms the basic shape of the surface. $B_i^n(s)$ is the Bernstein basis function, which weights the influence of each control point on a point on the surface and plays a role in forming a smooth surface. This polynomial is expressed as

\begin{equation}
B_i^n(s)=\binom{n}{i}s^i(1-s)^{(n-i)},
\end{equation}
where $\binom{n}{i}$ is a binomial coefficient. $s$ and $t$ are two parameters that determine the position on the surface, each ranging from 0 to 1. By varying these two parameters, the 3D coordinates and normal vector of any point on the surface can be calculated, making Bézier surfaces suitable for representing complex, smooth mirror surfaces.

\begin{figure}[ht]
\centering
\includegraphics[clip,width=10.0cm]{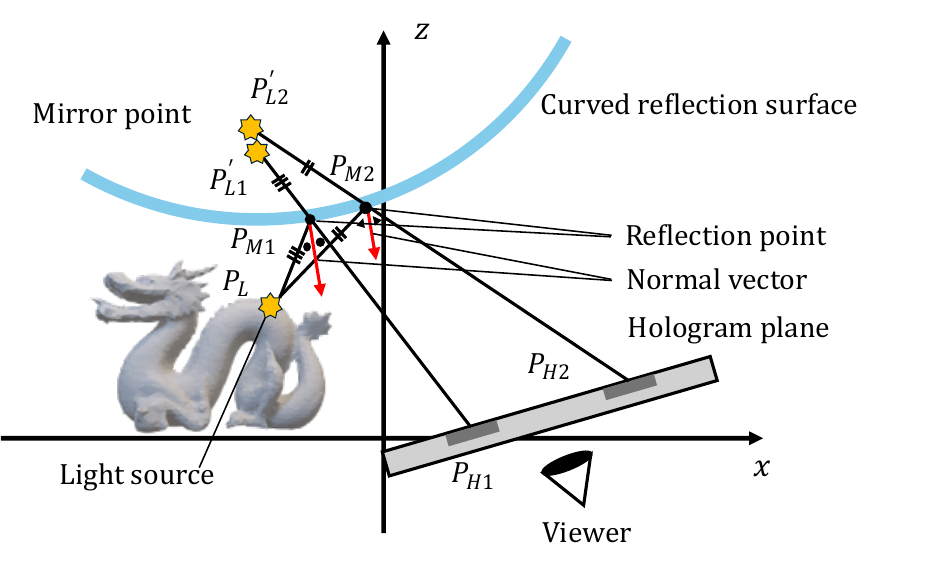}
\caption{Conceptual diagram of specular reflection calculation on a curved mirror using Bézier clipping. First, the error between the angle of incidence and the angle of reflection of the light ray from the point light source $P_L$ at the candidate point is calculated. Then, Bézier clipping is used to modify the candidate point in a direction that reduces the error, and the angle error is calculated again. This process is repeated to find a highly accurate reflection point $P_M$. After that, the distance from the point light source $P_L$ to the reflection point $P_M$ is calculated, and the specular image point $P^{'}_L$ is calculated by extending the vector from the viewpoint $P_H$ to the reflection point $P_M$ by that distance.}
\label{fig:Bezier_clipping}
\end{figure}

Arai et al.'s method searches for the reflection point based on the law of reflection, as shown in Fig. \ref{fig:Bezier_clipping}. A physically correct reflection point $P_M$ satisfies the condition that the normal vector of the mirror surface at that point bisects the angle between the vector to the point light source $P_L$ and the vector to the pixel $P_H$ on the hologram plane. Since it is difficult to analytically compute a point that satisfies this condition, Bézier clipping is applied to search numerically. First, the angular error between the angles of incidence and reflection at candidate points on the mirror surface is calculated. If the error is greater than a threshold, Bézier clipping is used to adjust the candidate point in a direction that reduces the error. This process is repeated until the error falls below the threshold, ultimately resulting in a highly accurate reflection point $P_M$. Once the reflection point $P_M$ is determined, the next step is to calculate the specular image point $P^{'}_L$. This is done by calculating the distance from the point light source $P_L$ to the reflection point $P_M$, and then extending the vector from viewpoint $P_H$ to the reflection point $P_M$ by that distance.

The above method enables the specular image point, which changes dynamically for each viewpoint, to be determined with high accuracy for each pixel without analytically solving higher-order equations. This makes it possible to express highly accurate mirror images using curved mirrors in CGH calculations, a task that was previously considered difficult.

\subsection{Calculation of specular reflection from Bézier surfaces by polygonal subdivision}

While the method by Arai et al. calculates reflection points with high accuracy using a recursive search, it suffers from an extremely high computational cost. For example, studies have shown that determining the reflection point can take over a second for a single point light source. Consequently, a significant reduction in computation time is essential to handle the thousands or tens of thousands of point light sources required for practical image quality.

\begin{figure}[ht]
\centering
\includegraphics[clip,width=10.0cm]{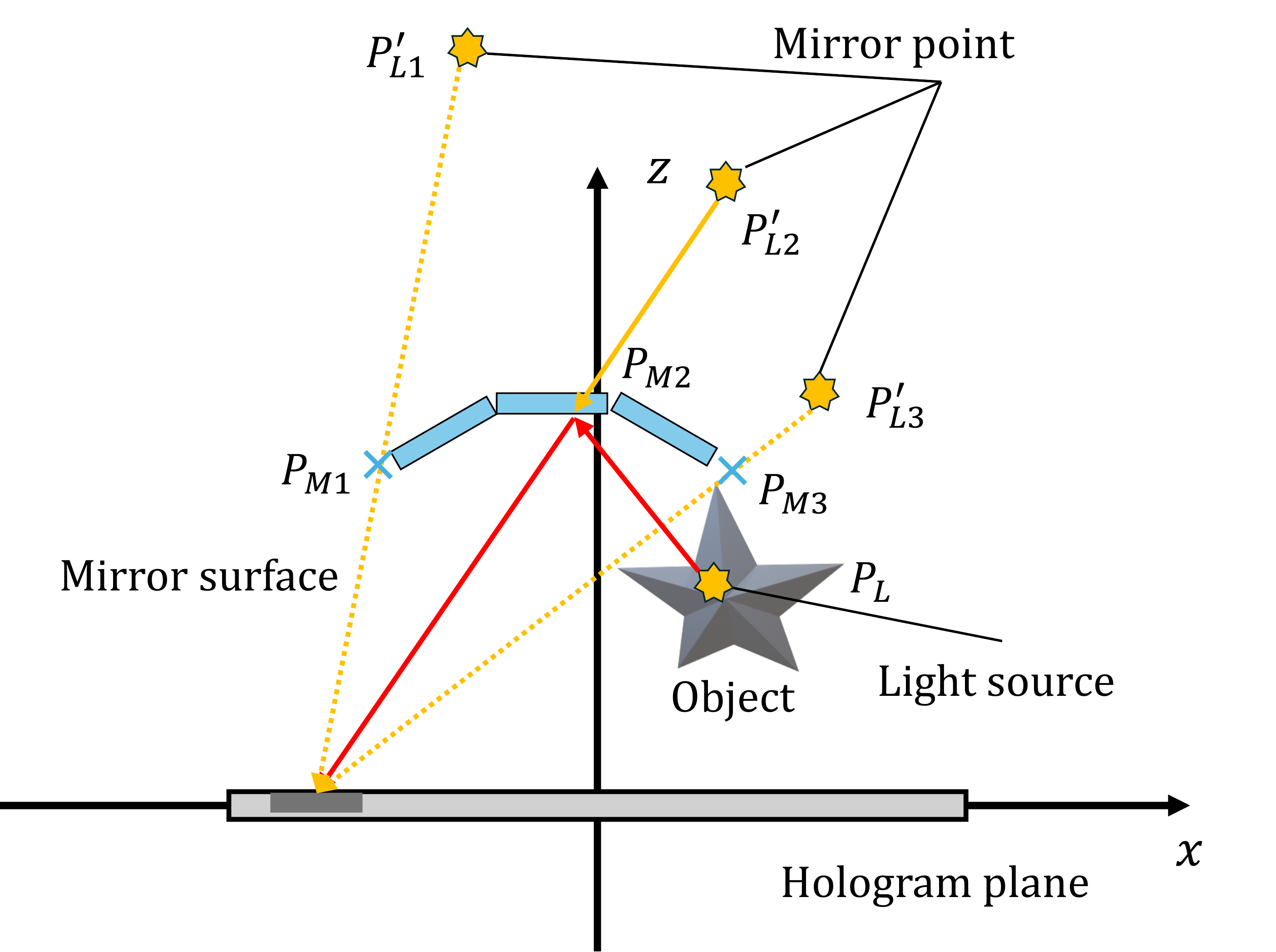}
\caption{Conceptual diagram of specular reflection calculation on a curved mirror using mirror surface subdivision method. The point light source $P_L$ is replicated as specular image points $P^{'}_{L1}$, $P^{'}_{L2}$, and $P^{'}_{L3}$ at positions symmetrical to each subdivided polygon plane. Next, it is checked whether the reflection points $P_{M1}$, $P_{M2}$, and $P_{M3}$ exist within the mirror polygon. If $P_{M1}$ and $P_{M3}$ do not exist on the mirror surface, their mirror images do not exist within the mirror, so light wave propagation calculations for $P^{'}_{L1}$ and $P^{'}_{L3}$ are not carried out.}
\label{fig:subdivision_method}
\end{figure}

One approach to resolve this computational time issue is the ``mirror surface subdivision method,'' which we proposed in a previous study \cite{10.1117/1.OE.64.7.075102}. As shown in Fig. \ref{fig:subdivision_method}, this method approximates a smooth, Bézier-defined mirror surface as a collection of numerous small triangular polygons, and then applies the planar mirror algorithm by Watanabe et al. to each polygon. The process for each polygon is as follows. First, the polygon is treated as a planar mirror, and the point light source is replicated as a specular image point at a position symmetric with respect to the polygon's plane. Next, the reflection point is geometrically determined as the intersection of this plane and the line segment connecting the specular image point to a hologram pixel. The process continues only if this reflection point lies within the boundaries of the triangular polygon. In the final step, a hidden surface removal check is carried out using ray tracing to determine if the complete optical path—from the original light source to the pixel via the reflection point—is occluded by other objects. The light wave propagation is calculated only if the path is clear. This entire sequence is carried out for all subdivided polygons.

This method has the advantage of significantly reducing computation time by replacing a complex recursive calculation with a multitude of simple planar geometric calculations. However, because this approach approximates a curved surface with a discrete set of planes, maintaining accuracy for mirrors with multi-directional curvature requires a large number of subdivisions. This, in turn, leads to a corresponding increase in computation time, presenting a clear trade-off between accuracy and speed.

\section{Proposed Method}
\subsection{Overview and basic concept}

In CGH mirror image calculation, the coordinates of the light ray's start point (a point light source) and end point (a hologram pixel) are known, while the intermediate reflection point on the mirror surface is unknown. The fundamental challenge, therefore, is how to find this unknown reflection point efficiently and accurately.

As discussed in the previous section, conventional approaches to this problem include the recursive search method using Bézier clipping and the approximation method of subdividing a curved surface into small planar polygons. However, each of these methods has unresolved limitations. The method by Arai et al., while capable of producing highly accurate results, suffers from the practical problem of requiring prohibitive computation time. The subdivision method, on the other hand, requires an increasing number of polygons to maintain accuracy for mirrors with multi-directional curvature, creating an inherent trade-off between precision and computation time.

\subsection{Formulation of optical path length based on Fermat's principle}

\begin{figure}[H]
\centering
\includegraphics[clip,width=10.0cm]{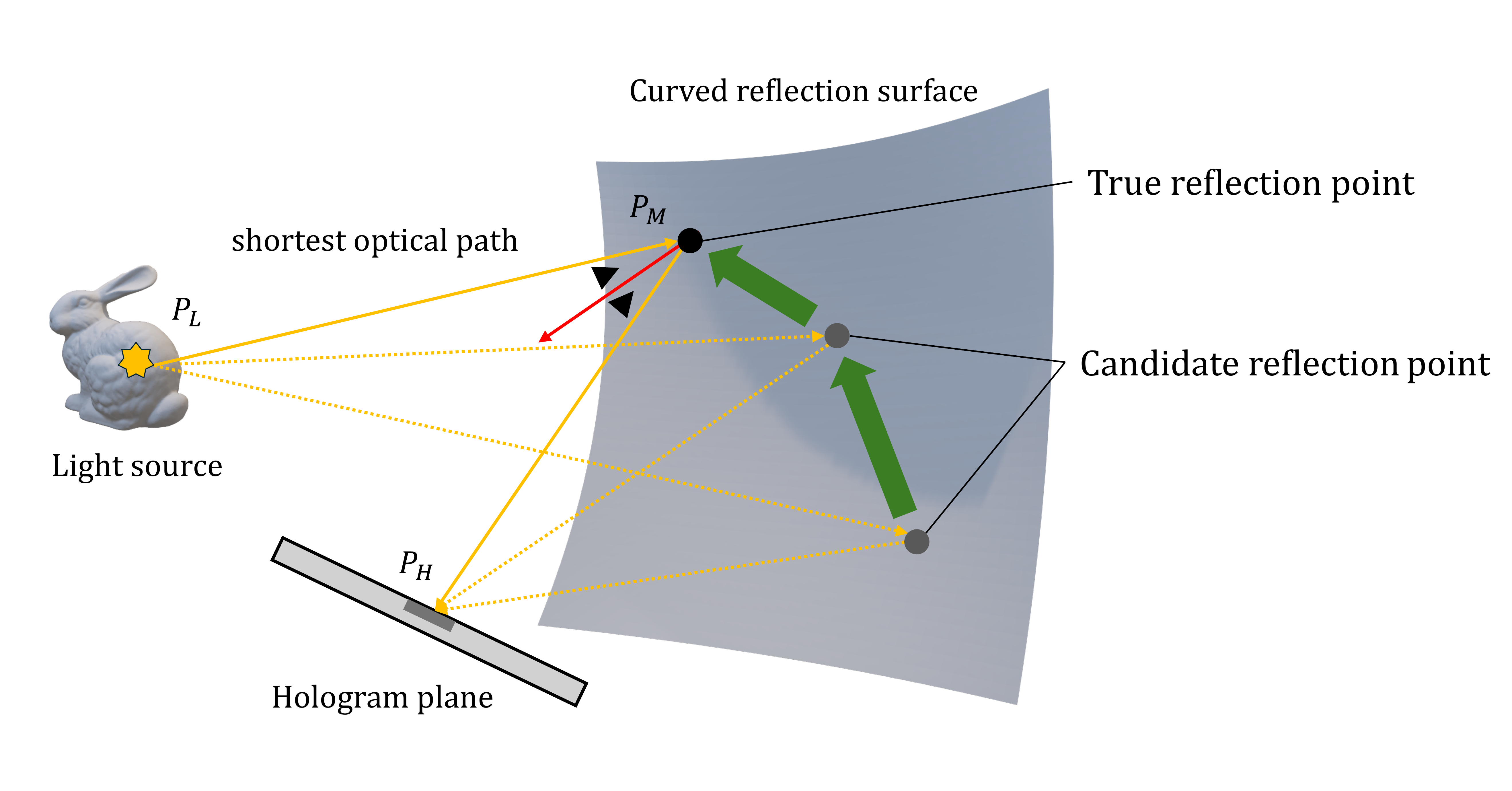}
\caption{Conceptual diagram of specular reflection calculation on a curved mirror using numerical calculation to minimize the optical path length. The reflection point that minimizes the optical path length connecting three points—the point light source $P_L$, the pixel $P_H$ on the hologram plane, and the reflection point $P_M$ on the curved mirror—is searched for using numerical optimization calculations. The search begins with an appropriate initial value on the mirror surface, and the candidate points are then updated in the direction of shortening the optical path length, utilizing factors such as the gradient of the optical path length function. The iterative calculation is completed when the optical path length converges to a local minimum value, and the candidate point at that point is adopted as the reflection point.}
\label{fig:proposed_method}
\end{figure}

Fermat's principle states that light travels between two points along the path that requires the least time \cite{goodman2005introduction}. In this study, we apply this principle to the optical path shown in Fig. \ref{fig:proposed_method}: the three-point path from a point light source $P_L$ on an object, via a reflection point $P_M(s,t)$ on the curved mirror, to a pixel $P_H$ on the hologram. Therefore, the correct reflection point can be found by determining the point on the mirror surface that minimizes this optical path length.

Specifically, let $P_L$ be the coordinates of the point light source, $P_H$ be the coordinates of the hologram pixel, and $P_M(s,t)$ be a point on the Bézier surface defining the mirror. The total optical path length $L$ is then expressed as a function of the surface parameters $s$ and $t$, as

\begin{equation*}
L(s,t)=\|P_M(s,t)-P_L\|+\|P_H-P_M(s,t)\|.
\end{equation*}

The exact reflection point $P_M(s,t)$ is determined by finding the set of parameters $(s,t)$ that minimizes the optical path length function $L(s,t)$. This reduces to the optimization problem of finding where the gradient of $L(s,t)$ is zero ($\nabla L(s,t) = 0$). In this study, we solve this optimization problem using a numerical iterative method.

A strict interpretation of the underlying principle must be considered. More precisely, Fermat's principle states that the actual path of light corresponds to a stationary point of the optical path length function, which is not necessarily only a local minimum \cite{Born_Wolf_Bhatia_Clemmow_Gabor_Stokes_Taylor_Wayman_Wilcock_1999}. Stationary points also include local maxima and saddle points. The numerical optimization method we utilize is an algorithm that finds the stationary point closest to a given initial value—a point that is, in many cases, a local minimum. Therefore, it is theoretically possible that the point found by the search is not the global minimum (the true, physically meaningful shortest path) but rather another local minimum or a saddle point. Additionally, since the algorithm is designed to find only a single stationary point (the one closest to the initial value), it does not account for cases where multiple stationary points may exist.

However, for smooth Bézier surfaces as used in CGH, and assuming a physically reasonable arrangement of the light source and viewpoint, the existence of multiple stationary points is extremely rare. Typically, only one physically meaningful path for reflected light exists. Thus, provided the search starts from a suitable initial value, the stationary point found by our method will almost certainly coincide with the single physically correct reflection point. For these reasons, we consider the proposed method sufficiently accurate for practical use in achieving high-fidelity mirror image calculations in CGH.

\subsection{Iterative search for reflection points using gradient method}
\label{subsec:iterative_search}
As stated in the previous section, the proposed method is reduces reduced to solving an optimization problem that minimizes the optical path length function $L(s,t)$. Because Since this function is based on the definition of a Bézier surface, it is smooth and differentiable. Therefore, numerical methods that utilize gradient information are well-suited for finding the solution. Such methods include the method of steepest descent, a fundamental technique applied in optics for problems like metasurface optimization and wavefront correction. In this study, we adopt Newton's method, which offers faster convergence \cite{nocedal2006numerical, Vorontsov:98, Xu:25}. Newton's method utilizes not only the first derivative (gradient) but also second-derivative information (curvature), enabling it to find high-precision solutions in fewer iterations.

To apply Newton's method, the gradient of the optical path length function, $\nabla L(s,t)$, must first be computed. The partial derivatives of $L(s,t)$ with respect to the parameters $s$ and $t$ can be expressed using the tangent vectors at point $P_M(s,t)$ on the Bézier surface, namely, $\frac{\partial P_M(s,t)}{\partial s}$ and $\frac{\partial P_M(s,t)}{\partial t}$.

\begin{align*}
\frac{\partial L}{\partial s}=\frac{(P_M(s,t)-P_L)\cdot \frac{\partial P_M (s,t))}{\partial s}}{\|P_M(s,t)-P_L\|}-\frac{(P_H-P_M(s,t))\cdot \frac{\partial P_M (s,t))}{\partial s}}{\|P_H-P_M(s,t)\|}\\\frac{\partial L}{\partial t}=\frac{(P_M(s,t)-P_L)\cdot \frac{\partial P_M (s,t))}{\partial t}}{\|P_M(s,t)-P_L\|}-\frac{(P_H-P_M(s,t))\cdot \frac{\partial P_M (s,t))}{\partial t}}{\|P_H-P_M(s,t)\|}
\end{align*}

The gradient vector $\nabla L(s,t)$ is then constructed from these partial derivatives. A point where this gradient is zero corresponds to a reflection that satisfies the law of reflection.

The update formula for the parameters $s$ and $t$ using Newton's method is given by the following equation, where $k$ is the iteration step:

\begin{equation*}
\begin{bmatrix}
 s_{k+1} \\
 t_{k+1}
\end{bmatrix}
=
\begin{bmatrix}
 s_{k} \\
 t_{k}
\end{bmatrix}
-\alpha_kH^{-1}_k
\begin{bmatrix}
 \frac{\partial L}{\partial s}(s_{k},t_{k}) \\
 \frac{\partial L}{\partial t}(s_{k},t_{k})
\end{bmatrix}
.
\end{equation*} 

Here, $\nabla L(s_k,t_k)$ is the gradient vector at step $k$, and $\alpha_k$ is the learning rate that adjusts the step size. $H_k$ is the Hessian matrix of the function $L(s_k,t_k)$ and is composed of the following second-order partial derivatives:

\begin{equation*}
H_k=\begin{pmatrix}
 \frac{\partial^2 L}{\partial s^2}(s_k,t_k) & \frac{\partial^2 L}{\partial s\partial t}(s_k,t_k) \\
 \frac{\partial^2 L}{\partial t\partial s}(s_k,t_k) & \frac{\partial^2 L}{\partial t^2}(s_k,t_k) 
\end{pmatrix}
.
\end{equation*} 

This update process is repeated until the difference between the current optical path length $L(s_k,t_k)$ and an estimated value for the true optical path length (as determined by the method detailed in the next section) falls below a predetermined threshold. The parameters $(s,t)$ from the iteration that satisfies this convergence condition are used to determine the final coordinates of the reflection point.

In an iterative algorithm like Newton's method, providing appropriate initial values is key to achieving fast and stable convergence. Therefore, for the initial point light source of an object, the proposed method sets the initial search value to $(s,t)=(0.5,0.5)$, the center of the mirror's parameter space. This is the most neutral choice when no prior information is available.

For subsequent point light sources, an initial value strategy that leverages the spatial proximity of the points is used to significantly improve computational efficiency. Since objects in the point-based method are typically represented by a dense cloud of point light sources, it we can be expected that reflection points for adjacent point light sources in physical space will be located very close to each other in the mirror's parameter space. Based on this property, the resulting parameters $(s,t)$ from the previously computed reflection point are reused as the initial search value for the current point light source.

This strategy allows the search for most point light sources to converge in just a few iterations, yielding a significant speedup in the overall computation. Furthermore, to maximize this effect, the proposed method pre-sorts all point light sources of the object into a spatially adjacent order before beginning the calculation.

\subsection{Stopping condition and allowable error for iterative calculation}
\label{subsec:stopping_condition}
Terminating the iterative process of Newton's method, as described in the previous section, requires a criterion to determine if the solution has converged sufficiently. Commonly utilized convergence criteria for Newton's method include checking if the change in parameters between steps or the norm of the gradient vector falls below a predefined threshold. However, these criteria do not allow for a direct evaluation of the remaining physical error in the optical path length.

Therefore, the proposed method instead estimates the optical path length error—that is, the difference between the current path length and the unknown true path length—and uses this value as its convergence criterion. This approach enables a highly reliable convergence test based directly on a physical quantity, thereby ensuring the accuracy of the calculation.

The path length error estimate utilized in this study is based on the idea that the path length function $L(s,t)$ can be approximated by a simple quadratic function in the vicinity of the true reflection point.

To derive this estimate, the problem is first simplified to one dimension. Let $s_t$ be the parameter of the true reflection point and $s_e$ be the parameter of the current candidate point. By applying a second-order Taylor expansion to the optical path length function $L(s)$ around the point $s_t$, we obtain

\begin{equation}
 L(s)\approx L(s_t)+L^{'}(s-s_t)+\frac{1}{2}L^{''}(s_t)(s-s_t)^2.
 \label{eq:optical_path_length_approximation_formula}
\end{equation}

Since the gradient is zero at the true reflection point (that is, $L'(s_t) = 0$), the preceding equation is simplified. By substituting the current parameter $s_e$, the error between the current optical path length $L_e$ and the true path length $L_t$ can be expressed in terms of the parameter error $(s_e - s_t)$, as 

\begin{equation*}
 (L_e-L_t)\approx \frac{1}{2}L^{''}(s_t)(s_e-s_t)^2.
\end{equation*}

Extending this relationship to the two-dimensional parameters $(s,t)$, the optical path length error can be expressed in terms of the parameter error and the Hessian matrix $H_t$ at the true reflection point, as 

\begin{equation*}
 (L_e-L_t)\approx \frac{1}{2}\begin{pmatrix}
 s_e-s_t\\
 t_e-t_t
\end{pmatrix}^T
H_t
\begin{pmatrix}
 s_e-s_t\\
 t_e-t_t
\end{pmatrix}
.
\end{equation*}

Differentiating Eq. (\ref{eq:optical_path_length_approximation_formula}) with respect to $s$ yields the following expression for the gradient, $L^{'}(s)$:

\begin{equation*}
 L^{'}(s) \approx L^{''}(s_t)(s-s_t).
\end{equation*}

Generalizing this equation to two dimensions and substituting the current parameters $(s_e,t_e)$, the current gradient vector $\nabla L_e$ is given as

\begin{equation*}
 \nabla L_e \approx H_t
 \begin{pmatrix}
 s_e-s_t\\
 t_e-t_t
\end{pmatrix}
.
\end{equation*}

From this equation, the parameter error is approximated in terms of the current gradient vector $\nabla L_e$, as
$\begin{pmatrix}
 s_e-s_t\\
 t_e-t_t
\end{pmatrix} \approx H^{-1}_t\nabla L_e$.

Combining these two approximations and using the gradient vector and Hessian matrix obtained at the current step yields the following theoretical formula to estimate the optical path length error $\Delta L=(L_e-L_t)$:

\begin{equation}
 \Delta L=\frac{1}{2}(\nabla L_e)^{T}H^{-1}_t(\nabla L_e).
 \label{eq:error_estimation}
\end{equation}

The preceding derivation is based on the approximation of the optical path length function as a quadratic function. Theoretically, this approximation is valid only in the immediate vicinity of the true reflection point. However, in the configurations typical of this study—where the hologram plane and the object are distant from the mirror along the optical axis, while their lateral spread is comparatively small—the overall optical path length function closely approximates a smooth paraboloid. Under these conditions, third- and higher-order derivative terms become negligible, which ensures the validity of the quadratic approximation over a wider range. This, in turn, justifies approximating the true Hessian $H_t$ with the Hessian $H_e$ calculated at the current search point. Therefore, the proposed convergence criterion can be considered highly reliable, particularly in typical CGH setups involving reflections from distant objects.

For practical implementation, a tolerance for the estimated error $\Delta L$ can be set based on established optical criteria. For example, using the Rayleigh limit ($\frac{\lambda}{4}$) or the more stringent Maréchal limit ($\frac{\lambda}{14}$)—both of which are widely used as indicators of diffraction-limited performance—is expected to yield mirror images of sufficiently high quality for visual observation.

\subsection{Overall algorithm of the proposed method}
\label{subsec:overall_algorithm}

In this section, we consolidate the process of the proposed method and discuss technical aspects of its implementation and its scalability to more complex phenomena such as multiple reflections. 

The proposed method is implemented using the NVIDIA OptiX ray-tracing framework \cite{10.1145/1778765.1778803, doi:10.2352/ISSN.2470-1173.2017.3.ERVR-095}. Our approach requires ray tracing to check for occlusions along the optical path from a point light source, via the reflection point on the mirror, to a pixel on the hologram plane. The computational load of this ray tracing increases proportionally with the number of polygons, $P$, in the virtual object, which can lead to prohibitive computation times for geometrically complex objects. To address this, our implementation leverages the dedicated ``RT Cores'' available on NVIDIA's RTX series GPUs. RT Cores are designed to accelerate Bounding Volume Hierarchy (BVH) traversal and ray-triangle intersection tests. By organizing polygons into a BVH tree structure, the search space for intersections is dramatically reduced. This optimization reduces the computational complexity from $O(P)$ to $O(\log P)$. Furthermore, as this acceleration is implemented at the hardware level, it achieves a significant speedup compared to pure software implementations.

\begin{figure}[H]
\centering
\includegraphics[clip,width=10.0cm]{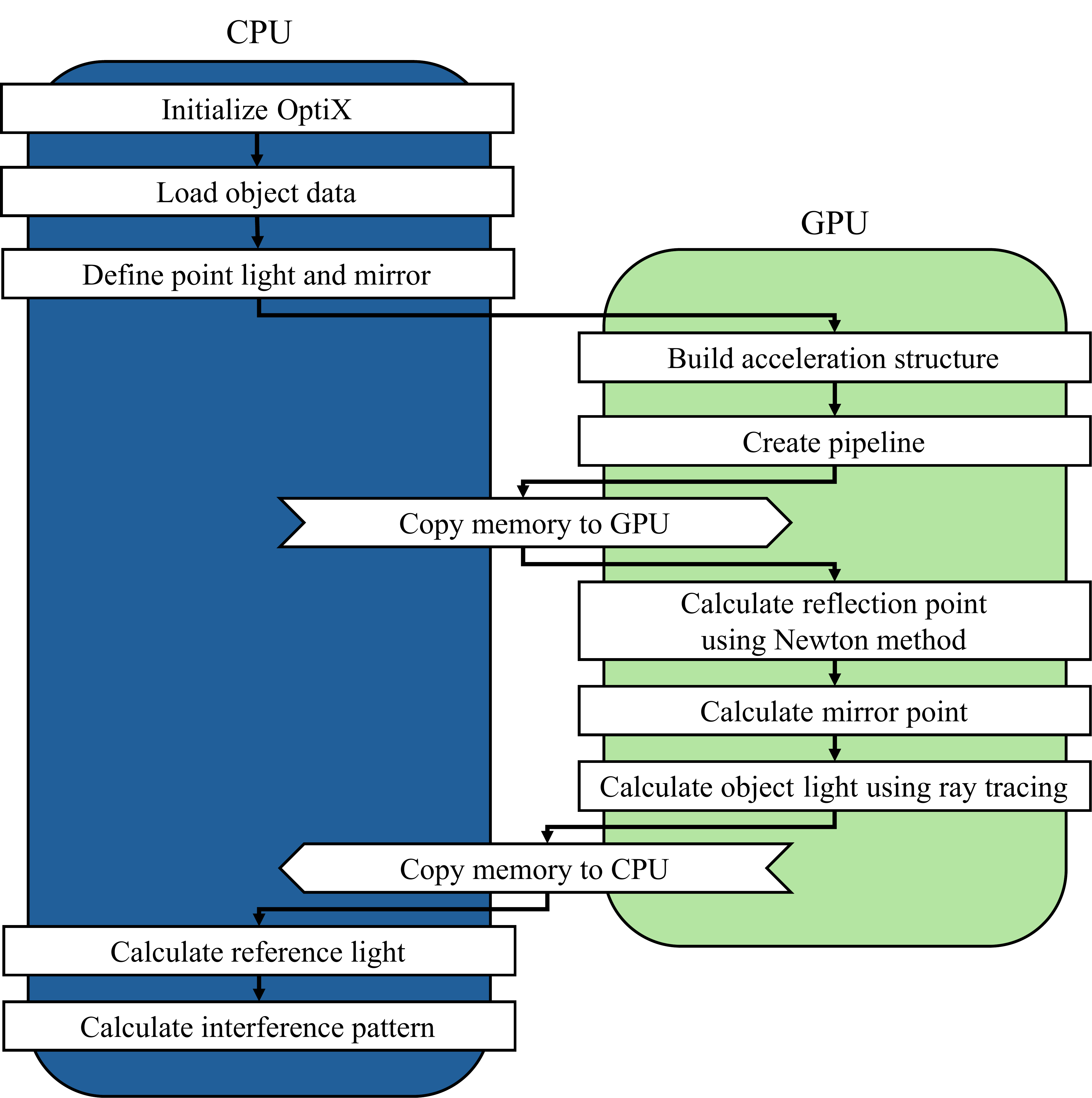}
\caption{Flowchart for implementing the proposed method. The blue blocks on the left represent the processing handled by the CPU, and the green blocks on the right represent the processing handled by the GPU. First, the CPU initializes OptiX and positions the virtual objects, then completes preparation of the acceleration structure and pipeline to be utilized by the GPU. This information is then copied to the GPU. The GPU searches for reflection points using Newton's method, calculates the specular image point, and carries out object light calculations using ray tracing. Finally, the calculated object light data is transferred to the CPU, where it undergoes interference pattern calculations with the reference light to generate the final hologram.}
\label{fig:flowchart}
\end{figure}

The flowchart of the implementation is shown in Fig. \ref{fig:flowchart}. First, OptiX is initialized on the CPU, and the polygon data for the virtual object is loaded from a file. Next, point light sources are distributed across the object's surface, and the mirror's shape and position are defined by its Bézier control points. Building an acceleration structure (AS) is essential for any OptiX-based ray tracing, as the AS enables efficient searching for ray-polygon intersections and dramatically improves performance. After building the AS and the associated processing pipeline, all necessary data—including point light sources, object geometry, and mirror control points—is transferred to the GPU.

Once the data is on the GPU, the iterative Newton's method search described in Section \ref{subsec:iterative_search} begins. Using the estimated optical path length error from Section \ref{subsec:stopping_condition} as the convergence criterion, the reflection point is calculated. Next, based on the found reflection point $P_M$, the corresponding specular image point $P_L'$ is computed. This is achieved by extending the vector from the viewpoint $P_H$ to the reflection point $P_M$ by a distance equal to the length of the incident path, $|P_L P_M|$. With this specular image point, hidden surface removal is carried out via ray tracing. If the path is unoccluded, the object light is calculated, and the resulting data is copied back to the CPU. Finally, the CPU computes the interference pattern with a reference light to generate the final hologram data.

The proposed method offers two primary advantages over conventional techniques. The first advantage is the significant acceleration of the reflection point calculation. The prior method by Arai et al. uses Bézier clipping to find the intersection of a ray and a Bézier surface, which requires a recursive subdivision of the surface's convex hull until a sufficiently small region is found. Repeating this computationally intensive process until convergence results in a very high cost for each reflection point. The mirror surface subdivision method, while fast for a small number of polygons, has its own scaling issues. To maintain accuracy for mirrors with multi-directional curvature, the number of subdivisions must be dramatically increased, causing the computation time to grow proportionally with the polygon count.

In contrast, each iteration in our proposed method primarily involves calculating the first and second derivatives of the optical path length function, which is computationally less demanding. Furthermore, Newton's method, which utilizes second-derivative information, exhibits a high rate of convergence and typically finds a solution in very few iterations. While the computation time of our method does increase with the number of control points defining a more complex Bézier surface, this increase is far more moderate than the exponential growth in polygons required by the subdivision method to achieve similar geometric fidelity. Therefore, the computation time of our method has a low dependency on mirror complexity, enabling substantial speedup over conventional methods, especially for intricate surfaces.

\begin{figure}[t]
 \centering
 \begin{subfigure}{0.48\textwidth}
 \includegraphics[width=\linewidth]{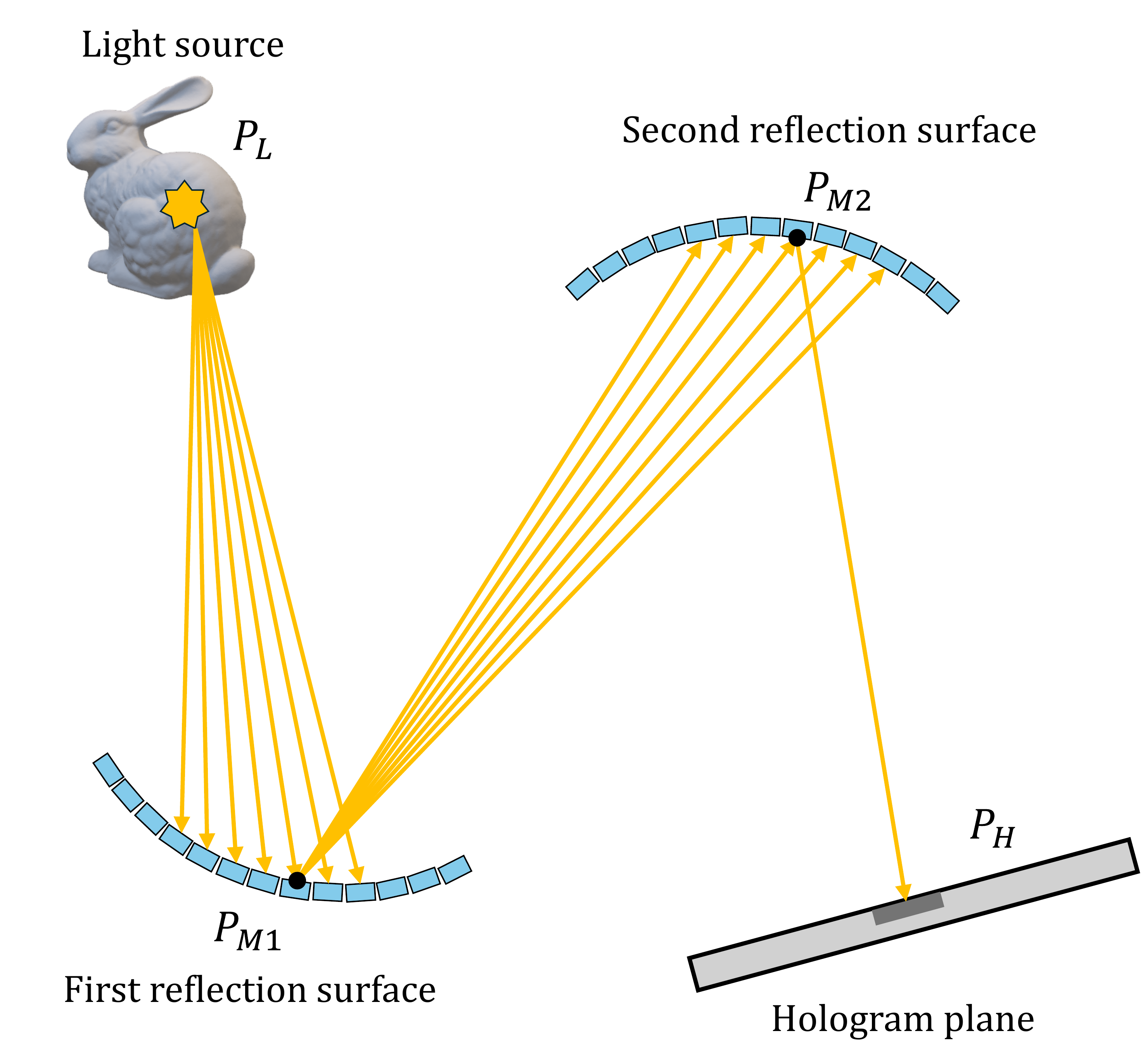}
 \caption{}
 \label{subfig:subdivision} % ラベル名を修正
 \end{subfigure}
 \hfill % 図と図の間のスペース
 \begin{subfigure}{0.48\textwidth}
 \includegraphics[width=\linewidth]{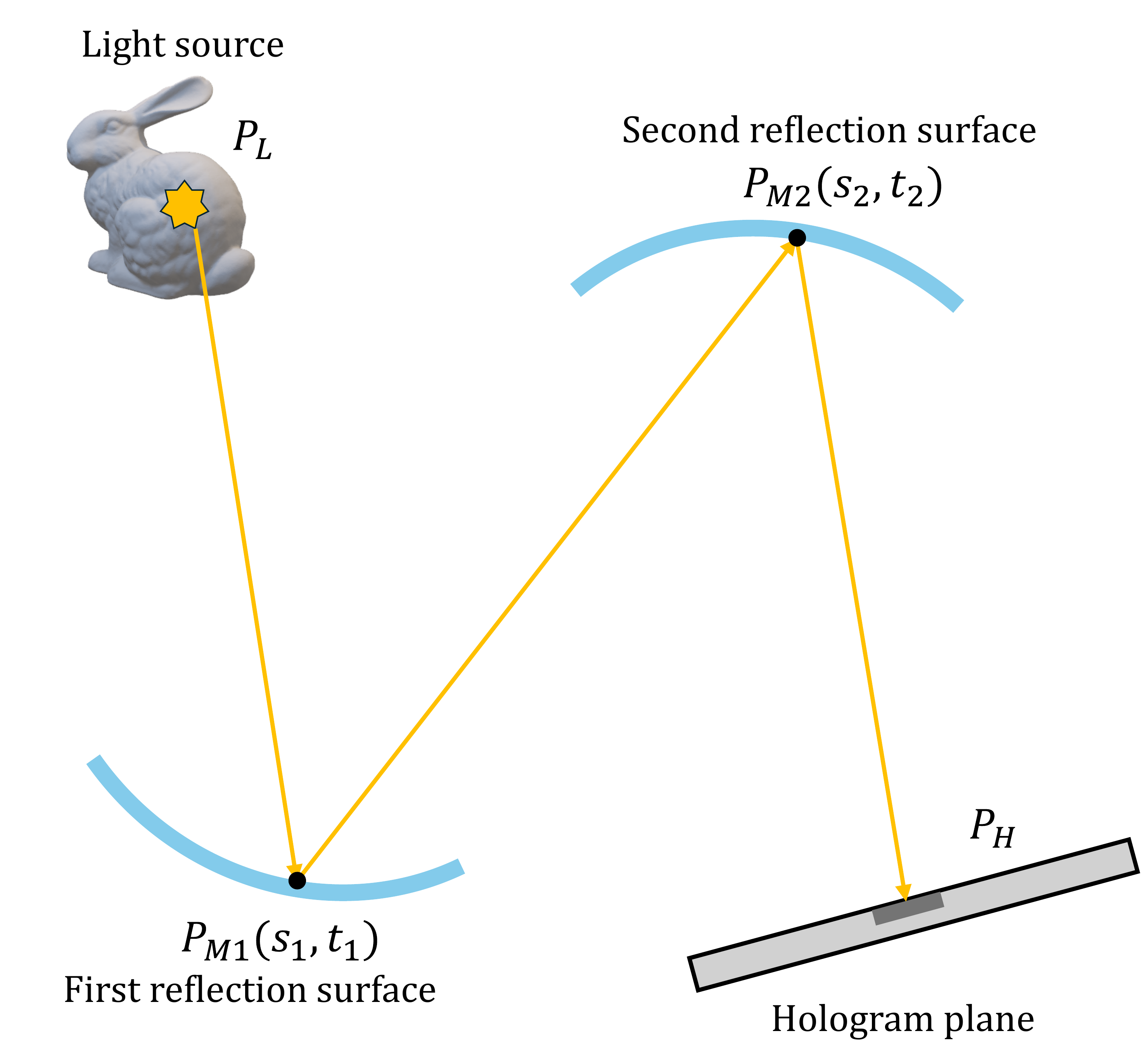}
 \caption{}
 \label{subfig:newton} % ラベル名を修正
 \end{subfigure}
\caption{Conceptual comparison of search problems in multiple reflection calculations. (a) Concept of the mirror surface subdivision method. Since the mirror surface is approximated as a collection of tiny polygons, the correct reflection path is treated as a search problem in a vast discrete space, determining which combination of polygons it should pass through. This method inherently suffers from the problem that the amount of calculations increases explosively with the number of polygons. (b) Concept of the proposed method. By directly treating the mirror surface as a smooth, continuous surface defined by parameters, the problem of multiple reflections is transformed into a low-dimensional continuous optimization problem of minimizing the optical path length. This fundamentally avoids the combinatorial explosion of conventional methods and makes calculations more realistic.}
\label{fig:multi_reflection}
\end{figure}

The second advantage is the method's practical scalability for computing multiple reflections. In general, multiple reflection calculations involve a vast number of possible mirror combinations for any given light ray, requiring the calculation for any single path to be extremely fast. Arai et al.'s method is already too computationally expensive for a single reflection, making it impractical for multiple reflections. The subdivision method, while fast for a single ray-polygon intersection, faces a different challenge: since the mirror itself is composed of thousands of polygons, the combinatorial explosion of possible reflection paths makes the total computation time unrealistic. Fig. \ref{fig:multi_reflection} conceptually illustrates the difference between the subdivision method and our proposed method for multiple reflections.


As shown in Fig. \ref{fig:multi_reflection}(\subref{subfig:newton}), the proposed method is fundamentally capable of solving these issues. Multiple reflections can be naturally formulated by simply increasing the number of variables in the optical path length function. For example, for a secondary reflection between two mirror surfaces, the optical path length $L$ becomes a function of four variables: the parameters $(s_1, t_1)$ and $(s_2, t_2)$ for the reflection points on each respective mirror.

\begin{equation*}
 L(s_1, t_1, s_2, t_2)=\|P_{M1}(s_1, t_1)-P_L\|+\|P_{M2}(s_2, t_2)-P_{M1}(s_1, t_1)\|+\|P_H-P_{M2}(s_2, t_2)\|
\end{equation*}

With this formulation, the calculation of multiple reflections is transformed into an optimization problem: finding the set of continuous parameters $(s_1, t_1, s_2, t_2, \dots)$ that minimizes the optical path length $L$.

The number of variables in the optimization problem for the proposed method increases linearly with the number of reflections. For example, even for a complex path involving ten reflections, the problem consists of only 20 variables. This is in stark contrast to the subdivision method, where the search space is defined by thousands of discrete polygons and grows combinatorially. Consequently, the complexity of our optimization problem is several orders of magnitude smaller.

Newton's method, the core of the proposed approach, is fundamentally suited for such multi-variable optimization problems. Even as the number of variables increases, the computational cost of the optimization remains well within a practical range. Therefore, by keeping the dimensionality of the problem inherently low, the proposed method makes the calculation of complex multiple reflections practical—a task that is virtually impossible for conventional methods.

\section{Experiment}

\begin{table}[ht]
    \centering
    \caption{Optical system parameters.}
    \label{tab:optical_system_parameters}
    \begin{tabular}{|l|l|l|}
        \hline
        % 「SLM」の行 (2行分を結合)
        \multirow{2}{*}{SLM} & Pixel pitch & 4.5 $\times$ 4.5 $\mu$m \\ \cline{2-3} 
                             & Number of pixels & 1,920 $\times$ 1,080 pixels \\
        \hline
        % 「Laser」の行 (3行分を結合)
        \multirow{3}{*}{Laser} & Wavelength (Red) & 638 nm \\ \cline{2-3}
                               & Wavelength (Green) & 512 nm \\ \cline{2-3}
                               & Wavelength (Blue) & 448 nm \\
        \hline
    \end{tabular}
\end{table}

To validate the proposed method, we conducted both optical reconstructions using computed hologram data and numerical simulations. Wave propagation for the simulations was calculated using the angular spectrum method \cite{10.1117/12.2609369}.

\begin{figure}[t]
\centering
\includegraphics[clip,width=8.0cm]{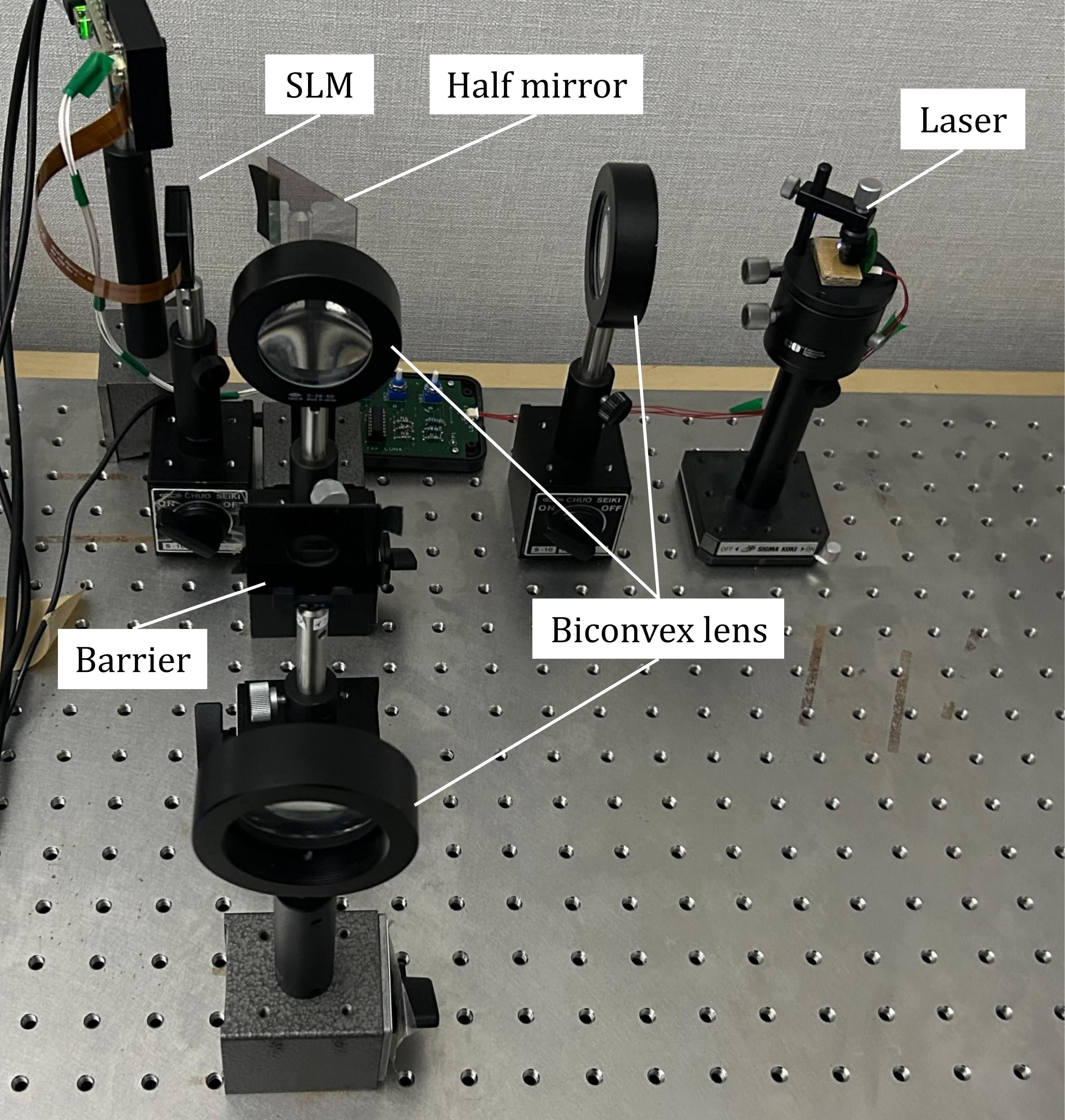}
\caption{Photograph of optical system used in the optical experiment.}
\label{fig:4f_system_picture}
\end{figure}

\begin{figure}[H]
\centering
\includegraphics[clip,width=10.0cm]{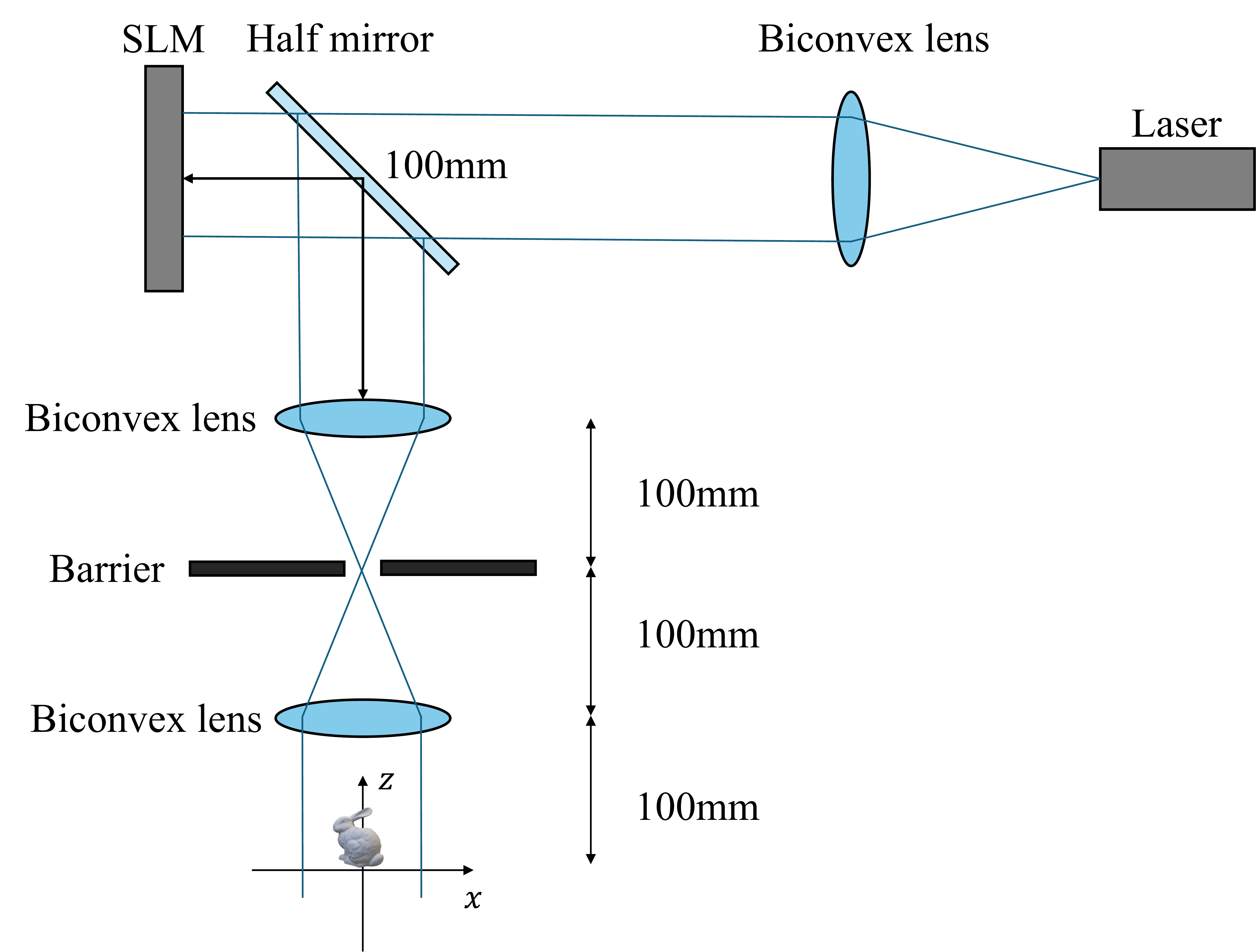}
\caption{Schematic diagram of optical system used in the optical experiment.}
\label{fig:4f_system_figure}
\end{figure}

Table \ref{tab:optical_system_parameters} lists the parameters of the experimental optical system. To achieve full-color reconstruction, three lasers corresponding to the primary colors of light were used. The laser wavelengths were 638 nm for red, 512 nm for green, and 448 nm for blue. The system utilizes a 4f optical configuration \cite{Leith:62}, as shown in Figs. \ref{fig:4f_system_picture} (photograph) and \ref{fig:4f_system_figure} (schematic diagram). The blocking aperture is a rectangle with a height of $f\tan\theta$ and a width of $2f\tan\theta$, where $f$ is the focal length of the lens and $\theta$ is the maximum diffraction angle. The aperture is offset vertically from the optical axis to filter out the zeroth-order and other unwanted higher-order diffracted light. In these experiments, a smooth Bézier surface with low curvature was used as the mirror to prevent intra-mirror reflections.

\subsection{Error threshold}

\begin{figure}[t]
\centering
\includegraphics[clip,width=10.0cm]{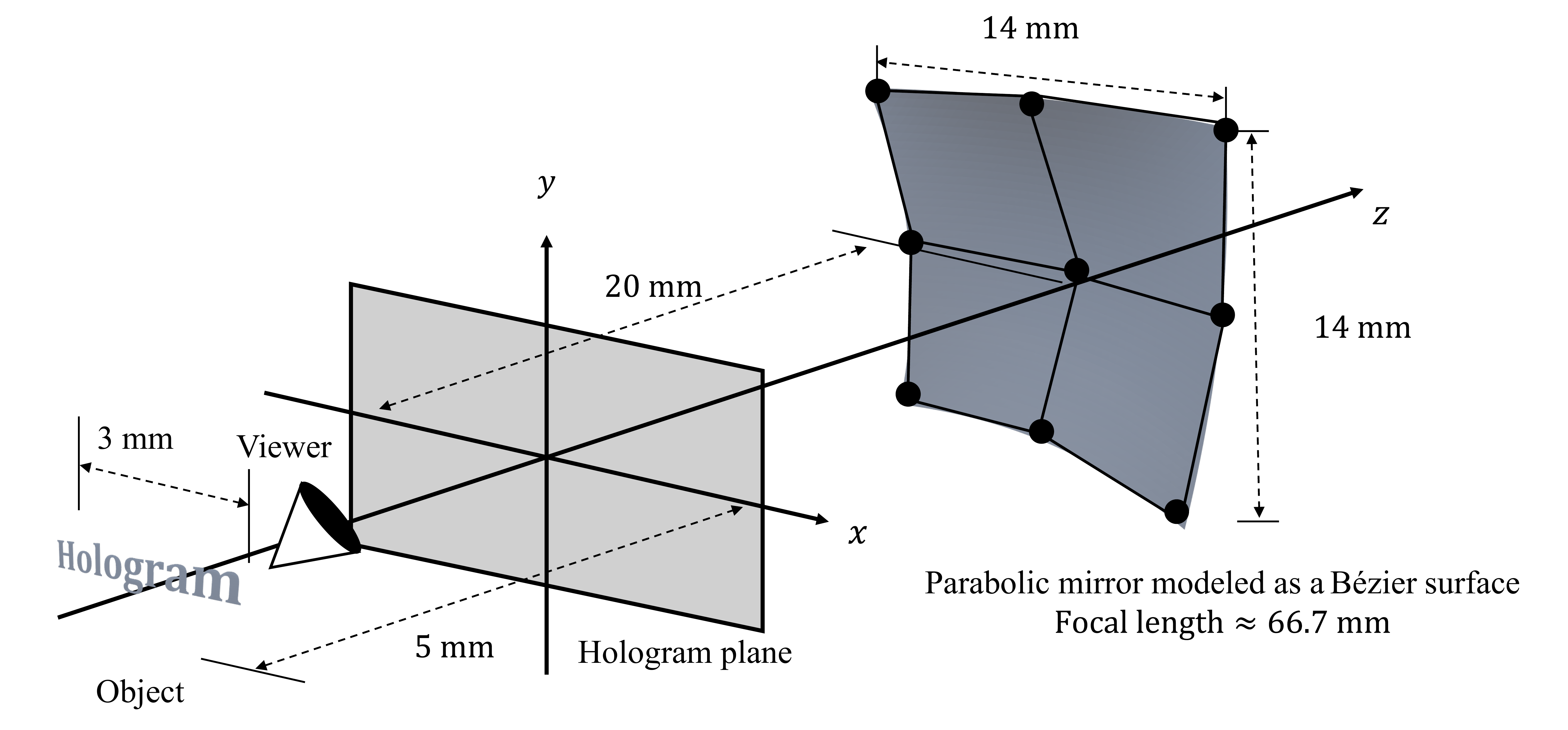}
\caption{Diagram of virtual object placement in the tolerance experiment. The virtual object is a 3-mm-wide text object called ``Hologram'' placed at center coordinates (0 mm, 0 mm, -5 mm). There are 30,000 point light sources placed on the surface of the object. The hologram plane is set to $z=0$. The mirror surface is a quadratic Bézier surface ($3\times3$ control points) measuring $\rm{14~mm} \times \rm{14~mm}$ that approximates a concave parabolic mirror with a focal length of approximately 66.7 mm, and its center coordinates are (0 mm, 0 mm, 20 mm). The object is placed behind the hologram plane, and only the mirror image is calculated.}
\label{fig:placement_error}
\end{figure}

In this section, we validate the termination condition for the iterative calculation proposed in Section \ref{subsec:stopping_condition}. Specifically, we evaluate via simulation how the tolerance for the estimated optical path length error $\Delta L$, defined in Eq. (\ref{eq:error_estimation}), affects the final mirror image accuracy. The diagram of the virtual object placement for this experiment is shown in Fig. \ref{fig:placement_error}. The hologram plane was set to $z = 0$, and a 3-mm-wide text object called ``Hologram'' was placed at its center coordinates (0 mm, 0 mm, –5 mm). The mirror was a $\rm{14~mm} \times \rm{14~mm}$ Bézier surface approximating a concave parabolic mirror with a focal length of approximately 66.7 mm, with its center coordinates set to (0 mm, 0 mm, 20 mm). The virtual object was shaded by a directional light source from (0.77, 0.45, 0.0). A total of 30,000 point light sources were distributed on the object's surface, with initial phases randomized between 0 and $2\pi$. These shading and phase conditions were used for all subsequent experiments in this paper. For the simulated reconstructions, the propagation distance was determined using the imaging equation, assuming an ideal concave mirror.

\begin{figure}[H]
 \centering
 \begin{subfigure}{0.48\textwidth}
 \includegraphics[width=\linewidth]{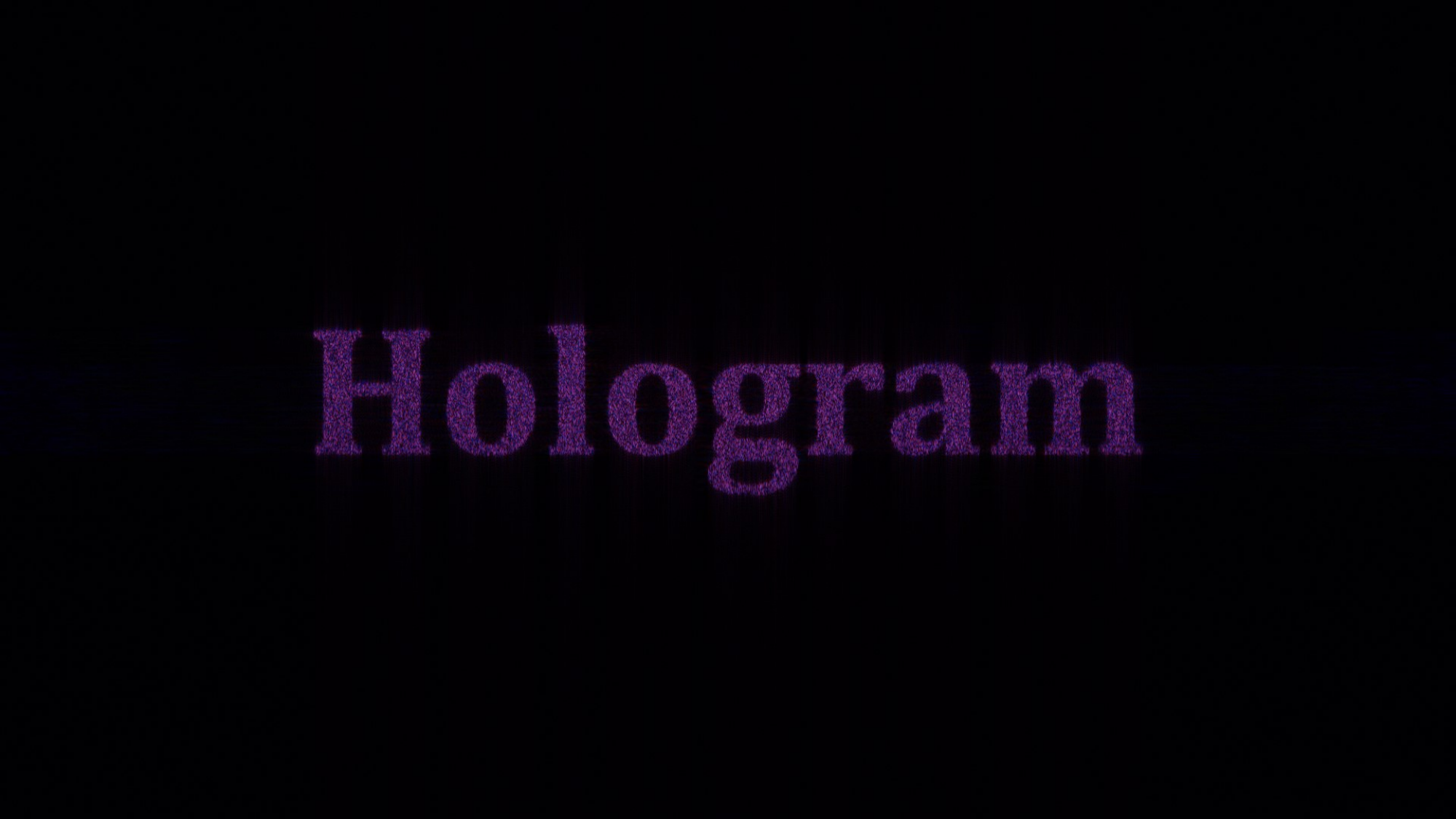}
 \caption{}
 \label{subfig:lambda_14}
 \end{subfigure}
 \hfill % 図と図の間のスペース
 \begin{subfigure}{0.48\textwidth}
 \includegraphics[width=\linewidth]{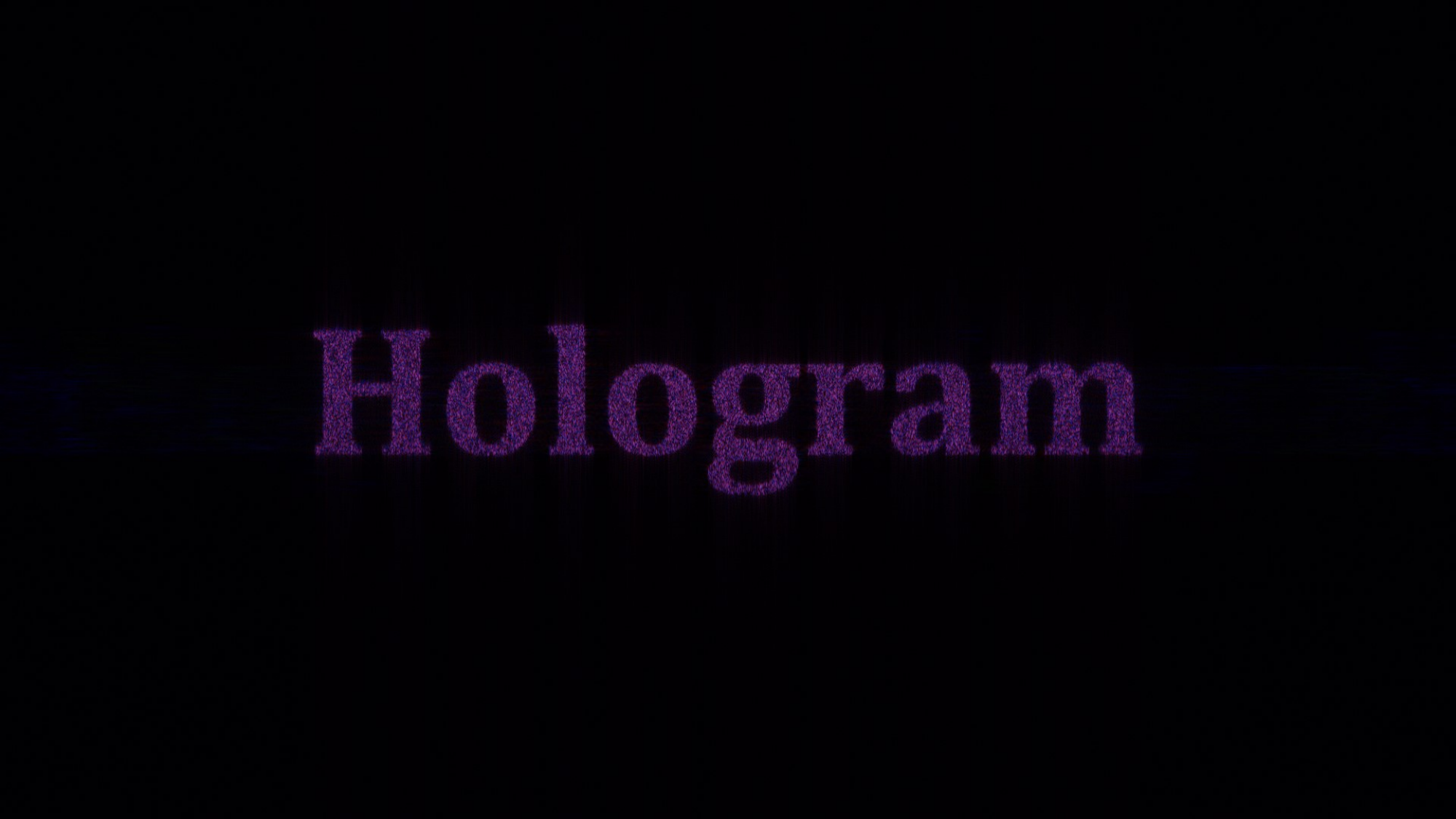}
 \caption{}
 \label{subfig:lambda_4}
 \end{subfigure}
 \vspace{5mm} % 1行目と2行目の間の縦方向のスペース(お好みで調整）
 \begin{subfigure}{0.48\textwidth}
 \includegraphics[width=\linewidth]{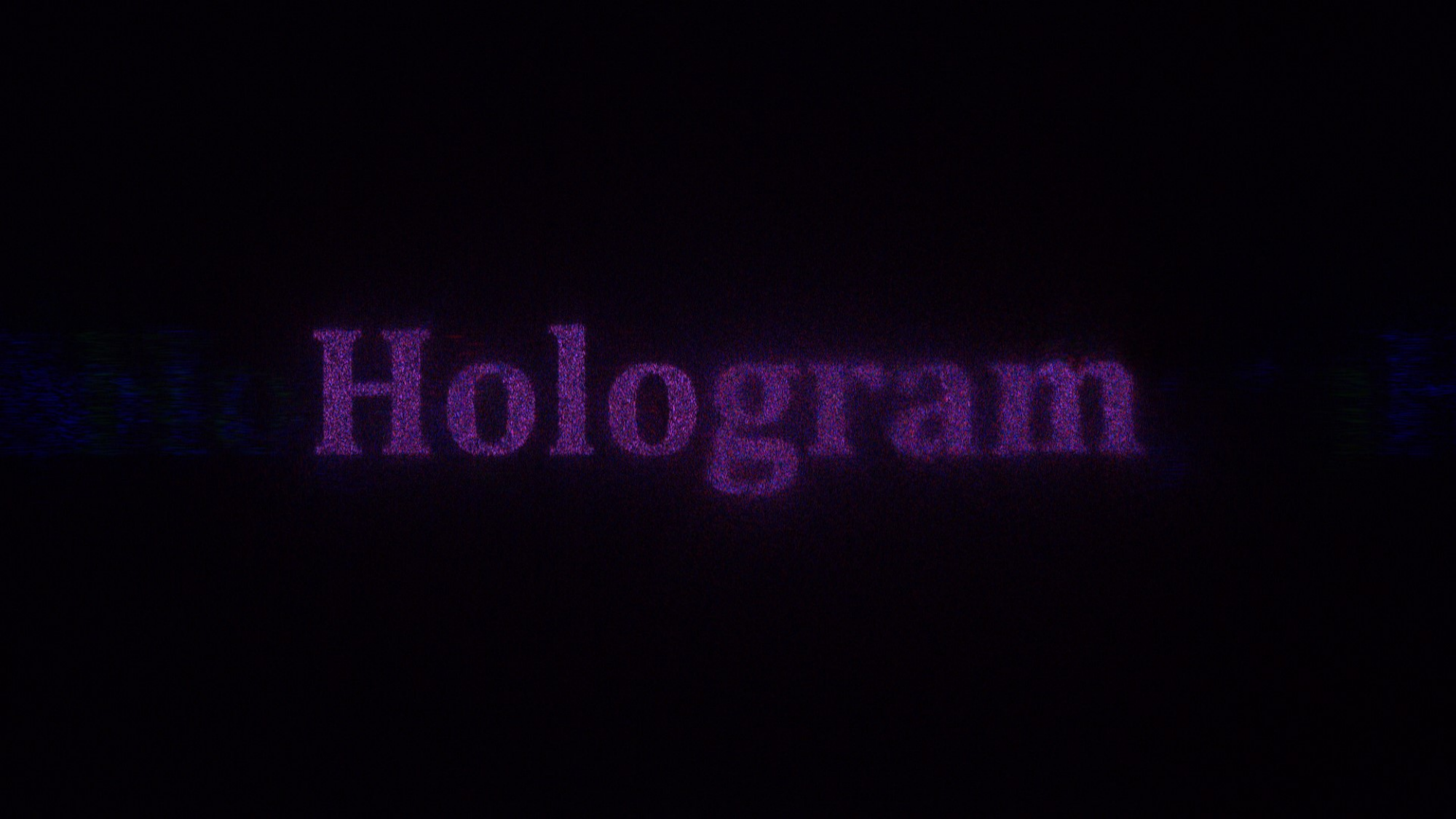}
 \caption{}
 \label{subfig:10_lambda_4}
 \end{subfigure}
 \hfill % 図と図の間のスペース
 \begin{subfigure}{0.48\textwidth}
 \includegraphics[width=\linewidth]{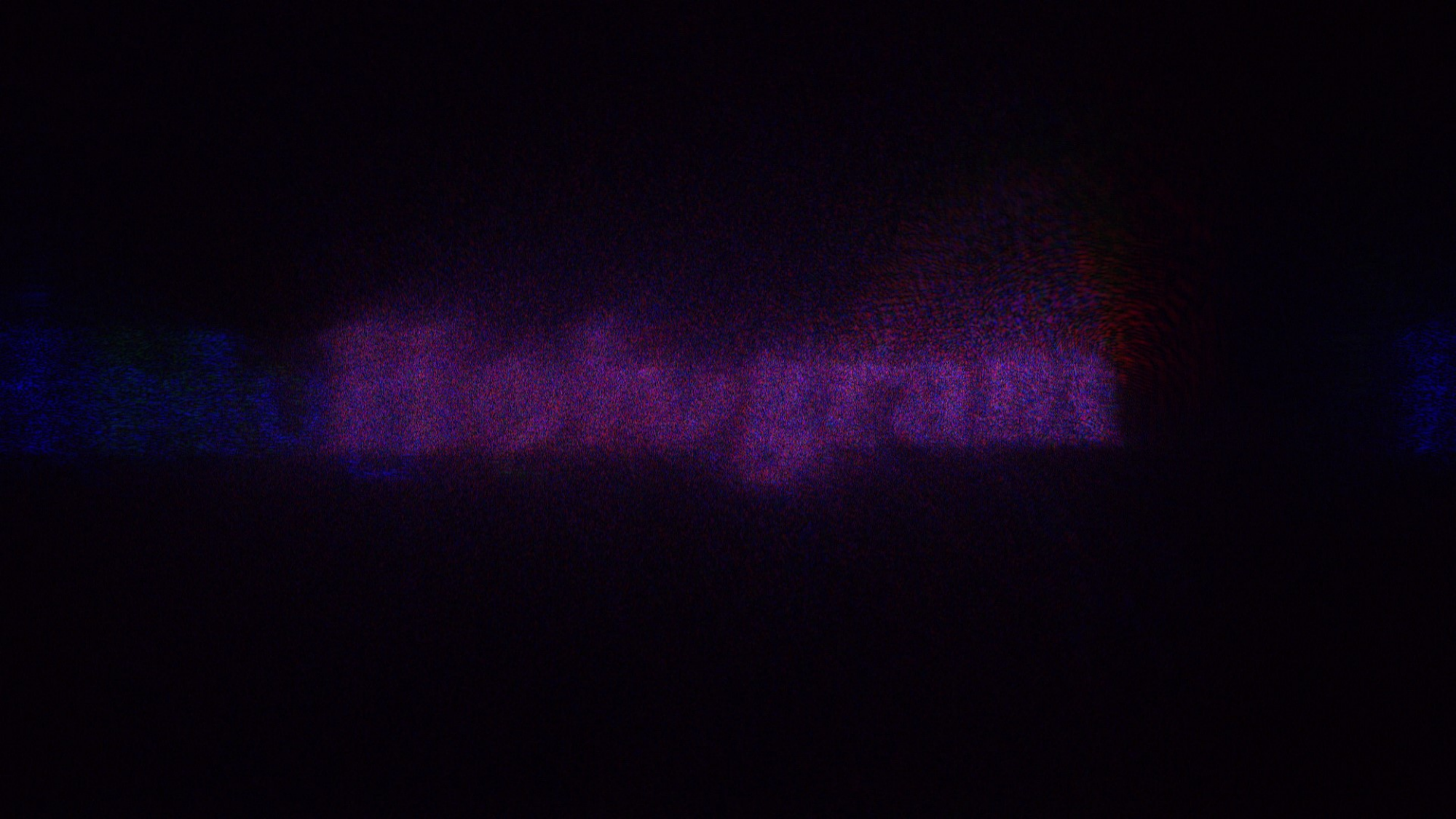}
 \caption{}
 \label{subfig:100_lambda_4}
 \end{subfigure}
\caption{Simulated mirror images calculated with different error tolerances. The tolerance $\Delta L$ for the estimated optical path length error is set to (a) the Maréchal limit ($\frac{\lambda}{14}$), (b) the Rayleigh limit ($\frac{\lambda}{4}$), (c) $10 \times \frac{\lambda}{4}$, and (d) $100 \times \frac{\lambda}{4}$. A visual comparison shows almost no discernible difference between (a) and (b). Image (c) exhibits noticeable blurring, particularly on the outline of the word ``gram,'' while the quality in (d) is degraded to the point of being illegible.}
 \label{fig:error_threshold} % 全体のラベル名を変更
\end{figure}

We tested four tolerance values for the estimated optical path length error $\Delta L$: the Maréchal limit ($\frac{\lambda}{14}$), a common criterion in optics; the Rayleigh limit ($\frac{\lambda}{4}$); and, for comparison, values ten and 100 times larger than the Rayleigh limit. Simulated mirror images were generated for each case. The optimization was carried out using the method of steepest descent, which converges more slowly than Newton's method. This was done to better illustrate the relationship between the tolerance and the final image accuracy, as the rapid convergence of Newton's method can obscure subtle degradations in quality when the tolerance is relaxed. Fig. \ref{fig:error_threshold} shows the simulated mirror images for each tolerance. We also measured the Dice coefficient \cite{ed278621-dc3e-343f-ae66-540d8990b60d} to quantify the shape similarity between the result from the strictest tolerance ($\frac{\lambda}{14}$, shown in (a)), which was treated as the correct image, and the results from the other tolerances (shown in (b)–(d)). The Dice coefficient measures the similarity between two sets and is widely utilized in fields such as image processing to compare object shapes. Here, we define the Dice coefficient for two sets, $X$ and $Y$, as

\begin{table}[H]
 \centering
\caption{Measurement results of Dice coefficient for each tolerance. The image in Fig. \ref{fig:error_threshold}(\subref{subfig:lambda_14}) (tolerance $\frac{\lambda}{14}$) was treated as the correct image.}
 \label{tab:error_dice}
 \begin{tabular}{|c|c|}
 \hline
 $\frac{\lambda}{4}$ & 0.988\\ 
 \hline
 $10\times \frac{\lambda}{4}$ & 0.822 \\
 \hline
 $100\times \frac{\lambda}{4}$ & 0.515\\
 \hline
 \end{tabular}
\end{table}

\begin{equation*}
 DC=\frac{2\left|X\bigcap Y\right|}{\left|X\right|+\left|Y\right|},
\end{equation*}
where $\left|X\bigcap Y\right|$ is the number of matching object pixels in the two images, and $\left|X\right|$ and $\left|Y\right|$ are the total number of pixels representing the object in each image. Table \ref{tab:error_dice} lists the measured Dice coefficients. A visual comparison of the simulated images in Fig. \ref{fig:error_threshold} reveals almost no discernible difference in quality between (a) $\frac{\lambda}{14}$ and (b) $\frac{\lambda}{4}$. In contrast, image (c), at $10\times\frac{\lambda}{4}$, exhibits blurring, particularly around the outline of the word ``gram,'' while image (d), at $100\times\frac{\lambda}{4}$, deteriorates to the point of being illegible. This visual evaluation is quantitatively supported by the Dice coefficient results in Table \ref{tab:error_dice}. The Dice coefficient for (b) $\frac{\lambda}{4}$ is very high at 0.988, indicating a high degree of shape similarity with the correct image. The coefficients then drop significantly to 0.822 for (c) $10\times\frac{\lambda}{4}$ and 0.515 for (d) $100\times\frac{\lambda}{4}$. These results clearly demonstrate that the tolerance for the estimated optical path length error $\Delta L$, as presented in Section \ref{subsec:stopping_condition}, directly impacts the accuracy of the mirror image. Specifically, setting the tolerance to the Rayleigh limit ($\frac{\lambda}{4}$) is sufficient to obtain a high-fidelity mirror image. Therefore, we conclude that setting $\Delta L < \frac{\lambda}{4}$ as the termination condition for the iterative calculation is appropriate in terms of both image quality and computational efficiency. On the basis of this finding, all subsequent experiments were conducted with the tolerance set to $\frac{\lambda}{4}$.

\subsection{Accuracy evaluation}
\subsubsection{Cylindrical mirror}

\begin{figure}[h]
\centering
\includegraphics[clip,width=10.0cm]{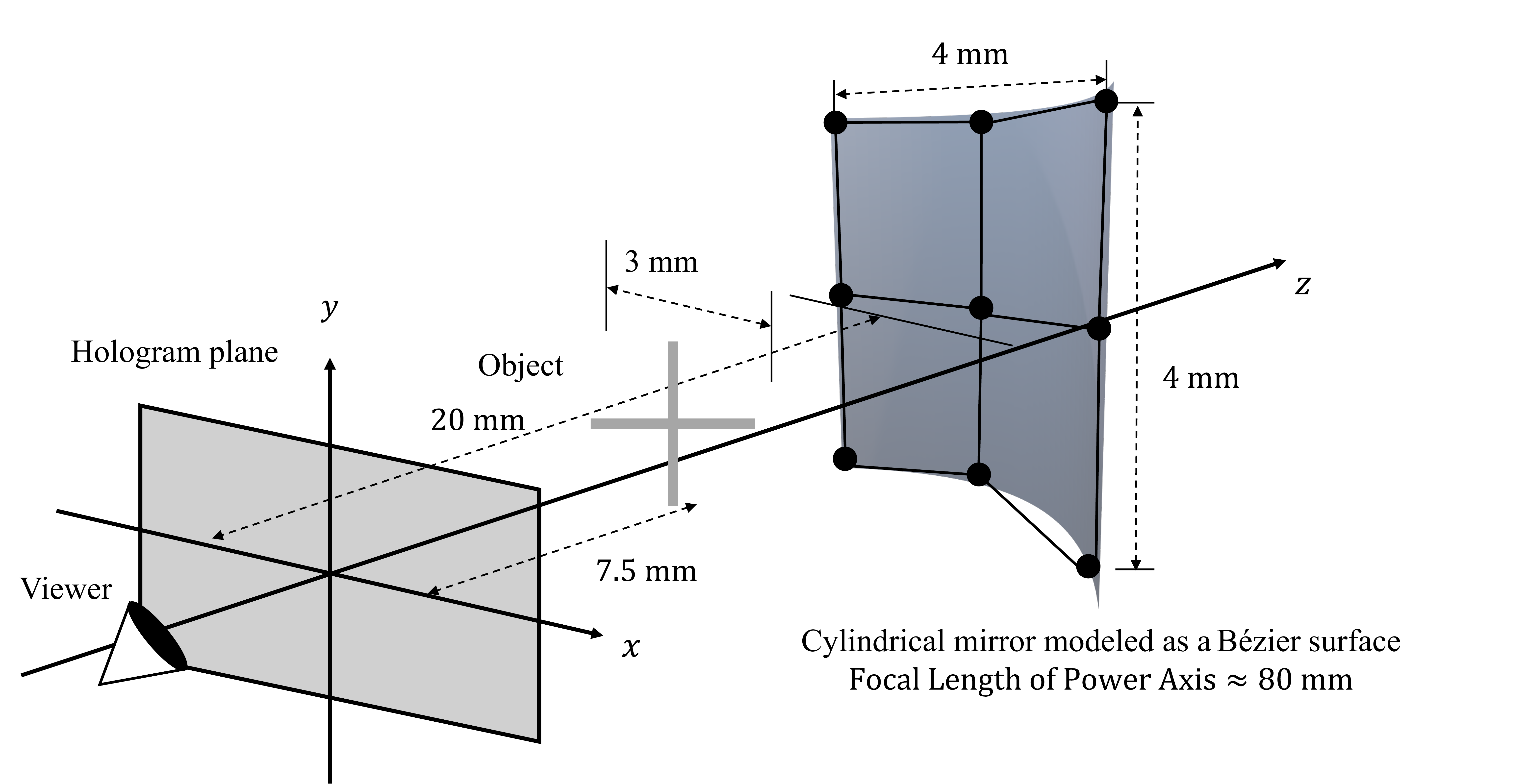}
\caption{Diagram of virtual object placement used in the experiments to verify the validity of mirror images using a cylindrical mirror. The virtual object is a 3-mm-wide cross-shaped object placed at center coordinates (0 mm, 0 mm, 7.5 mm). There are 200 point light sources placed on the surface of the object. The hologram plane is set to $z=0$. The mirror surface is a quadratic Bézier surface ($3\times3$ control points) measuring $\rm{4~mm} \times \rm{4~mm}$ that approximates a concave and convex cylindrical mirror with a focal length of approximately 80 mm, and its center coordinates are (0 mm, 0 mm, 20 mm). Although the actual object is positioned in a way that it can be seen, in this experiment only the mirror image is calculated.}
\label{fig:placement_cylindrical}
\end{figure}

\begin{figure}[t]
 \centering
 \begin{subfigure}{0.48\textwidth}
 \includegraphics[width=\linewidth]{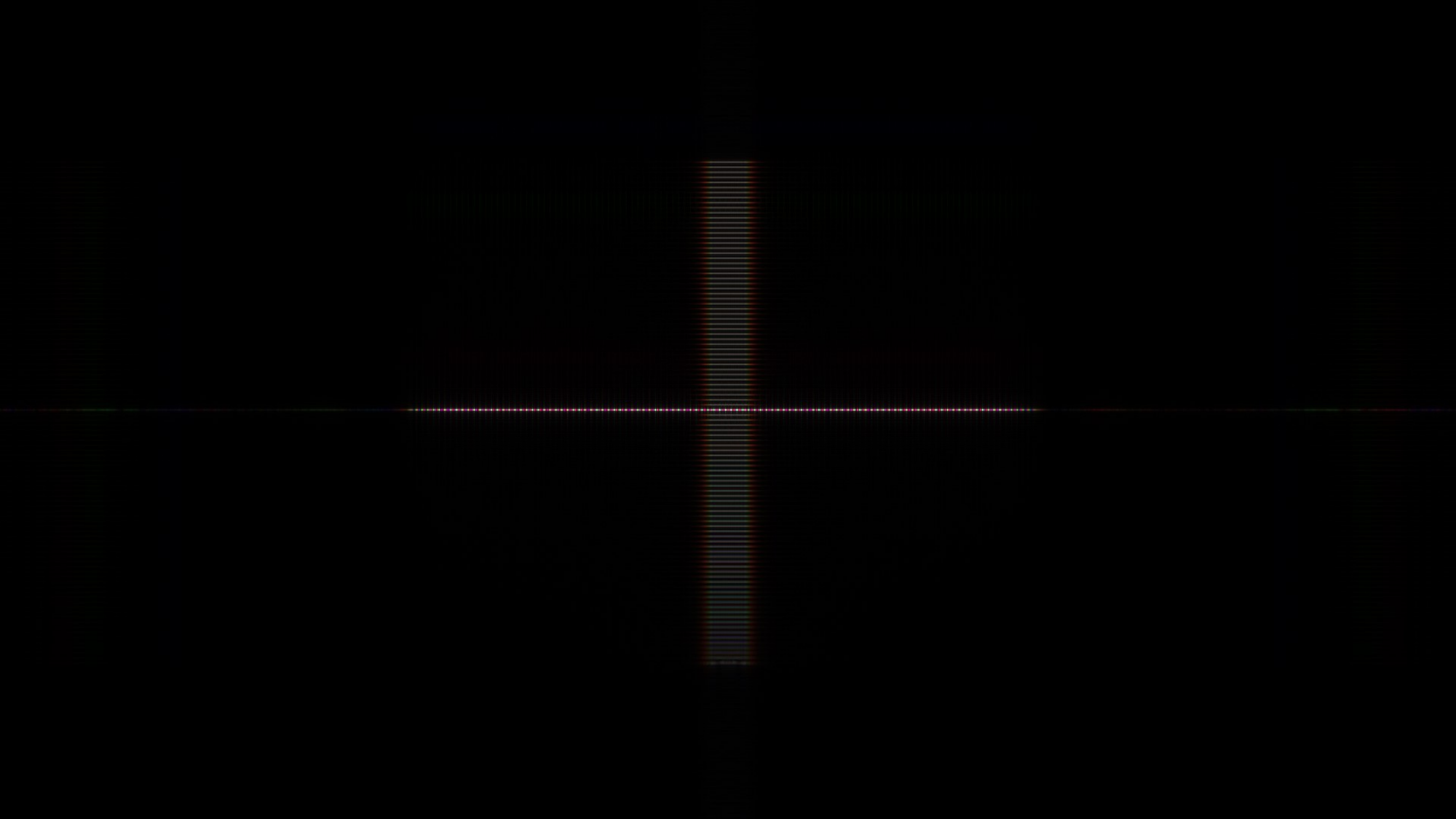}
 \caption{}
 \label{subfig:concave_horizontal}
 \end{subfigure}
 \hfill % 図と図の間のスペース
 \begin{subfigure}{0.48\textwidth}
 \includegraphics[width=\linewidth]{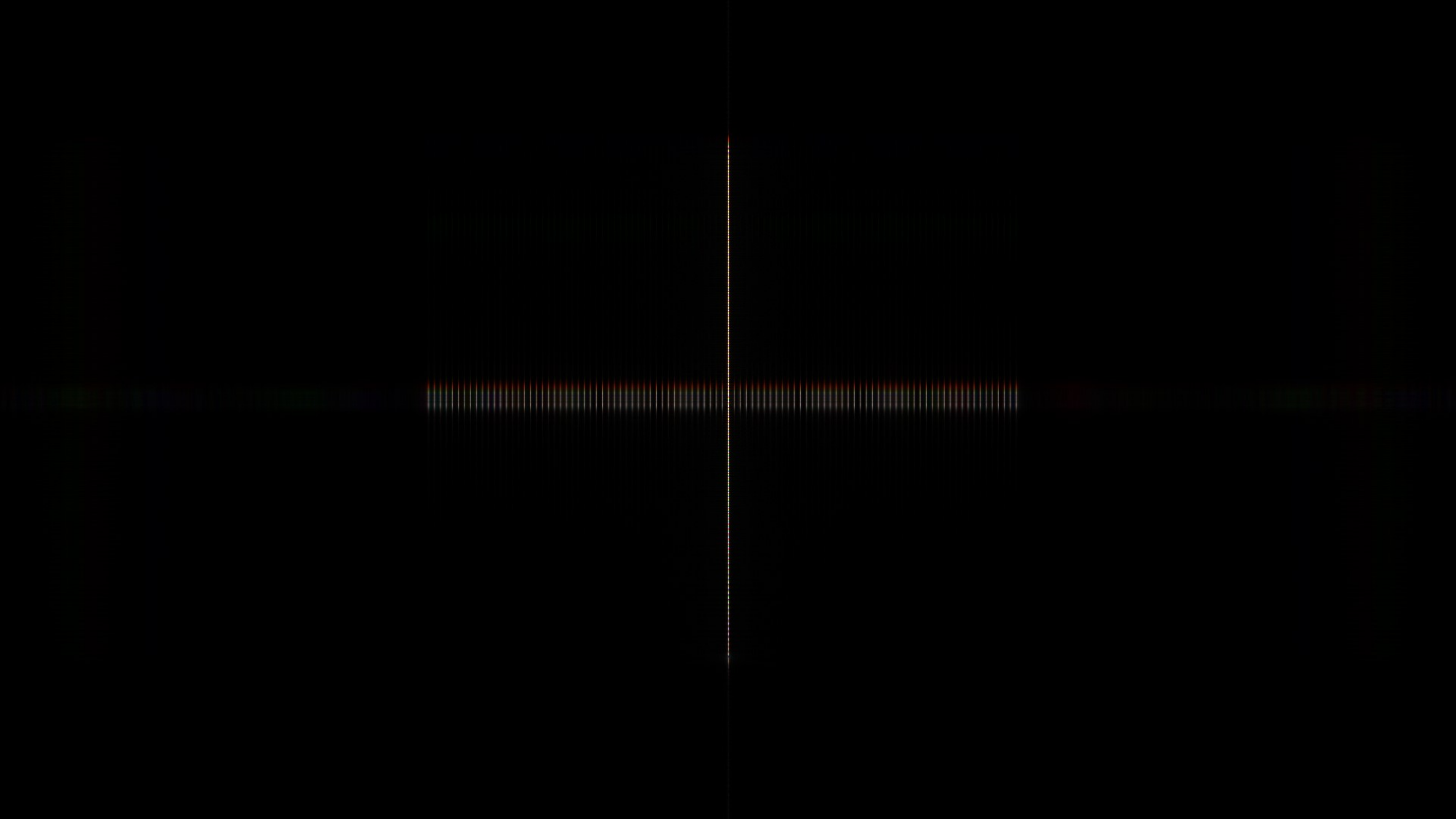}
 \caption{}
 \label{subfig:concave_vertical}
 \end{subfigure}
 \vspace{5mm} % 1行目と2行目の間の縦方向のスペース(お好みで調整）
 \begin{subfigure}{0.48\textwidth}
 \includegraphics[width=\linewidth]{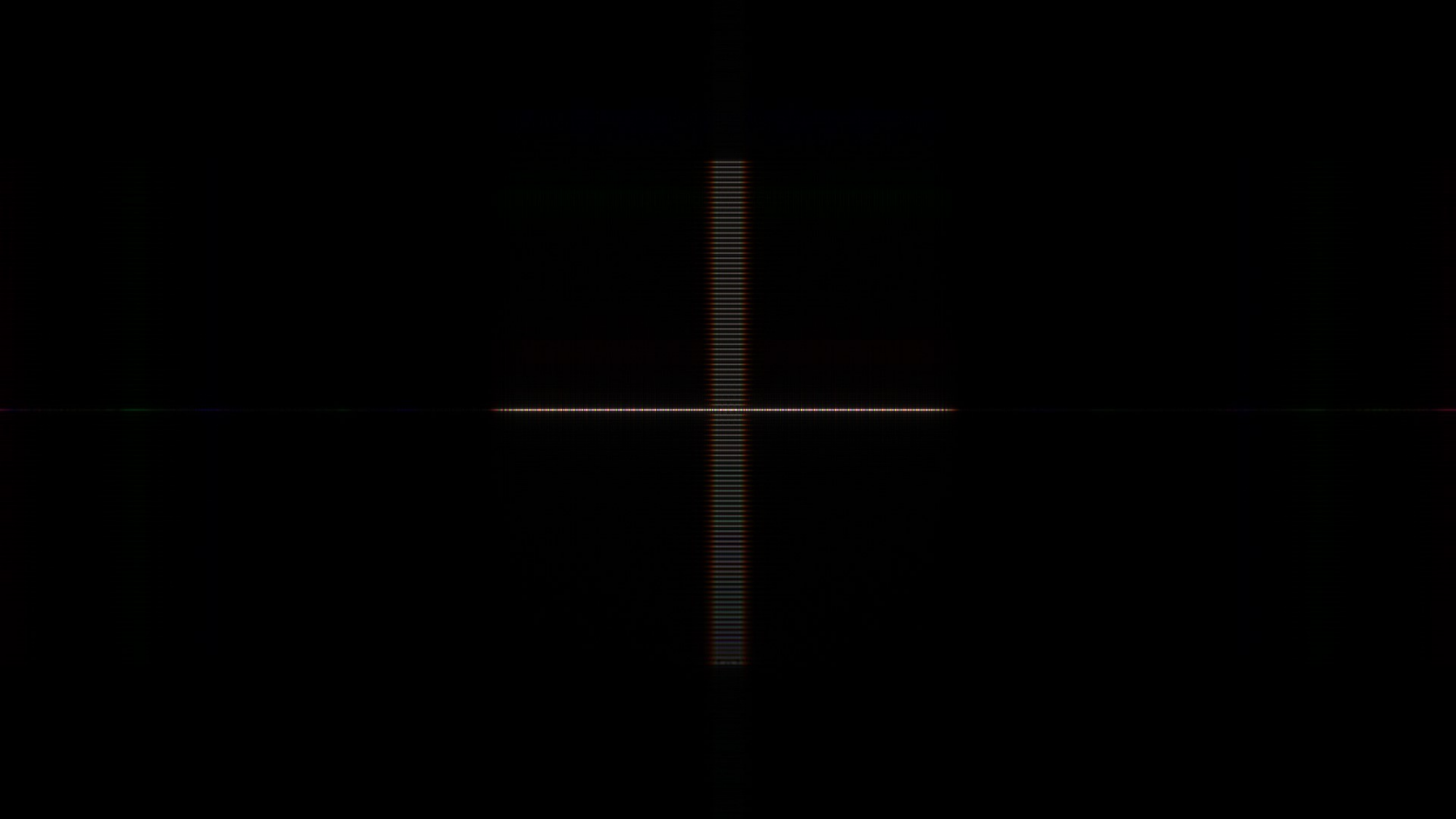}
 \caption{}
 \label{subfig:convex_horizontal}
 \end{subfigure}
 \hfill % 図と図の間のスペース
 \begin{subfigure}{0.48\textwidth}
 \includegraphics[width=\linewidth]{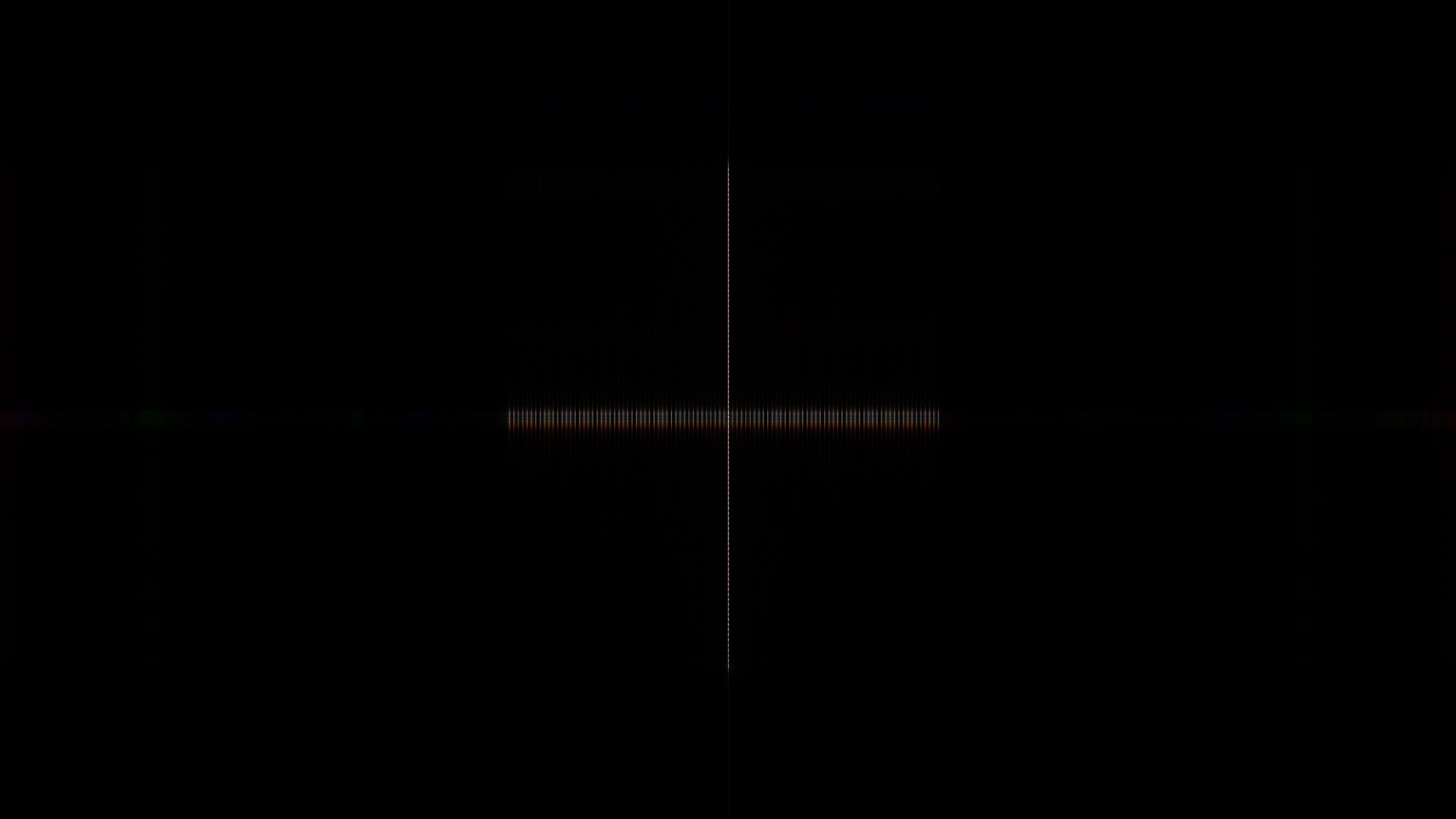}
 \caption{}
 \label{subfig:convex_vertical}
 \end{subfigure}
\caption{Simulated mirror images from concave and convex cylindrical mirrors, demonstrating the reconstruction of astigmatism. Images (a) and (b) are from a concave mirror; (c) and (d) are from a convex mirror. The reconstructions are focused at two different depths, each corresponding to an image distance calculated for one of the mirror's axes. For images (a) and (c), the image distance was calculated for the curved axis using the imaging equation. For images (b) and (d), the distance was calculated for the linear axis by treating the mirror as planar. As a result, the horizontal line of the cross is in sharp focus in (a) and (c), whereas the vertical line is in sharp focus in (b) and (d). This confirms that the proposed method correctly reconstructs astigmatism.}
 \label{fig:simulation_cylindrical} % 全体のラベル名を変更
\end{figure}

In this section, we conduct a simulation experiment using a cylindrical mirror, which has unique optical properties, to verify that the proposed method generates physically valid mirror images.

Cylindrical mirrors have an asymmetric shape, with curvature in only one axis and a linear profile in the other. Therefore, unlike spherical mirrors which have a single focal point, light converges differently along the curved axis and the linear axis. Specifically, light along the curved axis forms a focal line at a finite distance, while light along the linear axis remains parallel, focusing at infinity. This experiment verifies whether the proposed method can correctly reconstruct this phenomenon, known as astigmatism.

The diagram of virtual object placement used in this experiment is shown in Fig. \ref{fig:placement_cylindrical}. The hologram plane was set to $z = 0$, and a 3-mm-wide cross-shaped virtual object was placed at its center coordinates (0 mm, 0 mm, 7.5 mm). For the mirror, we used Bézier surfaces approximating concave and convex cylindrical mirrors, each measuring $\rm{4~mm} \times \rm{4~mm}$ with a focal length of approximately 80 mm, and their center coordinates were set to (0 mm, 0 mm, 20 mm). A total of 200 point light sources were placed on the object's surface with initial phases randomized between 0 and $2\pi$. In this experiment, only the mirror image was computed to generate the simulated reconstructions.

As shown in Fig. \ref{fig:simulation_cylindrical}, the reconstructed images were generated using two different propagation distances, each calculated to match one of the two distinct focal points of the cylindrical mirror. For the curved axis, the propagation distance was determined by applying the imaging equation with the mirror's finite focal length (Fig. \ref{fig:simulation_cylindrical}(\subref{subfig:concave_horizontal}) and (\subref{subfig:convex_horizontal})). For the linear axis (which has a focal point at infinity), the distance was calculated by treating the cylindrical mirror as planar and assuming the mirror image is located at a position symmetric to the object with respect to the mirror plane (Fig. \ref{fig:simulation_cylindrical}(\subref{subfig:concave_vertical}) and (\subref{subfig:convex_vertical})). In Fig. \ref{fig:simulation_cylindrical}(\subref{subfig:concave_horizontal}) and (\subref{subfig:convex_horizontal}), the horizontal line of the cross is sharply focused, indicating that light propagation along the curved axis was calculated correctly. Conversely, in Fig. \ref{fig:simulation_cylindrical}(\subref{subfig:concave_vertical}) and (\subref{subfig:convex_vertical}), the vertical line is sharply focused, confirming the correct calculation for the linear axis. This demonstrates that the proposed method accurately reconstructs the astigmatism of a cylindrical mirror—that is, the optical property of having different focal lengths in different directions.

\subsubsection{Parabolic Mirror}
\label{subsec:parabolic_mirror}
\begin{figure}[t]
\centering
\includegraphics[clip,width=10.0cm]{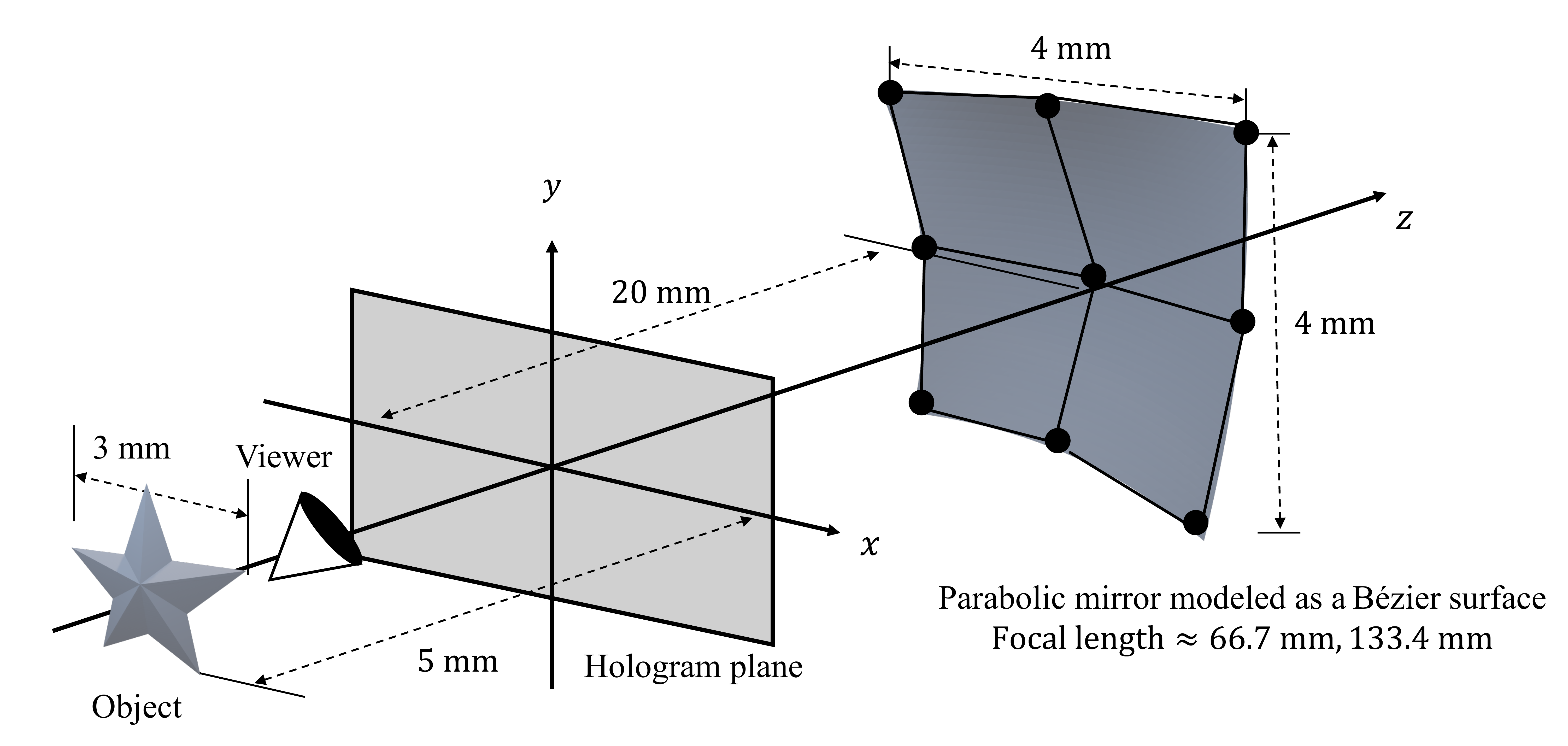}
\caption{Diagram of virtual object placement used in the experiments to verify the validity of mirror images using a parabolic mirror. The virtual object is a 3-mm-wide star-shaped object placed at center coordinates (0 mm, 0 mm, –5 mm). There are 30,000 point light sources placed on the surface of the object. The hologram plane is set to $z=0$. The mirror surface is a quadratic Bézier surface ($3\times3$ control points) measuring $\rm{4 ~mm} \times \rm{4~mm}$ that approximates a planar mirror and concave/convex parabolic mirror with a focal length of approximately 66.7 mm and 133.4 mm, and its center coordinates are (0 mm, 0 mm, 20 mm). The object is placed behind the hologram plane, and only the mirror image is calculated.}
\label{fig:placement_parabolic}
\end{figure}

\begin{table}[t]
 \centering
\caption{Measurement results of the Dice coefficient for each mirror surface shape. For the simulation images shown in Fig. \ref{fig:simulation_parabolic}, the results of the subdivision method (left column) were used as the correct images to calculate the shape similarity with the results of the proposed method (right column).}
 \label{tab:parabolic_dice}
 \begin{tabular}{|c|c|}
 \hline
 Concave mirror ($f\approx\rm{66.7~mm}$) & 0.951\\ 
 \hline
 Concave mirror ($f\approx\rm{133.4~mm}$) & 0.973\\ 
 \hline
 Planar mirror & 0.985\\ 
 \hline
 Convex mirror ($f\approx\rm{133.4~mm}$) & 0.987\\ 
 \hline
 Convex mirror ($f\approx\rm{66.7~mm}$) & 0.984\\ 
 \hline
 \end{tabular}
\end{table}

\begin{table}[t]
 \centering
\caption{Comparison of measured and theoretical values for the depth of the mirror image for each mirror surface shape. The depth of the optically reconstructed image shown in Fig. \ref{fig:reconstructed_parabolic} was measured and compared with the theoretical value calculated from the imaging equation.}
 \label{tab:parabolic_depth}
 \begin{tabular}{|c|c|c|}
 \hline
 & Measured value & Theoretical value\\
 \hline
 Concave mirror ($f\approx\rm{66.7~mm}$) & 62mm & 60.28mm\\ 
 \hline
 Concave mirror ($f\approx\rm{133.4~mm}$) & 48mm & 50.87mm\\ 
 \hline
 Planar mirror & 44mm & 45.00mm\\ 
 \hline
 Convex mirror ($f\approx\rm{133.4~mm}$) & 42mm & 41.12mm\\ 
 \hline
 Convex mirror ($f\approx\rm{66.7~mm}$) & 37mm & 38.30mm\\ 
 \hline
 \end{tabular}
\end{table}

In this section, we verify the accuracy of the mirror images generated by the proposed method using a more practical mirror shape: the parabolic mirror. The evaluation was conducted through both numerical simulations and optical experiments.

The diagram of virtual object placement used in this experiment is shown in Fig. \ref{fig:placement_parabolic}. The hologram plane was set to $z = 0$, and a 3-mm-wide star-shaped object was placed at its center coordinates (0 mm, 0 mm, –5 mm). Five types of mirrors were used: a planar mirror measuring 4 mm × 4 mm, and Bézier surfaces approximating concave and convex parabolic mirrors with focal lengths of approximately 66.7 mm and 133.4 mm. The center coordinates of each mirror were set to (0 mm, 0 mm, 20 mm). A total of 30,000 point light sources were placed on the object's surface with initial phases randomized between 0 and $2\pi$.

For the evaluation, the results of the proposed method were compared against those from the conventional mirror surface subdivision method, which served as the reference for this experiment. To generate the reference images, all five mirror types were discretized into 3,200 triangular polygons using the subdivision method, from which simulated and optically reconstructed images were obtained. The validity of using 3,200 subdivisions has been established in prior research with similar scene configurations, ensuring it provides sufficient accuracy for a comparative analysis.

\begin{figure}[H]
 \centering
 \begin{subfigure}{0.42\textwidth}
 \includegraphics[width=\linewidth]{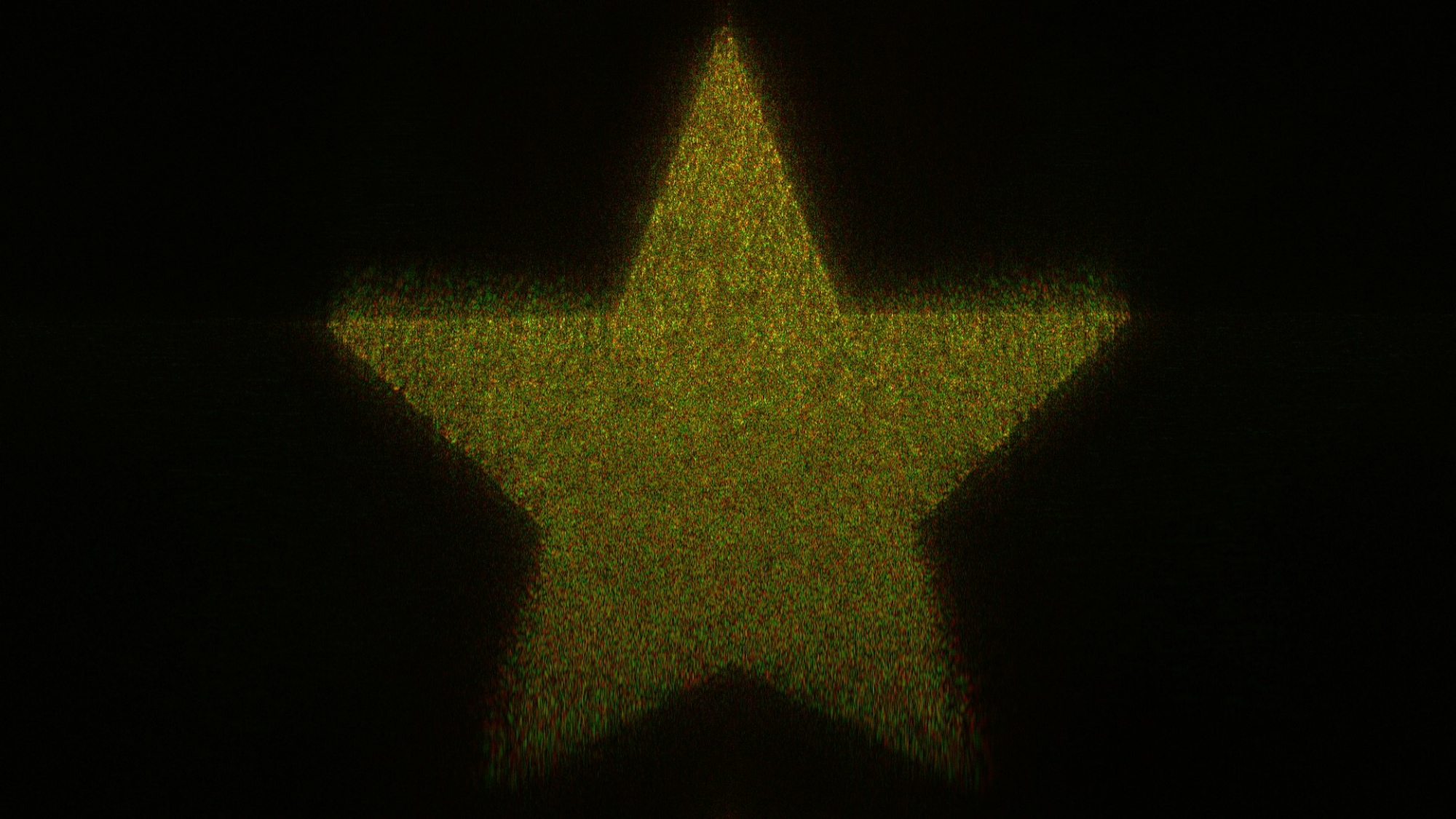}
 \caption{}
 \label{subfig:simulation_subdivision_concave}
 \end{subfigure}
 \begin{subfigure}{0.42\textwidth}
 \includegraphics[width=\linewidth]{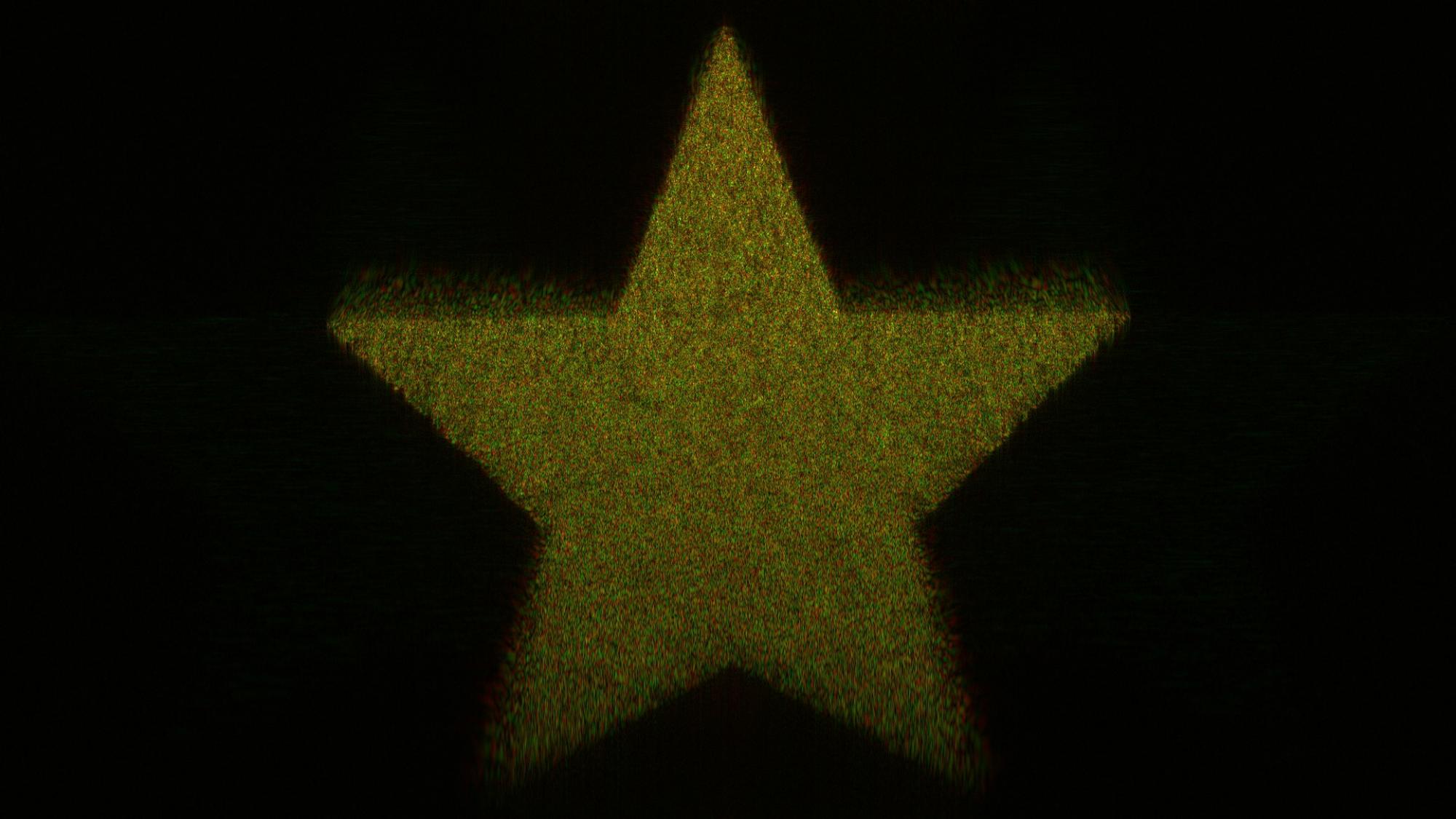}
 \caption{}
 \label{subfig:simulation_proposed_concave}
 \end{subfigure}
 \begin{subfigure}{0.42\textwidth}
 \includegraphics[width=\linewidth]{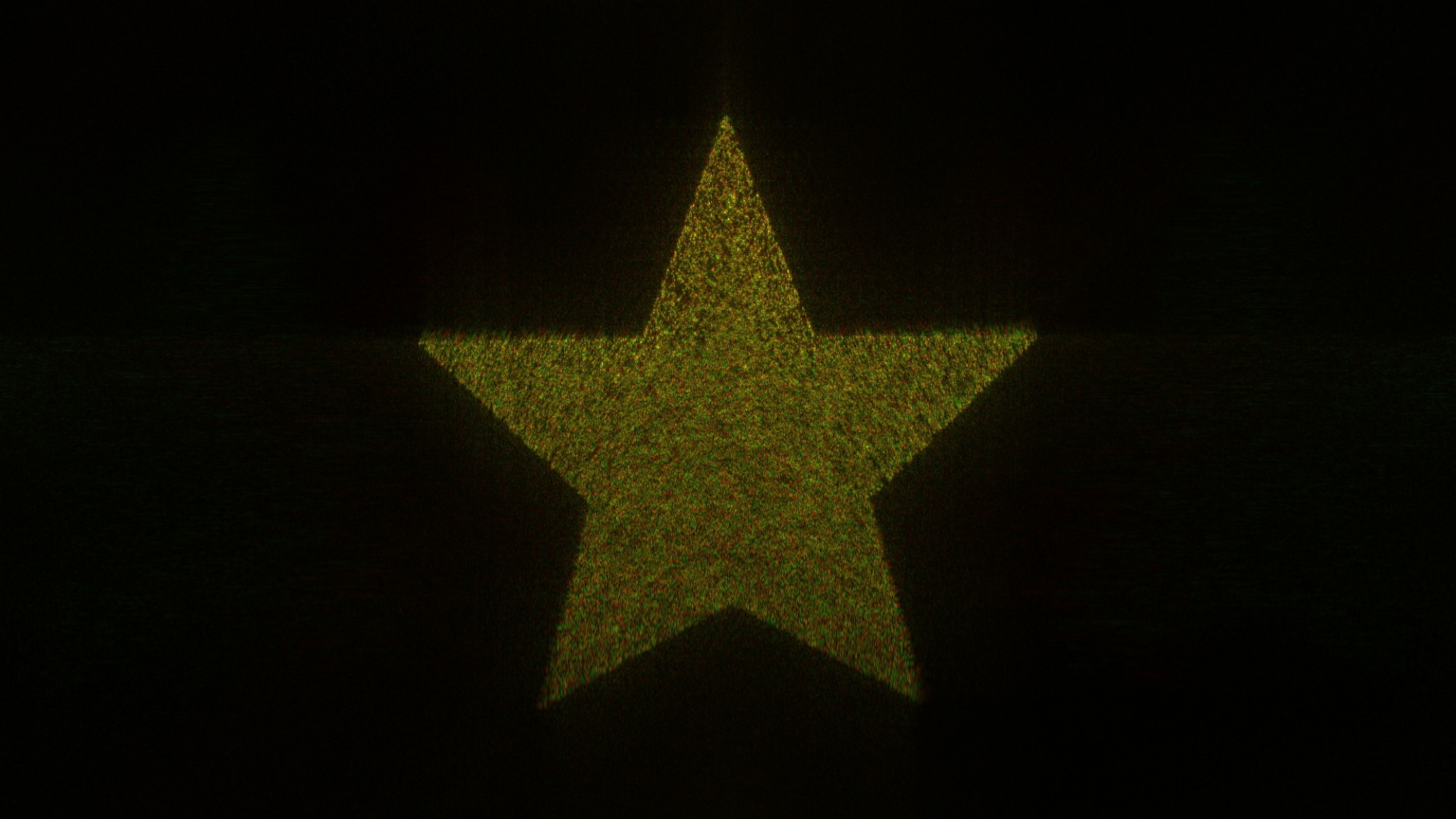}
 \caption{}
 \label{subfig:simulation_subdivision_weakly_concave}
 \end{subfigure}
 \begin{subfigure}{0.42\textwidth}
 \includegraphics[width=\linewidth]{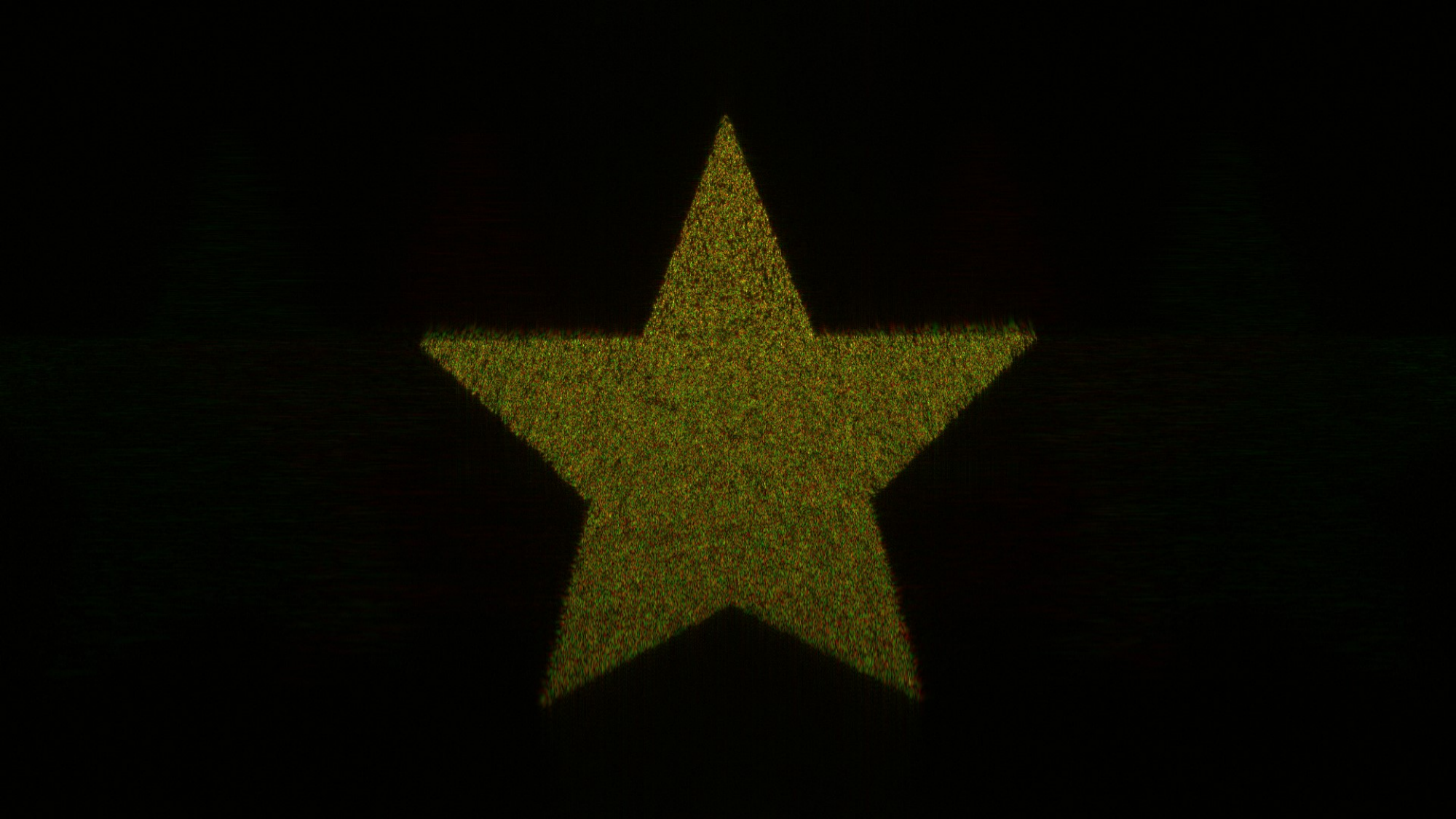}
 \caption{}
 \label{subfig:simulation_proposed_weakly_concave}
 \end{subfigure}
 \begin{subfigure}{0.42\textwidth}
 \includegraphics[width=\linewidth]{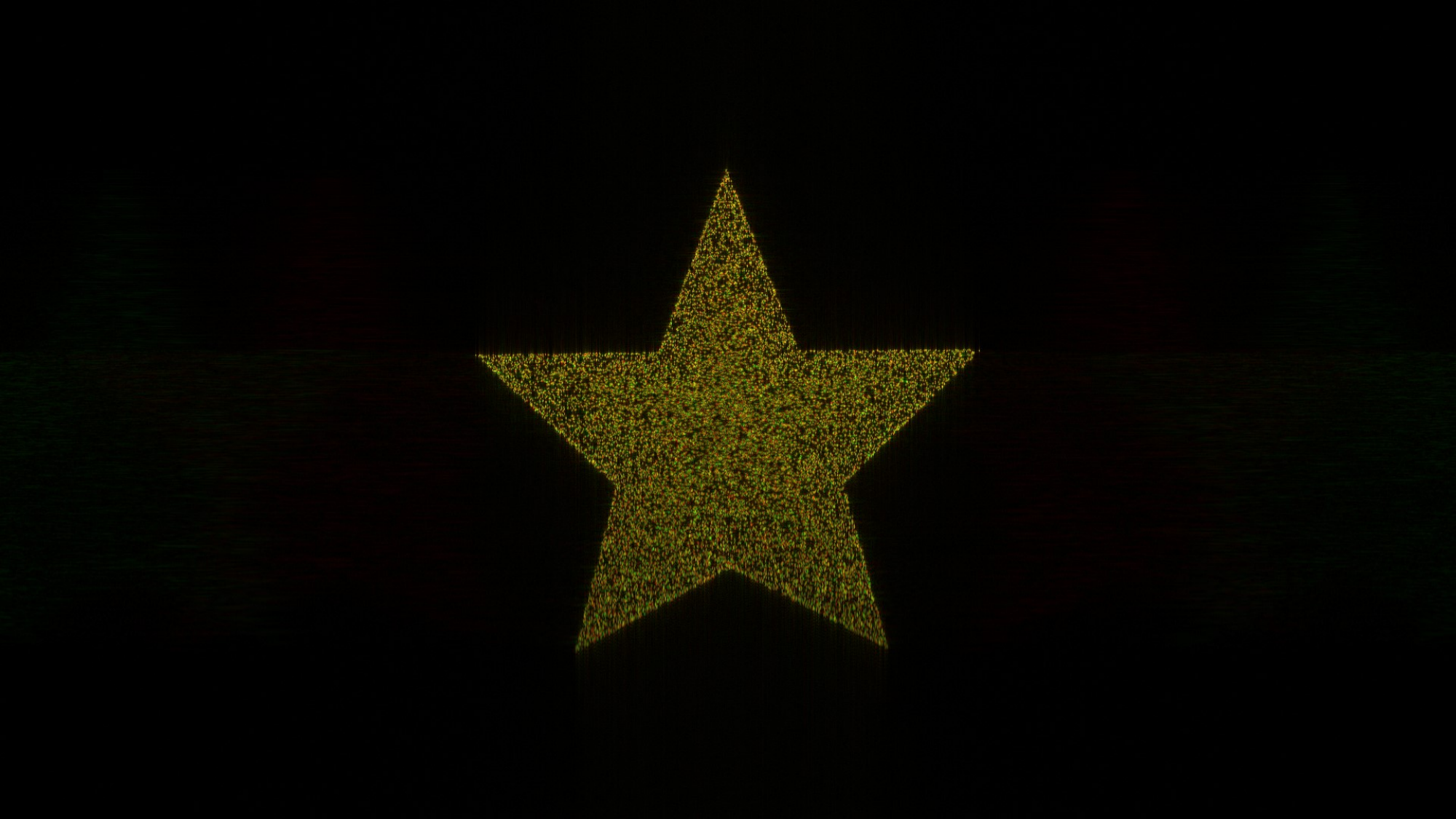}
 \caption{}
 \label{subfig:simulation_subdivision_planar}
 \end{subfigure}
 \begin{subfigure}{0.42\textwidth}
 \includegraphics[width=\linewidth]{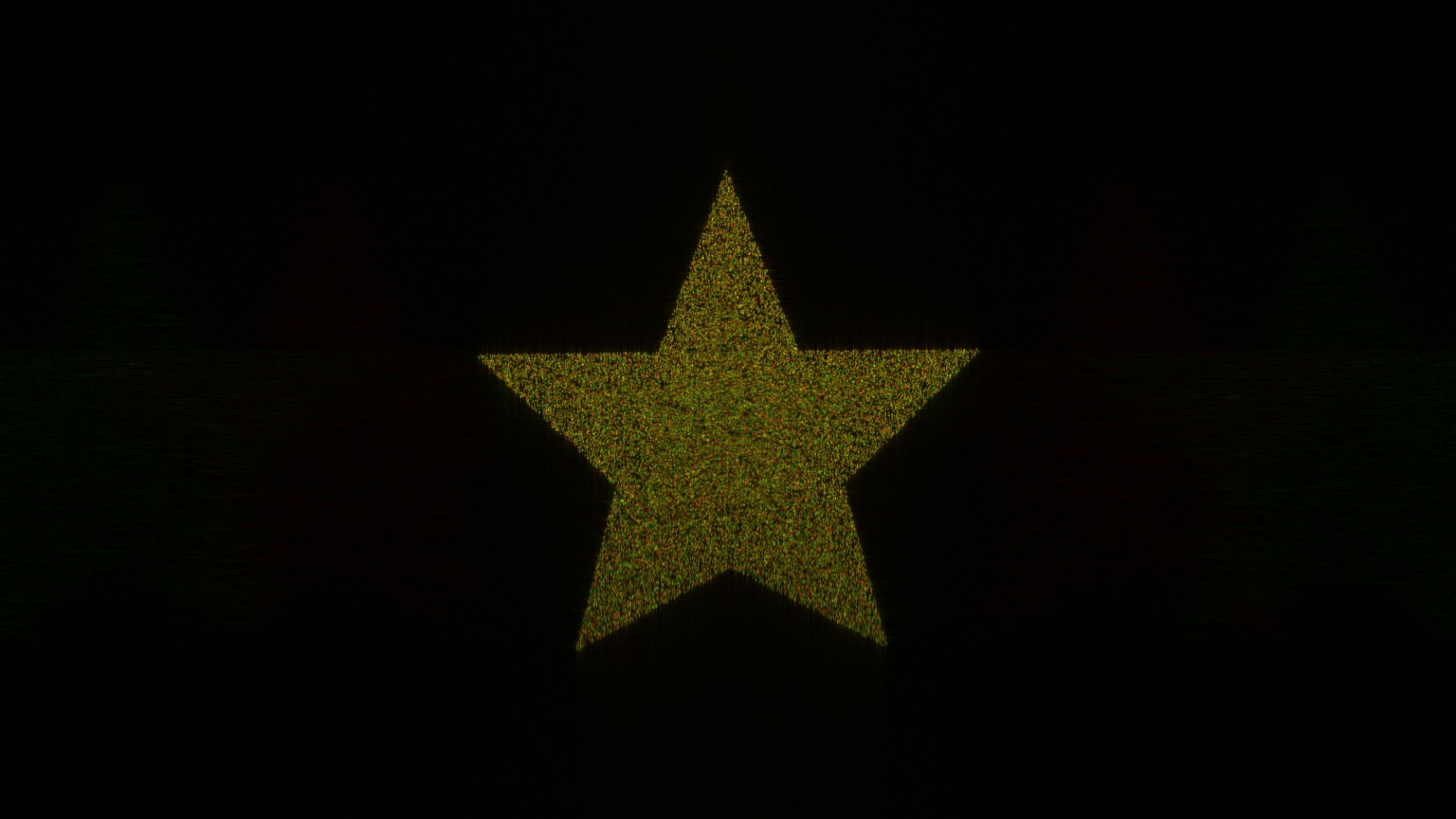}
 \caption{}
 \label{subfig:simulation_proposed_planar}
 \end{subfigure}
 \begin{subfigure}{0.42\textwidth}
 \includegraphics[width=\linewidth]{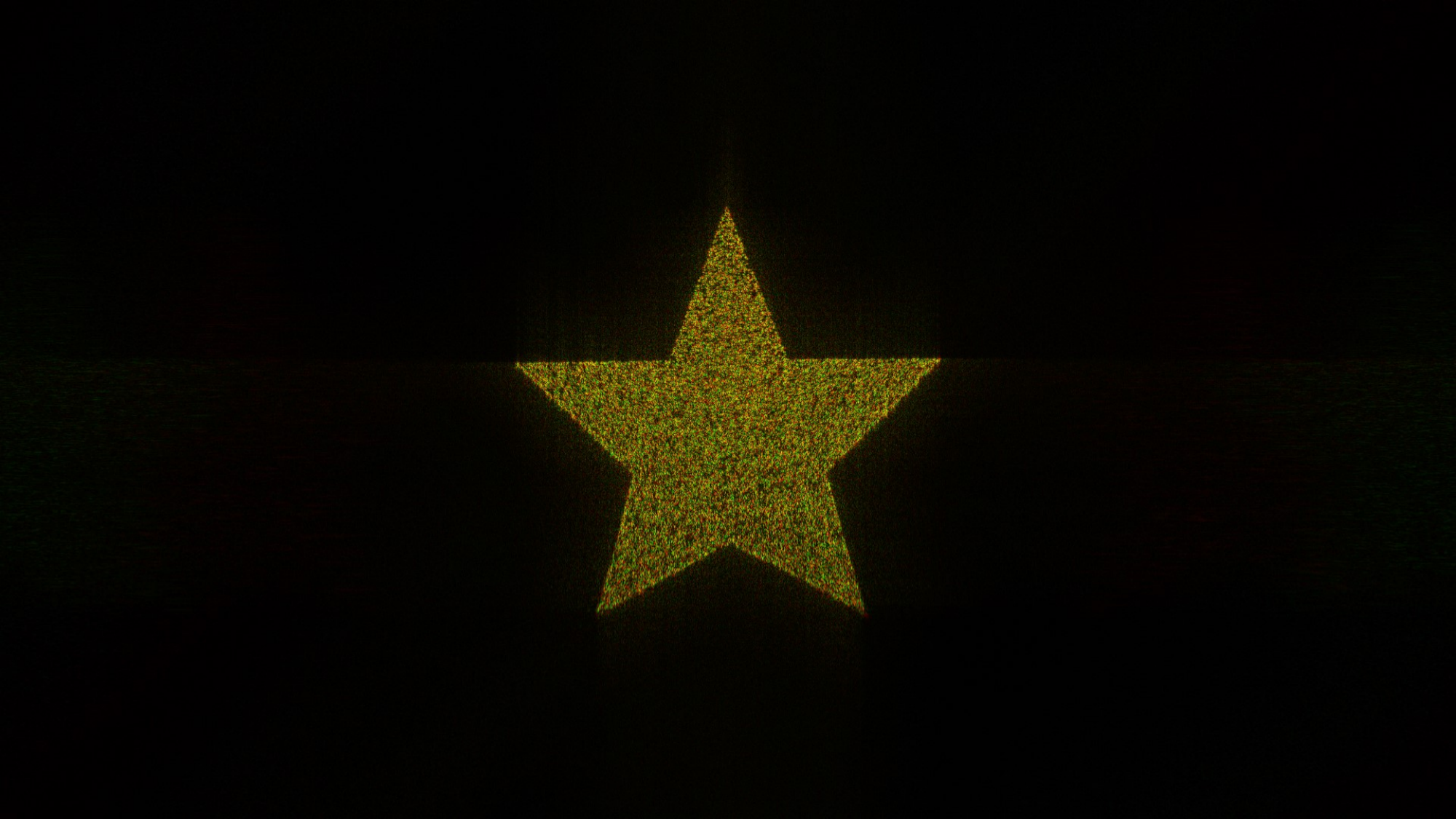}
 \caption{}
 \label{subfig:simulation_subdivision_weakly_convex}
 \end{subfigure}
 \begin{subfigure}{0.42\textwidth}
 \includegraphics[width=\linewidth]{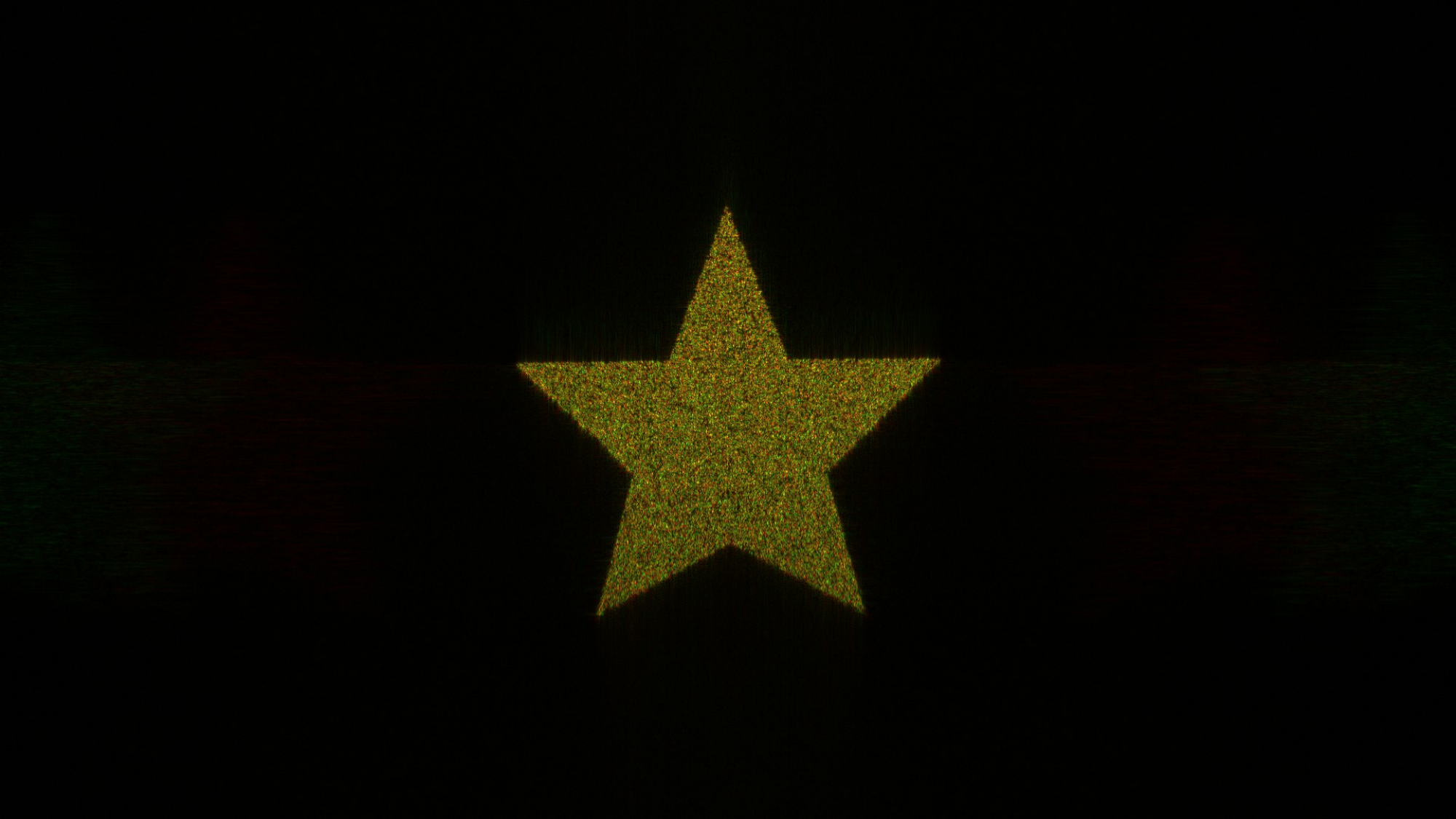}
 \caption{}
 \label{subfig:simulation_proposed_weakly_convex}
 \end{subfigure}
 \begin{subfigure}{0.42\textwidth}
 \includegraphics[width=\linewidth]{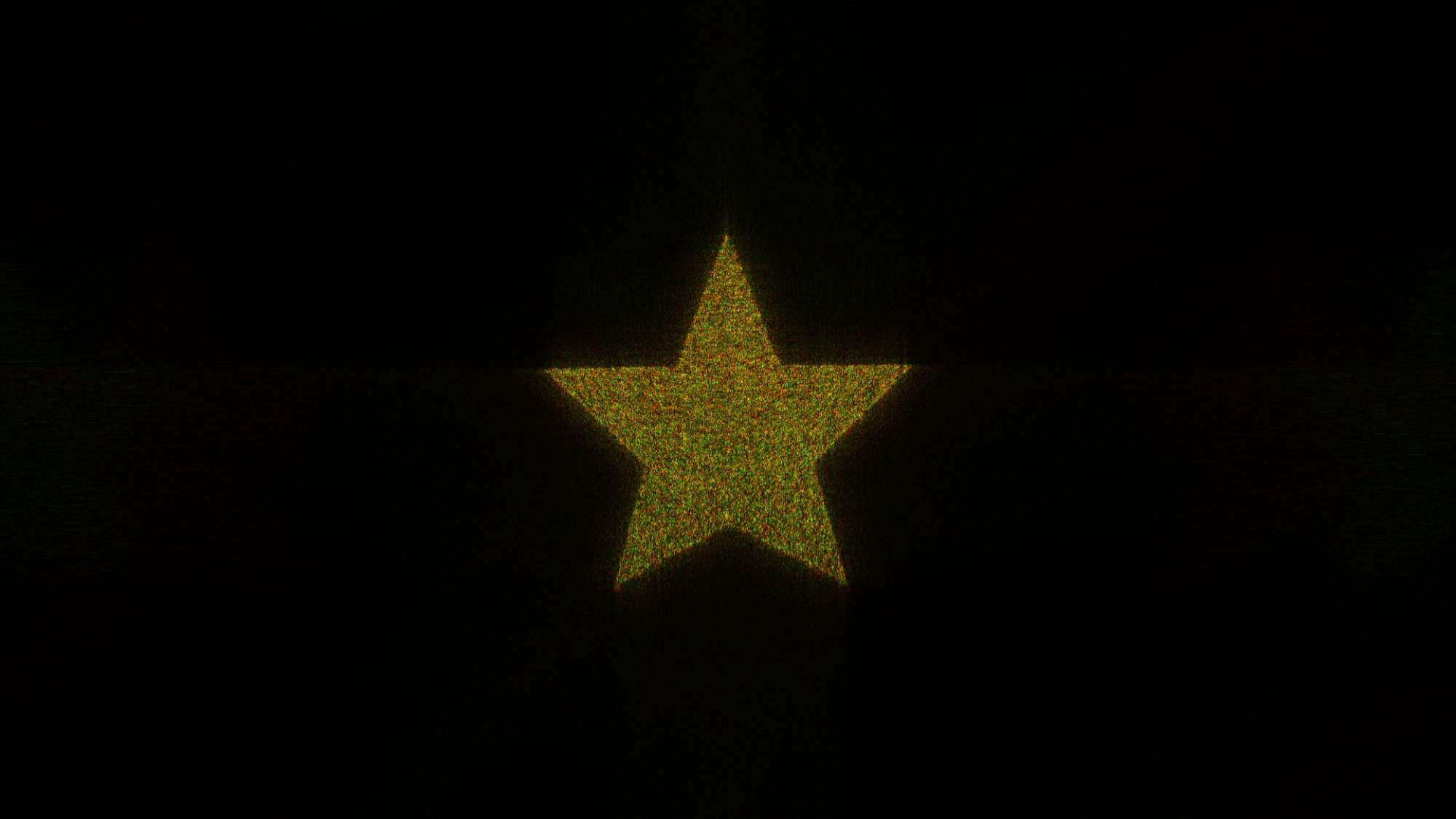}
 \caption{}
 \label{subfig:simulation_subdivision_convex}
 \end{subfigure}
 \begin{subfigure}{0.42\textwidth}
 \includegraphics[width=\linewidth]{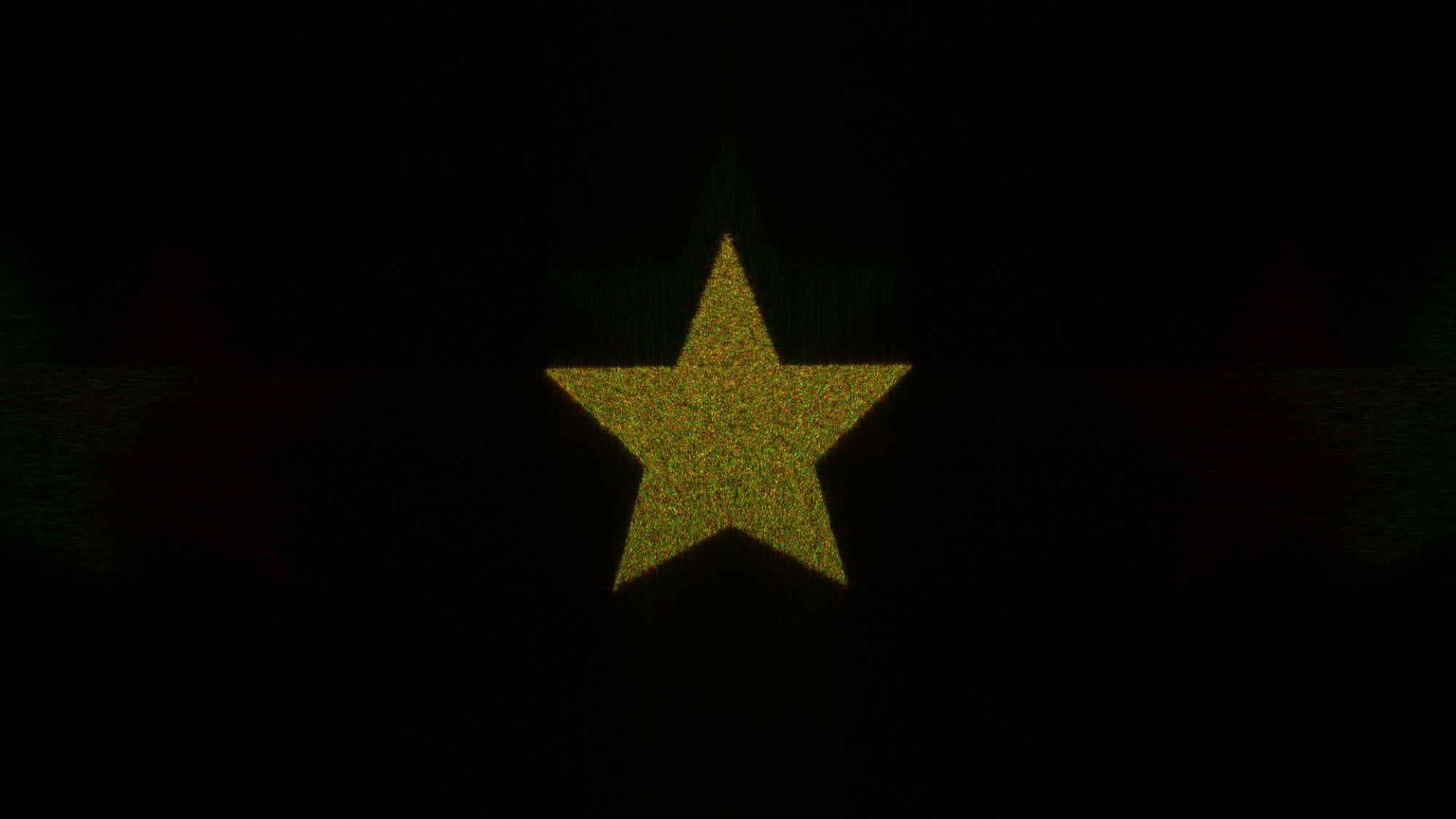}
 \caption{}
 \label{subfig:simulation_proposed_convex}
 \end{subfigure}
\caption{Simulated mirror images using parabolic mirrors. The left column shows the results of the subdivision method, and the right column shows the results of the proposed method. Each row corresponds to a different mirror shape: (a), (b) concave mirror ($f\approx\rm{66.7~mm}$), (c), (d) concave mirror ($f\approx\rm{133.4~mm}$), (e), (f) planar mirror, (g), (h) convex mirror ($f\approx\rm{133.4~mm}$), and (i), (j) convex mirror ($f\approx\rm{66.7~mm}$). A visual comparison revealed no differences between the simulated images obtained by the two methods, demonstrating that the proposed method is able to calculate mirror images with high accuracy.}
 \label{fig:simulation_parabolic} % 全体のラベル名を変更
\end{figure}

\begin{figure}[H]
 \centering
 \begin{subfigure}{0.42\textwidth}
 \includegraphics[width=\linewidth]{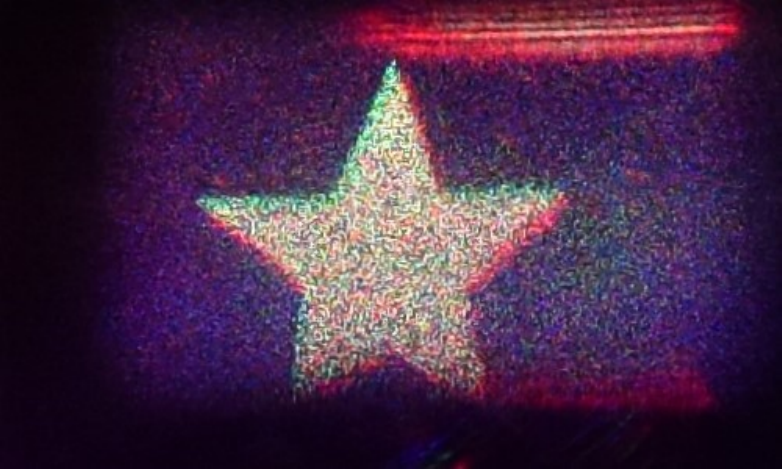}
 \caption{}
 \label{subfig:reconstructed_subdivision_concave}
 \end{subfigure}
 \begin{subfigure}{0.42\textwidth}
 \includegraphics[width=\linewidth]{styles/reconstructed_proposed_concave.pdf}
 \caption{}
 \label{subfig:reconstructed_proposed_concave}
 \end{subfigure}
 \begin{subfigure}{0.42\textwidth}
 \includegraphics[width=\linewidth]{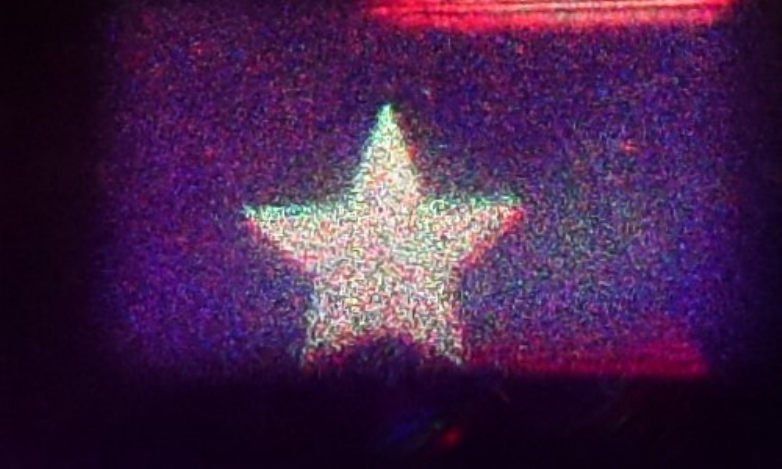}
 \caption{}
 \label{subfig:reconstructed_subdivision_weakly_concave}
 \end{subfigure}
 \begin{subfigure}{0.42\textwidth}
 \includegraphics[width=\linewidth]{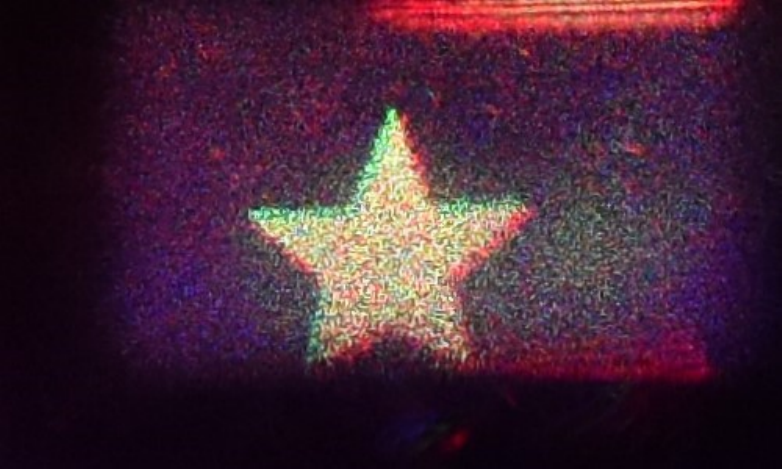}
 \caption{}
 \label{subfig:reconstructed_proposed_weakly_concave}
 \end{subfigure}
 \begin{subfigure}{0.42\textwidth}
 \includegraphics[width=\linewidth]{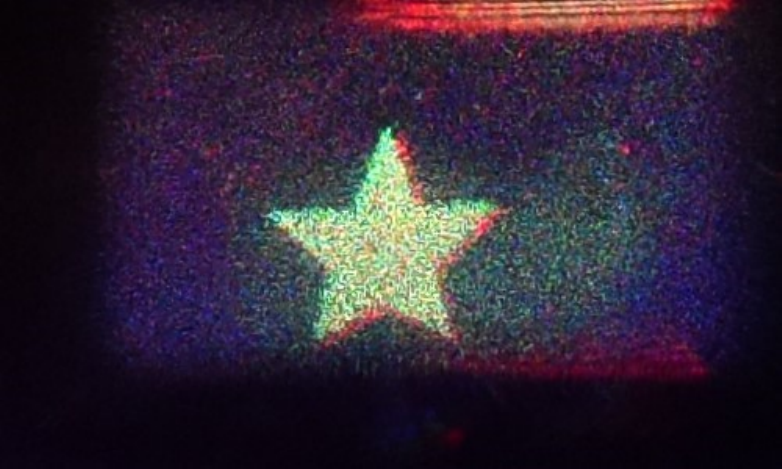}
 \caption{}
 \label{subfig:reconstructed_subdivision_planar}
 \end{subfigure}
 \begin{subfigure}{0.42\textwidth}
 \includegraphics[width=\linewidth]{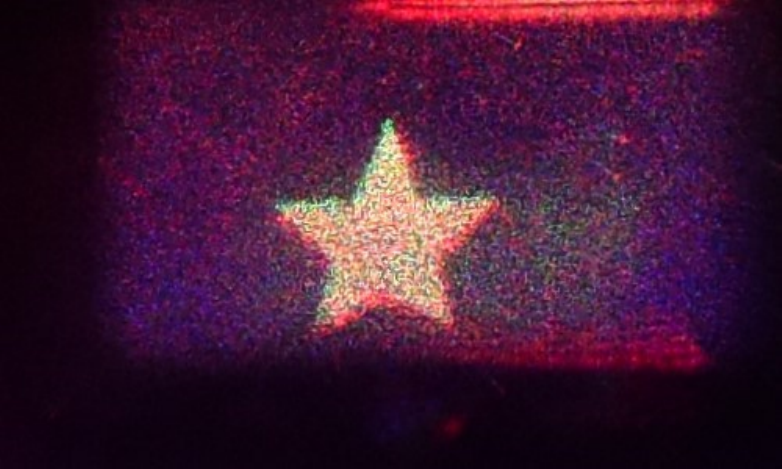}
 \caption{}
 \label{subfig:reconstructed_proposed_planar}
 \end{subfigure}
 \begin{subfigure}{0.42\textwidth}
 \includegraphics[width=\linewidth]{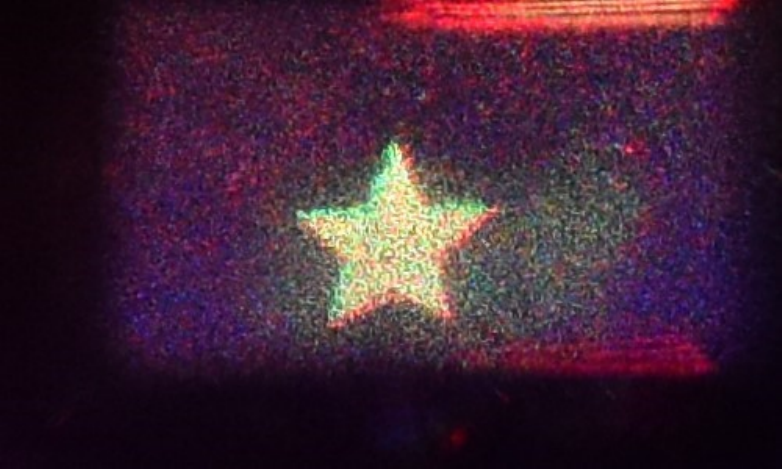}
 \caption{}
 \label{subfig:reconstructed_subdivision_weakly_convex}
 \end{subfigure}
 \begin{subfigure}{0.42\textwidth}
 \includegraphics[width=\linewidth]{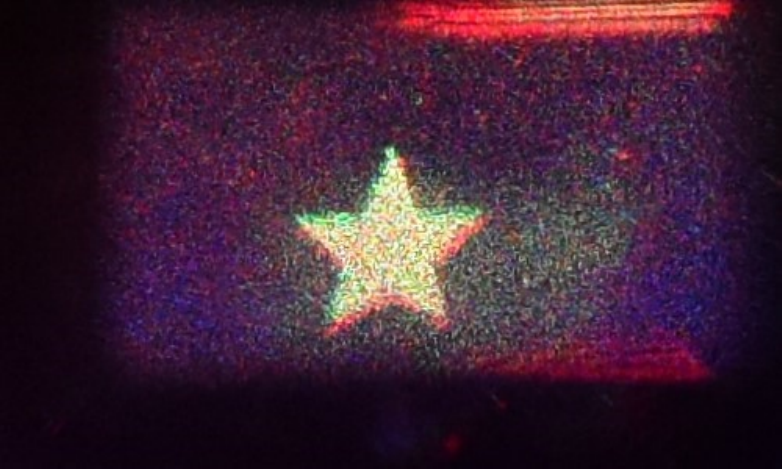}
 \caption{}
 \label{subfig:reconstructed_proposed_weakly_convex}
 \end{subfigure}
 \begin{subfigure}{0.42\textwidth}
 \includegraphics[width=\linewidth]{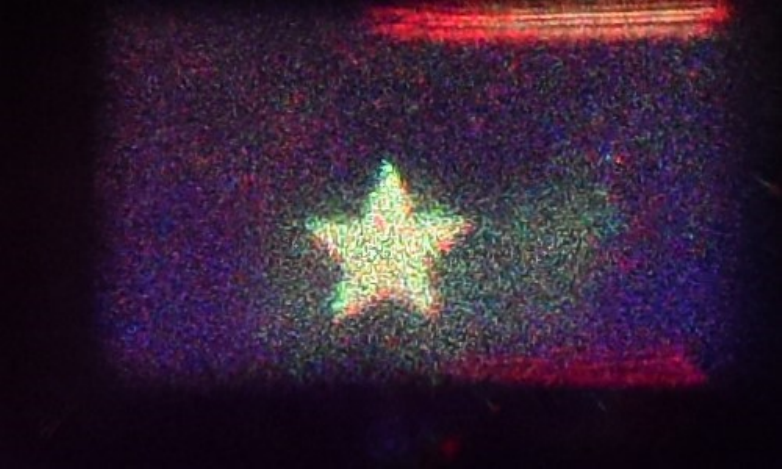}
 \caption{}
 \label{subfig:reconstructed_subdivision_convex}
 \end{subfigure}
 \begin{subfigure}{0.42\textwidth}
 \includegraphics[width=\linewidth]{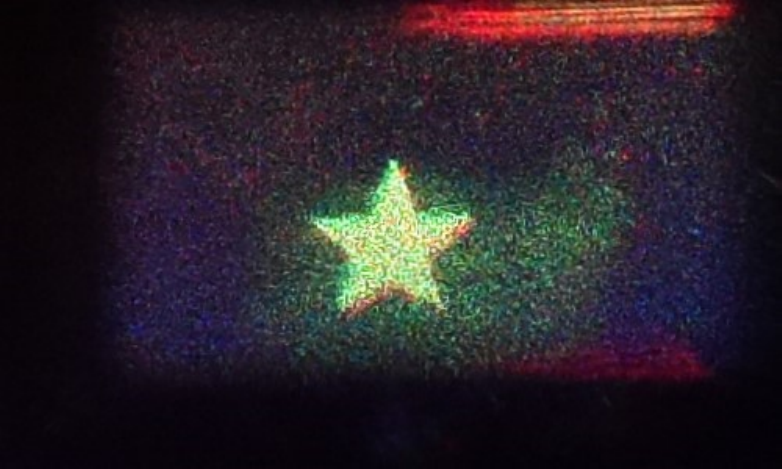}
 \caption{}
 \label{subfig:reconstructed_proposed_convex}
 \end{subfigure}
\caption{Optically reconstructed images of mirror images using parabolic mirrors. The left column shows the results of the subdivision method, and the right column shows the results of the proposed method. Each row corresponds to a different mirror shape: (a), (b) concave mirror ($f\approx\rm{66.7~mm}$), (c), (d) concave mirror ($f\approx\rm{133.4~mm}$), (e), (f) planar mirror, (g), (h) convex mirror ($f\approx\rm{133.4~mm}$), and (i), (j) convex mirror ($f\approx\rm{66.7~mm}$). A visual comparison revealed no differences between the reconstructed images obtained by the two methods, demonstrating that the proposed method is able to calculate mirror images with high accuracy.}
 \label{fig:reconstructed_parabolic} % 全体のラベル名を変更
\end{figure}

Figs. \ref{fig:simulation_parabolic} and \ref{fig:reconstructed_parabolic} show the simulated images and photographs of the optically reconstructed images, respectively. The propagation distance for the simulations was determined using the imaging equation, assuming ideal parabolic mirrors. A visual comparison revealed no discernible differences between the results of the proposed method and the subdivision method for either the simulated or the optically reconstructed images.

For a quantitative analysis, we first calculated the Dice coefficient using the simulated images to evaluate shape similarity. Here, the images generated by the subdivision method were treated as the correct images, against which the results of the proposed method were compared. Table \ref{tab:parabolic_dice} lists the calculated Dice coefficients for the five mirror shapes.

In addition to shape fidelity, we evaluated the depth of the optically reconstructed images. Table \ref{tab:parabolic_depth} compares the measured depths of the images shown in Fig. \ref{fig:reconstructed_parabolic} with their theoretical values calculated from the imaging equation. The resulting error was less than 3 mm for all mirror shapes. This minor discrepancy is likely attributable to shifts in the observer's viewpoint during measurement and the approximation of the ideal parabolic mirror with a Bézier surface.

In summary, no visible differences were observed between the results of the proposed method and the subdivision method in either simulations or optical reconstructions. This visual assessment is quantitatively confirmed by the high Dice coefficients, which all exceeded a value of 0.95, and the small depth errors, which were all less than 3 mm. These results demonstrate that the proposed method can generate high-fidelity mirror images with an accuracy that is highly consistent with the established subdivision method.

\subsection{Computation time}
\label{subsec:calculation_time}
\begin{table}[ht]
 \centering
 \caption{Computational environment.}
 \label{tab:calculation_environment}
 \begin{tabular}{|c|c|}
 \hline
 OS & Windows 10 Pro 64 bit\\ 
 \hline
 CPU & AMD Ryzen 7 5700G with Radeon Graphics 3.80Ghz\\ 
 \hline
 RAM & 32GB\\ 
 \hline
 GPU & NVIDIA GeForce RTX 3090\\ 
 \hline
 OptiX & NVIDIA OptiX 8.0\\ 
 \hline
 \end{tabular}
\end{table}

\begin{figure}[ht]
\centering
\includegraphics[clip,width=10.0cm]{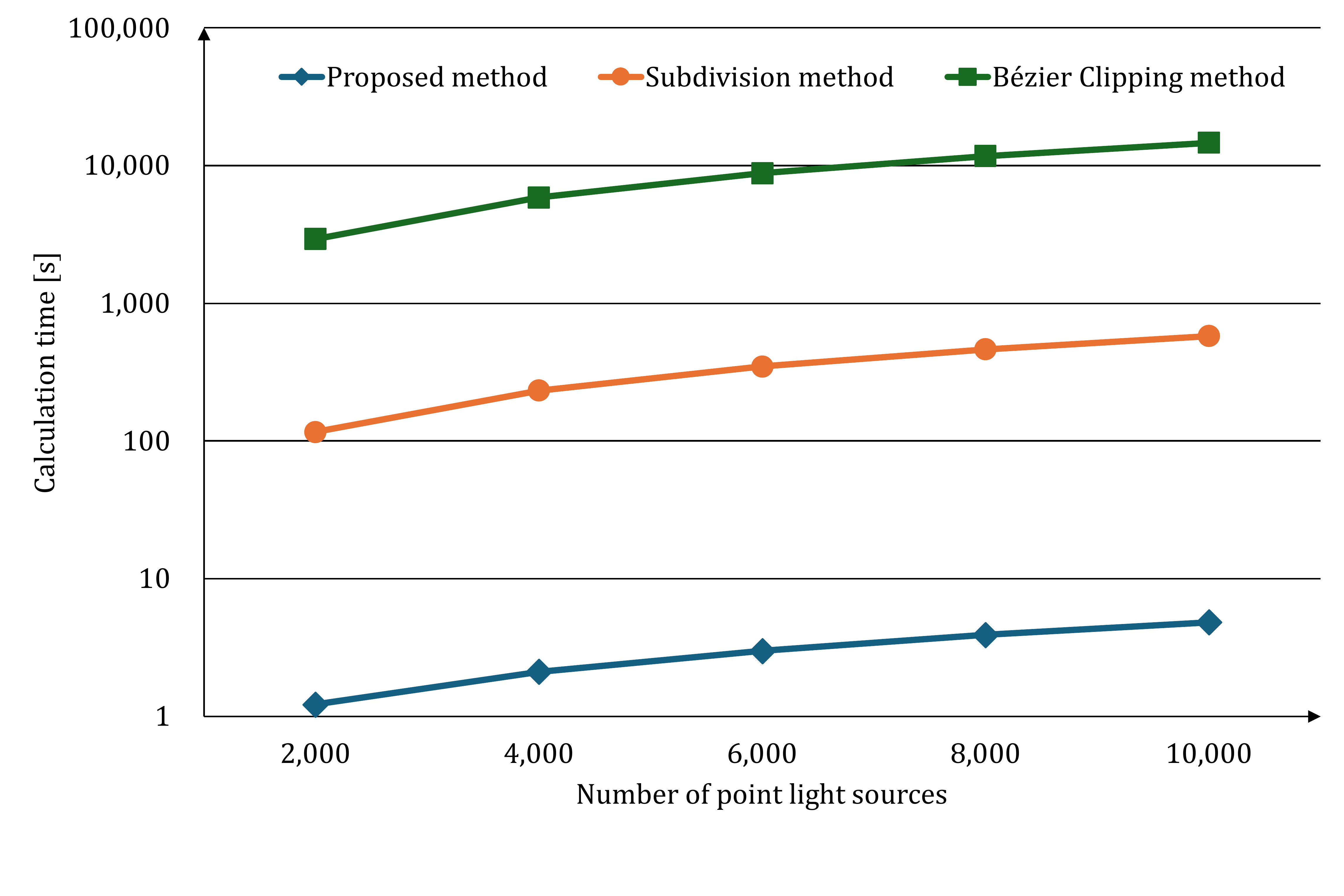}
\caption{Comparison of the computation time for each method. The number of point light sources was varied from 2,000 to 10,000, and the time it took to calculate hologram data for each method was measured (average value of five trials). When the number of point light sources was the maximum of 10,000, the computation time for the proposed method was less than 5 seconds. This is more than 100 times faster than the subdivision method (3,200 divisions). Furthermore, compared to the method used by Arai et al., which required approximately 14,000 seconds under the same conditions, this is more than 2,800 times faster, clearly demonstrating the superiority of the proposed method.}
\label{fig:calculation_time}
\end{figure}

We measured the computation times for the proposed method, the subdivision method, and the Bézier clipping method by Arai et al., using the same scene configuration as in the accuracy validation experiment in Section \ref{subsec:parabolic_mirror}. A concave parabolic mirror ($f \approx \rm{66.7~mm}$) was used for the test. For each method, the number of point light sources on the object's surface was varied from 2,000 to 10,000 in increments of 2,000. The total time from the start of the calculation to the output of the final hologram data was measured, with each reported value being the average of five trials. The details of the computational environment are listed in Table \ref{tab:calculation_environment}, and the results are plotted in Fig. \ref{fig:calculation_time}.

As we can see in Fig. \ref{fig:calculation_time}, even with the maximum of 10,000 point light sources, the computation time for the proposed method remained under 5 seconds. This is over 100 times faster than the subdivision method. Furthermore, compared to the approximately 14,000 seconds required by the method of Arai et al. under identical conditions, the proposed method achieves a speedup of over 2,800 times, and the difference is extremely significant. Furthermore, an analysis of the measurement data reveals that the computation time for the proposed method does not increase in a strictly linear fashion with the number of point light sources. This is attributed to the increased density of the point cloud on the object's surface at higher counts. This higher density enhances the effectiveness of the spatial proximity-based initialization strategy (described in Section \ref{subsec:iterative_search}), which in turn reduces the number of iterations required for Newton's method to converge.

\subsection{Motion parallax}

\begin{figure}[t]
\centering
\includegraphics[clip,width=9.0cm]{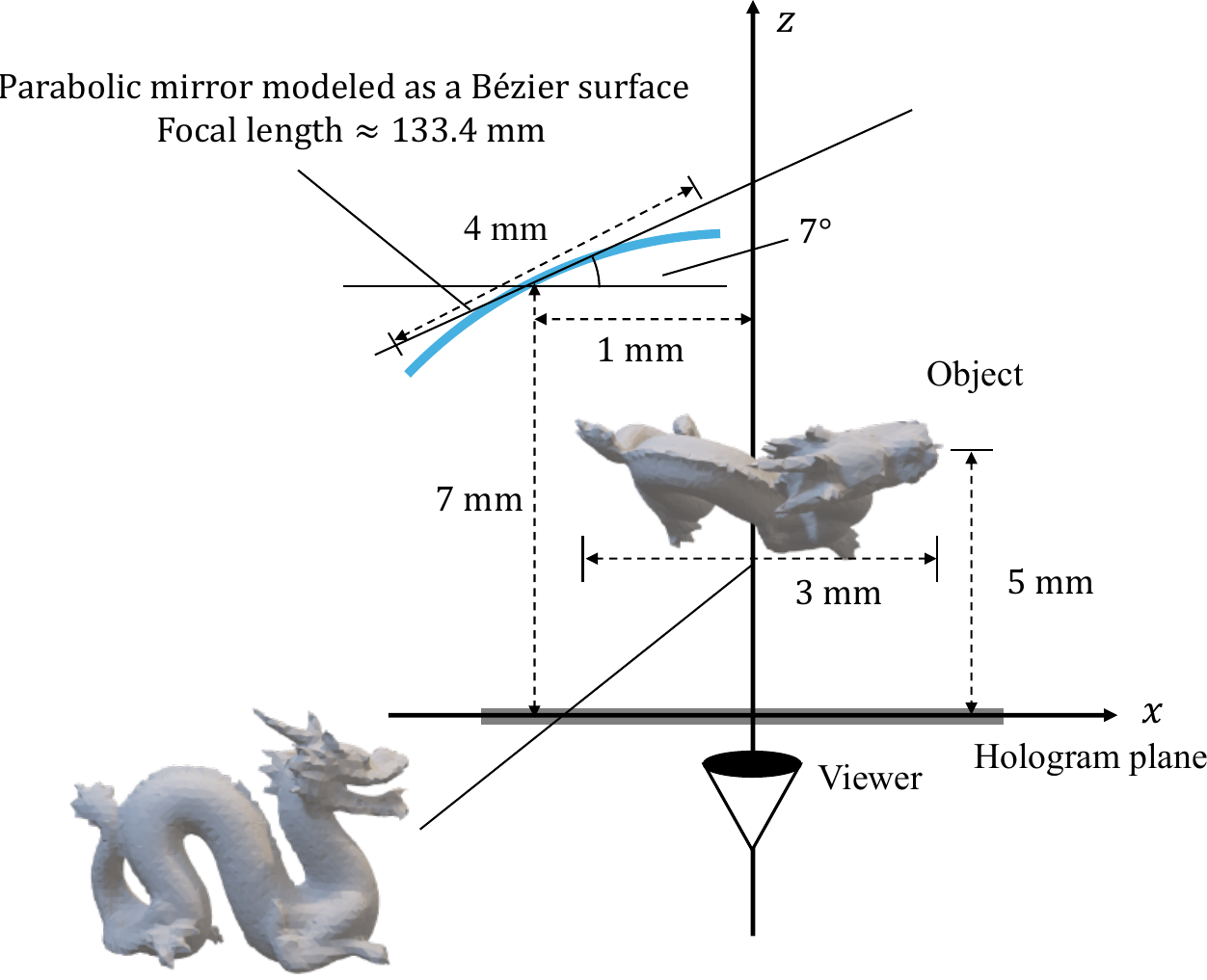}
\caption{Diagram of virtual object placement used in the experiment to confirm motion parallax of mirror images. The virtual object is a 3-mm-wide dragon-shaped object placed at center coordinates (0 mm, 0 mm, 5 mm). The hologram plane is set to $z=0$. The mirror surface is a quadratic Bézier surface ($3\times3$ control points) measuring $\rm{4~mm} \times \rm{4~mm}$ that approximates a concave parabolic mirror with a focal length of approximately 133.4 mm, and its center coordinates are (–1 mm, 0 mm, 7 mm). A frame is placed around the mirror surface, and 30,000 point light sources each are positioned on the dragon-shaped object and the frame.}
\label{fig:placement_motion_parallax}
\end{figure}

\begin{figure}[t]
 \centering
 \begin{subfigure}{0.48\textwidth}
 \includegraphics[width=\linewidth]{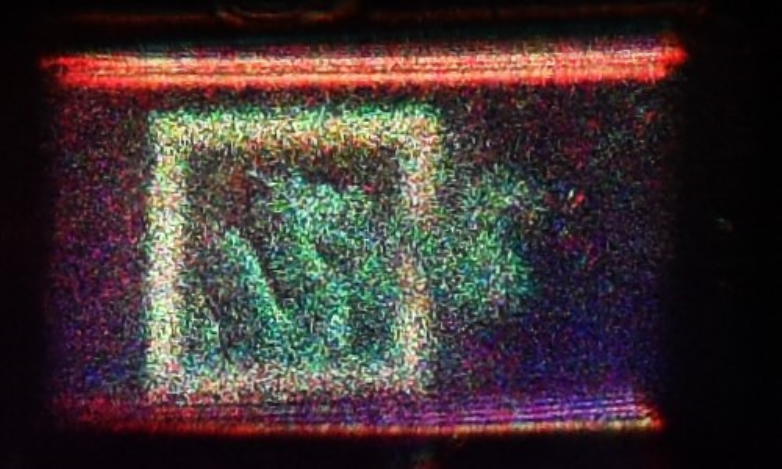}
 \caption{}
 \label{subfig:motion_parallax_left} % ラベル名を修正
 \end{subfigure}
 \hfill % 図と図の間のスペース
 \begin{subfigure}{0.48\textwidth}
 \includegraphics[width=\linewidth]{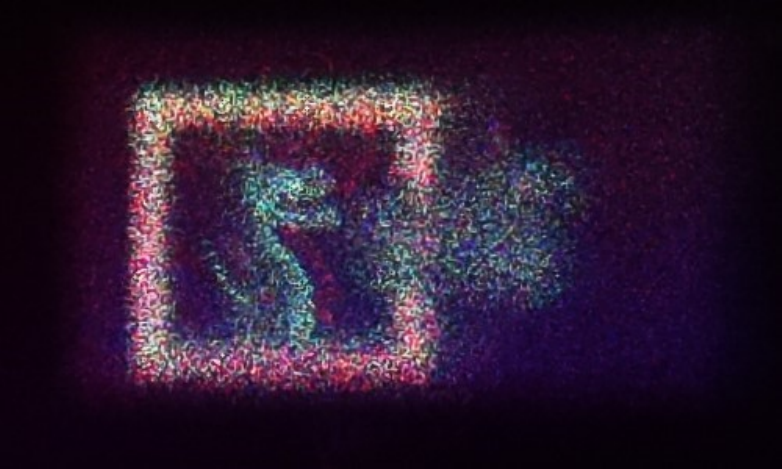}
 \caption{}
 \label{subfig:motion_parallax_right} % ラベル名を修正
 \end{subfigure}
\caption{Optically reconstructed images from the motion parallax experiment, captured from left and right viewpoints. Image (a) is the reconstruction from the left viewpoint, and (b) is from the right. The visible portion of the mirror image, occluded by the frame, changes according to the viewpoint's position. For example, a part of the dragon's torso hidden by the frame in the right view (b) becomes visible in the left view (a). This result confirms the correct reconstruction of motion parallax.}
\label{fig:motion_parallax_left_right}
\end{figure}

In this section, we evaluate the proposed method's ability to accurately reconstruct continuous motion parallax. To this end, we conducted an experiment in which a frame was placed around the mirror, and the change in the visible portion of the mirror image was observed as the viewpoint shifted.

The diagram of virtual object placement for this experiment is shown in Fig. \ref{fig:placement_motion_parallax}. The hologram plane was set to $z = 0$, and a 3-mm-wide dragon-shaped object was placed at its center coordinates (0 mm, 0 mm, 5 mm). The mirror used in the experiment was a Bézier surface approximating a concave parabolic mirror with a focal length of 133.4 mm, measuring $\rm{4~mm} \times \rm{4~mm}$, and was placed at its center coordinates (–1 mm, 0 mm, 7 mm). A total of 30,000 point light sources each were placed on the surfaces of the dragon-shaped object and the frame surrounding the mirror.

Fig. \ref{fig:motion_parallax_left_right} shows the optically reconstructed images captured from two different viewpoints. Comparing the image from the left viewpoint (\subref{subfig:motion_parallax_left}) with the image from the right viewpoint (\subref{subfig:motion_parallax_right}), the visible area of the mirror image, as occluded by the frame, clearly changes with the viewpoint's movement. For example, a portion of the dragon's torso, hidden by the frame in the right-view image (\subref{subfig:motion_parallax_right}), becomes visible within the frame in the left-view image (\subref{subfig:motion_parallax_left}).

These results show that the spatial relationship between the frame and the mirror image changes correctly in response to the observer's movement. This confirms that the proposed method can accurately represent continuous motion parallax.

\begin{figure}[t]
\centering
\includegraphics[clip,width=10.0cm]{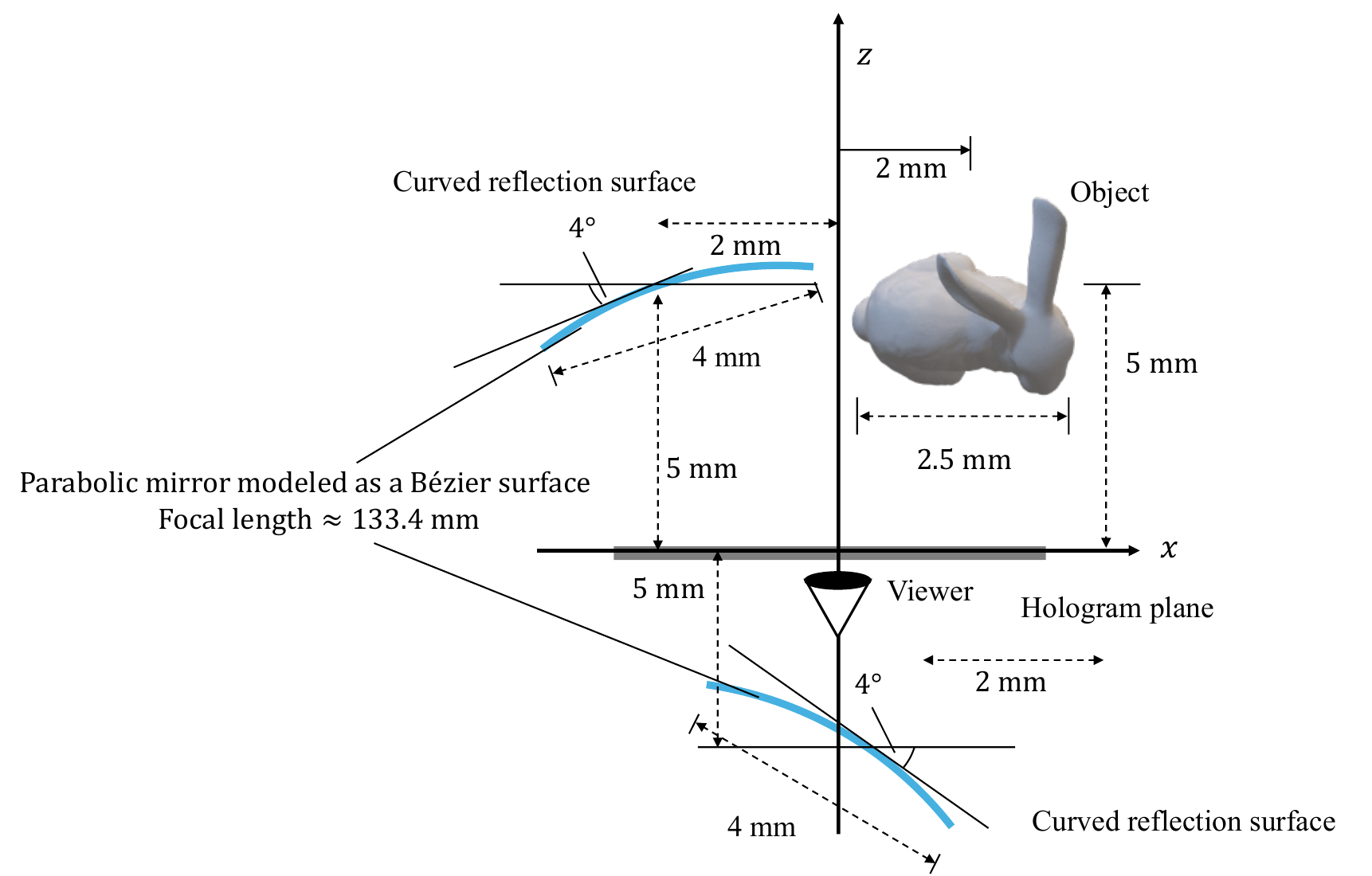}
\caption{Diagram of virtual object placement for the multiple reflection experiment. The virtual object is a 2.5-mm-wide rabbit-shaped object, placed at its center coordinates (2 mm, 0 mm, 5 mm). The hologram plane is set to $z=0$. The scene includes two quadratic Bézier surfaces ($3\times3$ control points) measuring $\rm{4~mm} \times \rm{4~mm}$, each approximating a parabolic mirror with a focal length of 133.4 mm. The first is a convex mirror centered at (0 mm, 0 mm, –5 mm), and the second is a concave mirror centered at (2 mm, 0 mm, 5 mm). A frame is placed around the second mirror, and 30,000 point light sources each are placed on the surfaces of the rabbit object and the frame.}
\label{fig:placement_multiple_reflection}
\end{figure}

\subsection{Multiple reflection}

\begin{figure}[h]
 \centering
 \begin{subfigure}{0.48\textwidth}
 \includegraphics[width=\linewidth]{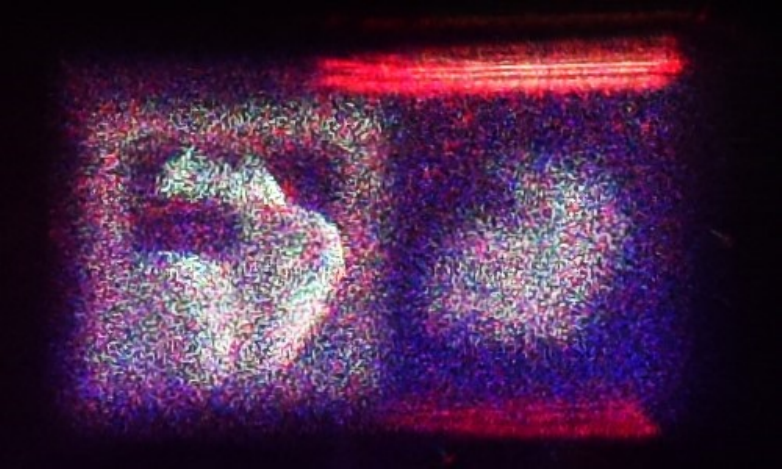}
 \caption{}
 \label{subfig:multiple_reflection_left} % ラベル名を修正
 \end{subfigure}
 \hfill % 図と図の間のスペース
 \begin{subfigure}{0.48\textwidth}
 \includegraphics[width=\linewidth]{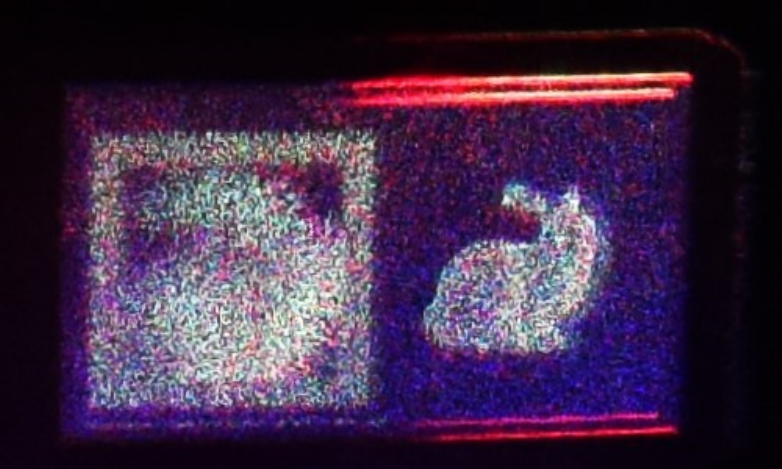}
 \caption{}
 \label{subfig:multiple_reflection_right} % ラベル名を修正
 \end{subfigure}
\caption{Optically reconstructed images from the multiple reflection experiment, captured at different focal depths. Image (a) is focused on the real rabbit object, while image (b) is focused on the secondary-reflection image formed by the two mirrors. The scene is configured such that a single reflection from the foreground mirror would not form an image of the rabbit. Therefore, the image visible in (b) definitively confirms the successful reconstruction of a secondary reflection, where light from the rabbit reflects first off the background convex mirror and then again off the foreground concave mirror.}
\label{fig:multiple_reflection_left_right}
\end{figure}

\begin{figure}[t]
\centering
\includegraphics[clip,width=10.0cm]{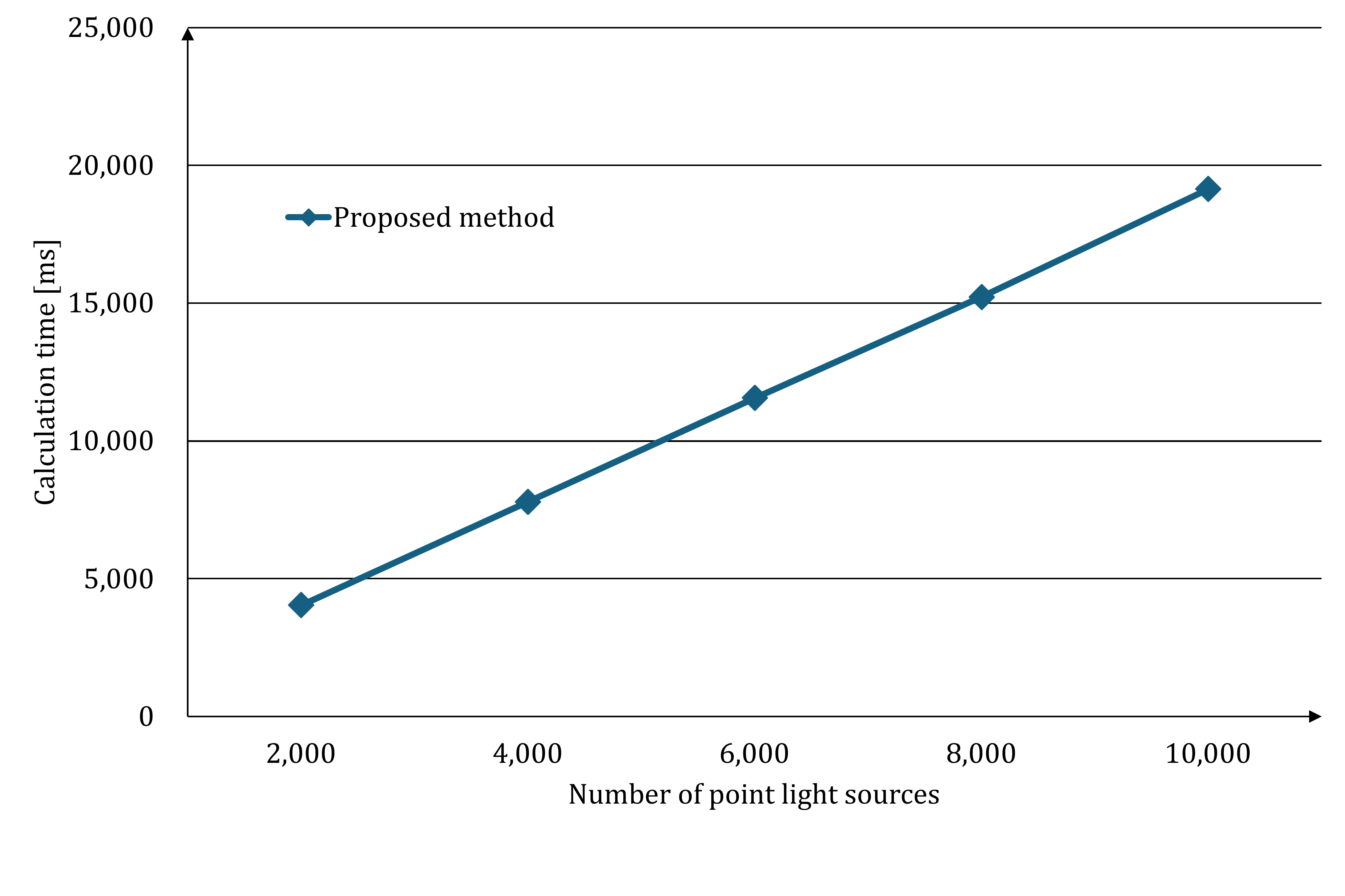}
\caption{Computation time for the multiple reflection calculation. The time required to compute the hologram data was measured as the number of point light sources was varied from 2,000 to 10,000 (values are an average of five trials). The measurement exclusively targeted the computation of the secondary-reflection image. At the maximum of 10,000 point light sources, the computation time was approximately 19 seconds. This is roughly four times the computation time for a single reflection measured in Section \ref{subsec:calculation_time} (with the same number of points), indicating a moderate increase in computation time.}
\label{fig:calculation_time_multi}
\end{figure}

In this section, we verify the scalability of the proposed method, as discussed in Section \ref{subsec:overall_algorithm}. We conducted an experiment to evaluate if the method can represent secondary reflections in a two-mirror system and if the corresponding computation time remains within a practical range.

The diagram of virtual object placement for this experiment is shown in Fig. \ref{fig:placement_multiple_reflection}. The hologram plane was set to $z = 0$, and a 2.5-mm-wide rabbit-shaped object was placed at its center coordinates (2 mm, 0 mm, 5 mm). Both mirrors used in the experiment were curved parabolic mirrors with dimensions of $\rm{4~mm} \times \rm{4~mm}$ and a focal length of 133.4 mm. The first was a convex mirror placed at its center coordinates (0 mm, 0 mm, –5 mm), and the second was a concave mirror placed at its center coordinates (2 mm, 0 mm, 5 mm). To improve the visibility of the reconstructed image, a frame was placed around the second mirror. A total of 30,000 point light sources each were placed on the surfaces of the rabbit-shaped object and the frame.

Fig. \ref{fig:multiple_reflection_left_right} shows the optically reconstructed image from this experimental setup, where we can see that the photograph clearly captures the intended secondary reflection. The image of the rabbit is first reflected by the convex mirror located in the background, and is then seen reflected again in the concave mirror positioned in the foreground. This result demonstrates that the proposed method can accurately represent multiple reflections.

Next, we measured the computation time for this secondary-reflection image. The measurement focused solely on the computation of the secondary-reflection image itself, excluding the calculations for the direct view of the rabbit object and the frame. The number of point light sources on the rabbit object was varied from 2,000 to 10,000 in 2,000-point increments. The results are shown in Fig. \ref{fig:calculation_time_multi}. For 10,000 point light sources, the computation time was approximately 19 seconds. Compared to the time for a single reflection with the same number of points (approx. 5 seconds), as measured in Section \ref{subsec:calculation_time}, doubling the reflections increased the computation time roughly fourfold. This result indicates that the computation time does not increase exponentially as reflections are added, but instead scales moderately and remains within a practical range.

These results confirm that the proposed method can accurately represent multiple reflections between several mirrors. Furthermore, the increase in computation time with each additional reflection is moderate, remaining well within a practical range.

\section{Conclusion}

In this study, we proposed a fast and accurate method for calculating mirror images from Bézier-defined surfaces in CGH calculations using the point-based method. The proposed method determines the reflection point by minimizing the optical path length—connecting a point light source on an object, a reflection point on the mirror surface, and a hologram pixel—using Newton's method.

Experimental evaluations demonstrated the proposed method's significant advantages in computation speed. While maintaining comparable accuracy, it achieved a speedup of over 100 times compared to the conventional mirror surface subdivision method and over 2,800 times compared to the Bézier clipping method of Arai et al. Furthermore, we confirmed that setting the convergence tolerance to the Rayleigh limit ($\frac{\lambda}{4}$) establishes an optimal balance between reconstruction quality and computational efficiency.

Optical experiments further confirmed that the shape, depth, and motion parallax of the generated mirror images are physically accurate. We also verified the method's scalability to multiple reflections. The computation time for secondary reflections was approximately four times that of a single reflection, indicating that the computational cost increases only moderately with the number of reflections.

These results demonstrate that the proposed method offers a comprehensive solution for rendering reflections from free-form mirrors in CGH, effectively addressing the challenges of accuracy, speed, and scalability.

We identify two primary directions for future work based on the results of this research.

The first is the extension to represent refraction. While this research focused on reflection, Fermat's principle of minimizing optical path length is equally applicable to refraction, where light passes through a transparent object like a lens. By extending the proposed optical path length function to handle refraction, it should be possible to open up new application areas, such as the high-speed computation of CGH for free-form lenses.

A second avenue is further optimization toward real-time processing. The computational speed achieved in this study enables applications not only in static scenes but also in dynamic situations where objects and mirrors move or deform. A significant and important future challenge is to apply the proposed method to the creation of interactive systems where holographic mirror images change without delay in response to user input.

\section*{Back matter}

\subsection*{Acknowledgment}
The authors would like to thank Professor Yoshinori Dobashi of the Graduate School of Information Science and Technology, Hokkaido University, for his invaluable guidance and insightful discussions throughout this research and the preparation of this manuscript.

These research results were obtained from the commissioned research(No.PJ012368C06801) by National Institute
of Information and Communications Technology (NICT) , Japan.

\subsection*{Disclosures}
There are no potential conflicts of interest, financial or otherwise, identified for this study.

\subsection*{Data availability} Data underlying the results presented in this paper are not publicly available at this time but may be obtained from the authors upon reasonable request.

\bibliographystyle{unsrt} % 引用順に並べる（工学系で一般的）
\bibliography{sample}

@article{gabor1948new,
  title={A new microscopic principle},
  author={Gabor, Dennis},
  journal={Nature},
  Volume={161},
  number={4098},
  pages={777--778},
  year={1948}
}

@article{waters1966holographic,
  title={Holographic image synthesis utilizing theoretical methods},
  author={Waters, James P},
  journal={Applied physics letters},
  volume={9},
  number={11},
  pages={405--407},
  year={1966},
  publisher={AIP Publishing}
}

@book{catmull1974subdivision,
  title={A subdivision algorithm for computer display of curved surfaces},
  author={Catmull, Edwin Earl},
  year={1974},
  publisher={The University of Utah}
}

@article{10.1145/360825.360839,
author = {Phong, Bui Tuong},
title = {Illumination for computer generated pictures},
year = {1975},
issue_date = {June 1975},
publisher = {Association for Computing Machinery},
address = {New York, NY, USA},
volume = {18},
number = {6},
issn = {0001-0782},
url = {https://doi.org/10.1145/360825.360839},
doi = {10.1145/360825.360839},
journal = {Commun. ACM},
month = jun,
pages = {311–317},
numpages = {7}
}

@inproceedings{dodik2022path,
  title={Path Guiding Using Spatio-Directional Mixture Models},
  author={Dodik, Ana and Papas, Marios and {\"O}ztireli, Cengiz and M{\"u}ller, Thomas},
  booktitle={Computer Graphics Forum},
  volume={41},
  pages={172--189},
  year={2022},
  organization={Wiley Online Library}
}

@article{koskela2019blockwise,
  title={Blockwise multi-order feature regression for real-time path-tracing reconstruction},
  author={Koskela, Matias and Immonen, Kalle and M{\"a}kitalo, Markku and Foi, Alessandro and Viitanen, Timo and J{\"a}{\"a}skel{\"a}inen, Pekka and Kultala, Heikki and Takala, Jarmo},
  journal={ACM Transactions on Graphics (TOG)},
  volume={38},
  number={5},
  pages={1--14},
  year={2019},
  publisher={ACM New York, NY, USA}
}

@article{10.1145/3503250,
author = {Mildenhall, Ben and Srinivasan, Pratul P. and Tancik, Matthew and Barron, Jonathan T. and Ramamoorthi, Ravi and Ng, Ren},
title = {NeRF: representing scenes as neural radiance fields for view synthesis},
year = {2021},
issue_date = {January 2022},
publisher = {Association for Computing Machinery},
address = {New York, NY, USA},
volume = {65},
number = {1},
issn = {0001-0782},
url = {https://doi.org/10.1145/3503250},
doi = {10.1145/3503250},
journal = {Commun. ACM},
month = dec,
pages = {99–106},
numpages = {8}
}

@article{HosseinEybposh:20,
author = {M. Hossein Eybposh and Nicholas W. Caira and Mathew Atisa and Praneeth Chakravarthula and Nicolas C. P\'{e}gard},
journal = {Opt. Express},
number = {18},
pages = {26636--26650},
publisher = {Optica Publishing Group},
title = {DeepCGH: 3D computer-generated holography using deep learning},
volume = {28},
month = {Aug},
year = {2020},
url = {https://opg.optica.org/oe/abstract.cfm?URI=oe-28-18-26636},
doi = {10.1364/OE.399624},
}

@article{Yamaguchi:09,
author = {Kazuhiro Yamaguchi and Yuji Sakamoto},
journal = {Appl. Opt.},
number = {34},
pages = {H203--H211},
publisher = {Optica Publishing Group},
title = {Computer generated hologram with characteristics of reflection: reflectance distributions and reflected images},
volume = {48},
month = {Dec},
year = {2009},
url = {https://opg.optica.org/ao/abstract.cfm?URI=ao-48-34-H203},
doi = {10.1364/AO.48.00H203},
}

@article{Nishi:25,
author = {Hirohito Nishi and Kyoji Matsushima},
journal = {Opt. Express},
number = {1},
pages = {704--716},
publisher = {Optica Publishing Group},
title = {Rendering of transparent objects in large-scale full-parallax polygon-based computer holography},
volume = {33},
month = {Jan},
year = {2025},
url = {https://opg.optica.org/oe/abstract.cfm?URI=oe-33-1-704},
doi = {10.1364/OE.541610},
}

@article{Ichikawa:13,
author = {Tsubasa Ichikawa and Kazuhiro Yamaguchi and Yuji Sakamoto},
journal = {Appl. Opt.},
number = {1},
pages = {A201--A209},
publisher = {Optica Publishing Group},
title = {Realistic expression for full-parallax computer-generated holograms with the ray-tracing method},
volume = {52},
month = {Jan},
year = {2013},
url = {https://opg.optica.org/ao/abstract.cfm?URI=ao-52-1-A201},
doi = {10.1364/AO.52.00A201},
}

@article{Watanabe:24,
author = {Keita Watanabe and Keigo Yamauchi and Yuji Sakamoto},
journal = {Appl. Opt.},
number = {7},
pages = {B126--B133},
publisher = {Optica Publishing Group},
title = {Realistic rendering method for specular reflections with continuous motion parallax in a computer-generated hologram},
volume = {63},
month = {Mar},
year = {2024},
url = {https://opg.optica.org/ao/abstract.cfm?URI=ao-63-7-B126},
doi = {10.1364/AO.506341},
}

@inproceedings{10.1145/1198555.1198743,
author = {Whitted, Turner},
title = {An improved illumination model for shaded display},
year = {2005},
isbn = {9781450378338},
publisher = {Association for Computing Machinery},
address = {New York, NY, USA},
url = {https://doi.org/10.1145/1198555.1198743},
doi = {10.1145/1198555.1198743},
booktitle = {ACM SIGGRAPH 2005 Courses},
pages = {4–es},
location = {Los Angeles, California},
series = {SIGGRAPH '05}
}

@article{10.1145/97880.97916,
author = {Nishita, Tomoyuki and Sederberg, Thomas W. and Kakimoto, Masanori},
title = {Ray tracing trimmed rational surface patches},
year = {1990},
issue_date = {Aug. 1990},
publisher = {Association for Computing Machinery},
address = {New York, NY, USA},
volume = {24},
number = {4},
issn = {0097-8930},
url = {https://doi.org/10.1145/97880.97916},
doi = {10.1145/97880.97916},
journal = {SIGGRAPH Comput. Graph.},
month = sep,
pages = {337–345},
numpages = {9}
}

@article{Arai:24,
author = {Hiroya Arai and Kodai Ono and Yuji Sakamoto},
journal = {Opt. Express},
number = {21},
pages = {36469--36488},
publisher = {Optica Publishing Group},
title = {CGH calculation algorithm for expressing reflection on a curved mirror surface},
volume = {32},
month = {Oct},
year = {2024},
url = {https://opg.optica.org/oe/abstract.cfm?URI=oe-32-21-36469},
doi = {10.1364/OE.532849},
}

@article{Lee:14,
author = {Yoon-Hyuk Lee and Young-Ho Seo and Ji-Sang Yoo and Dong-Wook Kim},
journal = {J. Opt. Soc. Korea},
number = {6},
pages = {698--705},
publisher = {Optica Publishing Group},
title = {High-Performance Computer-Generated Hologram by Optimized Implementation of Parallel GPGPUs},
volume = {18},
month = {Dec},
year = {2014},
url = {https://opg.optica.org/josk/abstract.cfm?URI=josk-18-6-698},
}

@article{10.1117/1.OE.64.7.075102,
author = {Kodai Ono and Seok Kang and Yuji Sakamoto},
title = {{Fast computer-generated hologram calculation algorithm for mirror images reflected on mirror surface of Bézier surfaces using subdivision}},
volume = {64},
journal = {Optical Engineering},
number = {7},
publisher = {SPIE},
pages = {075102},
year = {2025},
doi = {10.1117/1.OE.64.7.075102},
URL = {https://doi.org/10.1117/1.OE.64.7.075102}
}

@book{goodman2005introduction,
  title={Introduction to Fourier optics},
  author={Goodman, Joseph W},
  year={2005},
  publisher={Roberts and Company publishers}
}

@book{Born_Wolf_Bhatia_Clemmow_Gabor_Stokes_Taylor_Wayman_Wilcock_1999,
  place={Cambridge}, 
  edition={7}, 
  title={Principles of Optics: Electromagnetic Theory of Propagation, Interference and Diffraction of Light},
  publisher={Cambridge University Press},
  author={Born, Max and Wolf, Emil and Bhatia, A. B. and Clemmow, P. C. and Gabor, D. and Stokes, A. R. and Taylor, A. M. and Wayman, P. A. and Wilcock, W. L.},
  year={1999}
}

@book{nocedal2006numerical,
  title={Numerical Optimization},
  author={Nocedal, J. and Wright, S.},
  isbn={9780387400655},
  lccn={2006923897},
  series={Springer Series in Operations Research and Financial Engineering},
  url={https://books.google.co.jp/books?id=VbHYoSyelFcC},
  year={2006},
  publisher={Springer New York}
}

@article{Vorontsov:98,
author = {M. A. Vorontsov and V. P. Sivokon},
journal = {J. Opt. Soc. Am. A},
number = {10},
pages = {2745--2758},
publisher = {Optica Publishing Group},
title = {Stochastic parallel-gradient-descent technique for high-resolution wave-front phase-distortion correction},
volume = {15},
month = {Oct},
year = {1998},
url = {https://opg.optica.org/josaa/abstract.cfm?URI=josaa-15-10-2745},
doi = {10.1364/JOSAA.15.002745},
}

@article{Xu:25,
author = {Yi Xu and Fu Li and Jianqiang Gu and Quan Xu and Zhen Tian and Jiaguang Han and Weili Zhang},
journal = {Chin. Opt. Lett.},
number = {8},
pages = {083601},
publisher = {Optica Publishing Group},
title = {Gradient-descent optimization of metasurfaces based on one deep-enhanced RseNet},
volume = {23},
month = {Aug},
year = {2025},
url = {https://opg.optica.org/col/abstract.cfm?URI=col-23-8-083601},
}

@article{10.1145/1778765.1778803,
author = {Parker, Steven G. and Bigler, James and Dietrich, Andreas and Friedrich, Heiko and Hoberock, Jared and Luebke, David and McAllister, David and McGuire, Morgan and Morley, Keith and Robison, Austin and Stich, Martin},
title = {OptiX: a general purpose ray tracing engine},
year = {2010},
issue_date = {July 2010},
publisher = {Association for Computing Machinery},
address = {New York, NY, USA},
volume = {29},
number = {4},
issn = {0730-0301},
url = {https://doi.org/10.1145/1778765.1778803},
doi = {10.1145/1778765.1778803},
journal = {ACM Trans. Graph.},
month = jul,
articleno = {66},
numpages = {13}
}

@article{doi:10.2352/ISSN.2470-1173.2017.3.ERVR-095,
author = {Dylan McCarthy and J¨ Urgen P. Schulze},
title = {Distributed VR Rendering Using NVIDIA OptiX},
journal = {Electronic Imaging},
volume = {29},
number = {3},
pages = {36--36},
doi = {10.2352/ISSN.2470-1173.2017.3.ERVR-095},
url = {https://library.imaging.org/ei/articles/29/3/art00007},
year = {2017}, 
}

@inproceedings{10.1117/12.2609369,
author = {Zehao He and Kexuan Liu and Xiaomeng Sui and Liangcai Cao},
title = {{Angular-spectrum algorithm for holographic 3D display based on 2D-to-3D approach}},
volume = {12025},
booktitle = {Ultra-High-Definition Imaging Systems V},
editor = {Seizo Miyata and Toyohiko Yatagai and Yasuhiro Koike},
organization = {International Society for Optics and Photonics},
publisher = {SPIE},
pages = {1202509},
year = {2022},
doi = {10.1117/12.2609369},
URL = {https://doi.org/10.1117/12.2609369}
}

@article{Leith:62,
author = {Emmett N. Leith and Juris Upatnieks},
journal = {J. Opt. Soc. Am.},
number = {10},
pages = {1123--1130},
publisher = {Optica Publishing Group},
title = {Reconstructed Wavefronts and Communication Theory$\ast$},
volume = {52},
month = {Oct},
year = {1962},
url = {https://opg.optica.org/abstract.cfm?URI=josa-52-10-1123},
doi = {10.1364/JOSA.52.001123},
}

@article{ed278621-dc3e-343f-ae66-540d8990b60d,
 ISSN = {00129658, 19399170},
 URL = {http://www.jstor.org/stable/1932409},
 author = {Lee R. Dice},
 journal = {Ecology},
 number = {3},
 pages = {297--302},
 publisher = {[Wiley, Ecological Society of America]},
 title = {Measures of the Amount of Ecologic Association Between Species},
 urldate = {2025-08-28},
 volume = {26},
 year = {1945}
}

\end{document}